\documentstyle[12pt,epsfig,psfig]{article}
\newcommand {\ignore}[1]{}

\newcommand{\bc}{\begin{center}}
\newcommand{\ec}{\end{center}}

\def\ifmath#1{\relax\ifmmode #1\else $#1$\fi}

\def\3quarter{{\textstyle{3 \over 4}}}

\def\ra{\rightarrow}

\def\lf{\leaders\hbox to 1em{\hss.\hss}\hfill}
\def\21{$SU(2) \ot U(1)$}
\def\321{$SU(3) \ot SU(2) \ot U(1)$}
\def\ne{\hbox{$\nu_e$ }}
\def\nm{\hbox{$\nu_\mu$ }}
\def\nt{\hbox{$\nu_\tau$ }}
\def\ns{\hbox{$\nu_{s}$ }}

\def\O{\hbox{$\cal O$ }}

\def\mne{\hbox{$m_{\nu_e}$ }}
\def\mnm{\hbox{$m_{\nu_\mu}$ }}
\def\mnt{\hbox{$m_{\nu_\tau}$ }}

\def\ie{\hbox{\it i.e., }}        \def\etc{\hbox{\it etc. }}
\def\eg{\hbox{\it e.g., }}        
\def\etal{\hbox{\it et al., }}
\def\neus{\hbox{neutrinos }}
\def\gau{\hbox{gauge }}
\def\neu{\hbox{neutrino }}
\def\c{\mathop{\cos \theta }}
\def\s{\mathop{\sin \theta }}

\def\eq#1{{eq. (\ref{#1})}}
\def\Eq#1{{Eq. (\ref{#1})}}

\def\fig#1{{Fig. (\ref{#1})}}

\def\VEV#1{\left\langle #1\right\rangle}

\def\abs#1{\left| #1\right|}

\def\ltap{\raisebox{-.4ex}{\rlap{$\sim$}} \raisebox{.4ex}{$<$}}
\def\gtap{\raisebox{-.4ex}{\rlap{$\sim$}} \raisebox{.4ex}{$>$}}
\def\lsim{\raise0.3ex\hbox{$\;<$\kern-0.75em\raise-1.1ex\hbox{$\sim\;$}}}
\def\gsim{\raise0.3ex\hbox{$\;>$\kern-0.75em\raise-1.1ex\hbox{$\sim\;$}}}

\def\beq{\begin{equation}}
\def\eeq{\end{equation}}
\def\bef{\begin{figure}}
\def\eef{\end{figure}}
\def\bet{\begin{table}}
\def\eet{\end{table}}
\def\bea{\begin{eqnarray}}
\def\ba{\begin{array}}
\def\ea{\end{array}}
\def\bi{\begin{itemize}}
\def\ei{\end{itemize}}
\def\ben{\begin{enumerate}}
\def\een{\end{enumerate}}
\def\ra{\rightarrow}
\def\ot{\otimes}
\def\eea{\end{eqnarray}}
\def\apj#1#2#3{          {\it Astrophys. J. }{\bf #1} (19#2) #3}

\def\ib#1#2#3{           {\it ibid. }{\bf #1} (19#2) #3}
\def\nat#1#2#3{          {\it Nature }{\bf #1} (19#2) #3}
\def\nps#1#2#3{        {\it Nucl. Phys. B (Proc. Suppl.) }{\bf #1} (19#2) #3} 
\def\np#1#2#3{           {\it Nucl. Phys. }{\bf #1} (19#2) #3}
\def\pl#1#2#3{           {\it Phys. Lett. }{\bf #1} (19#2) #3}
\def\pr#1#2#3{           {\it Phys. Rev. }{\bf #1} (19#2) #3}
\def\prep#1#2#3{         {\it Phys. Rep. }{\bf #1} (19#2) #3}
\def\prl#1#2#3{          {\it Phys. Rev. Lett. }{\bf #1} (19#2) #3}

\def\ptp#1#2#3{          {\it Prog. Theor. Phys. }{\bf #1} (19#2) #3}

\def\rmp#1#2#3{          {\it Rev. Mod. Phys. }{\bf #1} (19#2) #3}
\def\zp#1#2#3{           {\it Zeit. fur Physik }{\bf #1} (19#2) #3}

\def\Sci#1#2#3{          {\it Science }{\bf #1} (19#2) #3}
\def\n.c.#1#2#3{         {\it Nuovo Cim. }{\bf #1} (19#2) #3}
\def\r.n.c.#1#2#3{       {\it Riv. del Nuovo Cim. }{\bf #1} (19#2) #3}
\def\sjnp#1#2#3{         {\it Sov. J. Nucl. Phys. }{\bf #1} (19#2) #3}

\def\mpl#1#2#3{          {\it Mod. Phys. Lett. }{\bf #1} (19#2) #3}

\def\ppnp#1#2#3{           {\it Prog. Part. Nucl. Phys. }{\bf #1} (19#2) #3}

\begin{document}
\newcommand{\dis}{\displaystyle}
\begin{titlepage}
\begin{center}
\hfill hep-ph/9603307\\
\ \\
{\Large \bf Physics Beyond the Standard Model
\footnote{These lectures were given at the {\sl VIII 
Jorge Andre Swieca Summer School} (Rio de Janeiro, February 1995) 
and at the {\sl V Taller Latinoamericano de Fenomenologia de las 
Interacciones Fundamentales} (Puebla, Mexico, October 1995).
For convenience the material of the lectures has been
reorganized and updated.}
}\\[0.42cm]
{\bf \large J. W. F. Valle}
\footnote{E-mail: valle@flamenco.uv.es, URL http://neutrinos.uv.es}
\\[0.4333cm]
{\it Instituto de F\'{\i}sica Corpuscular - C.S.I.C.\\
Departament de F\'{\i}sica Te\`orica, Universitat de Val\`encia\\
46100 Burjassot, Val\`encia, SPAIN}\\[1cm]
{\large \bf Abstract}
\begin{quotation}
\noindent
We discuss some of the signatures associated with 
extensions of the Standard Model related to the neutrino and 
electroweak symmetry breaking sectors, with and without 
supersymmetry. The topics include a basic discussion of the 
theory of \neu mass and the corresponding extensions of the 
Standard Model that incorporate massive neutrinos; 
an overview of the present observational status of neutrino 
mass searches, with emphasis on solar neutrinos, as well the 
as cosmological data on the amplitude of primordial density 
fluctuations; the implications of neutrino mass in cosmological 
nucleosynthesis, non-accelerator, as well as in high energy 
particle collider experiments. 
Turning to the electroweak breaking sector, we discuss the
physics potential for Higgs boson searches at LEP200, 
including Majoron extensions of the Standard Model, and the 
physics of invisibly decaying Higgs bosons. We discuss the
minimal supersymmetric Standard Model phenomenology, as well
as some of the laboratory signatures that would be associated 
to models with R parity violation, especially in Z and scalar 
boson decays.

\end{quotation}
\end{center}

\end{titlepage}
\tableofcontents
\newpage
\setcounter{page}{1}
 
\section{Introduction}
\label{intro}

There is a wide consensus that, although very successful wherever 
it has been tested, our present Standard Model leaves open
too many fundamental issues in particle physics to be an ultimate 
theory of nature. These lectures will focus on two of these that 
have to do with the understanding of what lies behind the mechanism 
of mass generation in general, and with the masses and the properties 
of neutrinos in particular.

A basic assumption of the Standard Model is the Higgs 
mechanism, which is introduced as a way to generate the
masses of all the fundamental particles. This mechanism
implies the existence of a fundamental scalar boson \cite{HIGGS}. 
If such an elementary boson exists it is widely believed 
that some stabilising principle - e.g. supersymmetry (SUSY) - 
should be operative at the electroweak scale in order to 
explain the stability of its mass scale against quantum 
corrections associated with physics at super-high energies. 
The observed joining of the three \gau coupling strengths 
as they are evolved from the presently accessible energies 
up to a common scale of $\sim 10^{16}$ GeV provides 
circumstantial evidence that supersymmetry seems to set 
in somewhere around $M_{SUSY} \sim 10^3$ GeV. Probing the
details of this rich structure constitutes one of the main 
goals in the agenda of the next generation of elementary
particle colliders.

Another fundamental issue in the Standard Model refers to the 
properties of neutrinos. Apart from being a theoretical puzzle, 
in the sense that there is no principle that dictates that neutrinos 
are massless, as postulated in the Standard Model, nonzero masses may 
in fact be required in order to account for the data on solar and on
atmospheric neutrinos, as well as for an explanation of the dark 
matter in the universe. The possible detection of nonzero neutrino 
masses would have far-reaching implications for the understanding 
of fundamental issues in particle physics, astrophysics, as well 
as for the large-scale structure of our universe.

The above two different types of extensions of the Standard Model 
may be inter-connected. An example is provided by supersymmetric 
models with spontaneously broken R parity, which necessarily imply 
non-vanishing neutrino masses. As a result in some of these models
there are novel processes that could be observed at high energy
colliders whose magnitude would be correlated to the mass of the
neutrinos. One interesting aspect of these models is that they 
may affect the physics of the electroweak sector in such a 
remarkable way, that can be probed in various present and future
experiments, both at accelerator as well as underground installations,
as we will describe.

\subsection{Standard Model Basics}
\label{321}
 
The Standard Model is a Yang-Mills theory based on the
\321 gauge group, and described by the field representations
in table 1, where all fermions are left-handed.
\bet
\caption{Multiplets of the Standard Model}
\begin{center}
\begin{math}
\begin{array}{|c|c|} \hline
& \ \ \ {\mbox SU(3)\otimes SU(2)\otimes U(1)} \\
\hline
W     & (1,3,0)\\
B     & (1,1,0)\\
\hline
\phi  & (1,2,1)\\
\hline
\ell_a  & (1,2,-1)\\
e_a^c   & (1,1,2)\\
Q_a     & (3,2,1/3)\\
u_a^c   & (\bar{3},1,-4/3)\\
d_a^c   & (\bar{3},1,2/3)\\
\hline
\end{array}
\end{math}
\end{center}
\label{SM}
\eet
The fundamental constituents of matter - quarks and leptons -
interact mainly due to the exchange of the gauge bosons. The
Lagrangian describing the Standard Model can be found in many 
textbooks and reviews \cite{nubooks}. In order to comply with the fact 
that the weak interaction is mediated by massive vector bosons, 
the $W$ and the $Z$, the gauge symmetry has to be broken. The
best way to do this in a way consistent with renormalizability
and unitarity is through the nonzero vacuum expectation value 
(VEV) 
\beq
v/\sqrt{2} = \VEV{\phi^0}
\label{WEAKSCALE}
\eeq
of the neutral component of a complex Higgs scalar doublet $\phi$ 
\beq
\phi = \left(\begin{array}{ccccc}
\phi^+\\
\phi^0
\end{array}\right).
\label{DOUBLET}
\eeq
Through the term $|D \phi |^2$, where $D$ denotes the covariant 
derivative that follows from the quantum numbers in table 1,
this gives mass to all gauge bosons except one. This corresponds to
the surviving $U(1)_{em}$ gauge symmetry corresponding to the 
unbroken QED electro-magnetic gauge invariance, whose generator 
is the electric charge, identified as
\beq
\label{Q}
Q = T_3 + Y/2
\eeq
where $T_3$ and $Y$ denote the diagonal isospin and hyper-charge
generators. Diagonalizing the neutral gauge boson mass matrix
leads to a mixing between the gauge bosons $B_{\mu}$ and ${W_3}_{\mu}$
associated to the generators $Y$ and $T_3$,
\beq
\begin{array}{ccccc}
Z_{\mu}  = \cos \theta_W {W_3}_{\mu} + \sin \theta_W B_{\mu}\\
A_{\mu}  = -\sin \theta_W {W_3}_{\mu} + \cos \theta_W B_{\mu}
\label{Weinberg1}
\end{array}
\eeq
where 
\beq
\theta_W = \arctan \frac {g'}{g}
\label{Weinberg2}
\eeq
is the electroweak mixing angle, $g$ and $g'$ are the $SU(2)$ and
$U(1)$ gauge coupling strengths, respectively. The massless vector 
boson $A_{\mu}$ is the photon, while $Z_{\mu}$ has a mass
\beq
m_Z = \frac {g v}{2 \cos \theta_W}
\label{Z}
\eeq
while the charged gauge bosons $W^{\pm} = \frac{1}{\sqrt2} (W_1 \mp iW_2)$ 
have a mass
\beq
m_W = \frac {g v}{2}.
\label{W}
\eeq
The $W$ and the $Z$ gauge bosons have been discovered in 1983 by the
UA1 and UA2 experiments at CERN. The properties of the $Z$ have now been 
very precisely determined by the LEP experiments \cite{Martinez}, while 
those of the $W$ mostly come from UA2 and the Fermilab CDF and D0 
experiments \cite{top}. The measured \gau boson mass values agree well with 
the electroweak theory predictions, once radiative corrections are included. 
 
\subsection{The Fermion Sector}

The Standard Model contains three generations of quarks and leptons 
($a=1,2,3$), also given in table 1. Different generations 
are identical in their gauge properties, but differ in mass. 
The fermion assignment is clearly not invariant under parity, 
since only the left handed fermions (not the anti-fermions) transform 
as $SU(2)$ doublets. As a result the fermions can not be given gauge
invariant masses. Fermion masses arise from Yukawa-type interactions 
with the Higgs doublet scalar, 
\bea
\begin{array}{ccc}
{h_u}_{ab} \: u_a^c \: Q_b \: \tau_2 \phi^* +
{h_d}_{ab}\: d_a^c \: Q_b \: \phi +
{h_e}_{ab} \: e_a^c \: \ell_b \: \phi \: .
\label{YUKAWA}
\end{array}
\eea
which generate masses to all charged fermions $f=u,d,e$
\bea
h_f v / \sqrt2 = M_f
\eea
after electroweak breaking. Thus in the Standard Model 
the weak gauge boson as well as the charged fermion masses are 
generated spontaneously, by the Higgs mechanism. Note that the
coupling matrices $h_u$, $h_d$ and $h_e$ are arbitrary complex
non-symmetric matrices in generation space. This gives rise to 
a vast proliferation of parameters in the model. The understanding 
the pattern of fermion masses and mixing is one of the main challenges
of the theory and clearly lies beyond the Standard Model.
 
The currents describing the interaction of quarks and leptons with
the standard electroweak \gau fields can be written simply from the
gauge covariant derivatives that follow from table 1. For example
the terms in the Lagrangian describing the interaction of quarks and
leptons with the charged gauge bosons $W^{\pm}$ are
\beq
\label{CC}
\frac{ig}{\sqrt{2}} W_{\mu}^+ \sum_{a=1}^{n}
\bar{u}_{La} \ \gamma_{\mu} \ d_{La} +
\frac{ig}{\sqrt{2}} W_{\mu}^+ \sum_{a=1}^{n}
\bar{\nu}_{La} \ \gamma_{\mu} \ e_{La} + H.C.
\eeq
where $u_a$ and $d_a$ are weak-eigenstate quarks ($u,c,t$ and $d,s,b$),
and $\nu_a$ and $e_a$ denote the three \neus and the three charged leptons,
$e,\mu ,\tau $. The contact with the old four-Fermi weak interaction 
theory is established through
\beq
\frac {G_F}{\sqrt{2}} = \frac {g^2}{8 m_W^2}\:.
\label{fermi}
\eeq
where $G_F$ is the Fermi constant, well determined from $\mu$ decay.
In addition to the charged current weak interactions, quarks and
leptons also have neutral current weak interactions, described by
\beq
\label{NC}
\frac{ig'}{\sin \theta_W}\
Z_{\mu} \ \sum_A \bar{\Psi}_{A} \
\ \gamma_{\mu} \
(T_3 - \mbox{sin}^2 \theta_W Q) \
\Psi_{A}
\eeq
where $g \sin \theta_W = g' \cos \theta_W = e = \sqrt{4\pi\alpha}\:$ 
and $\Psi_A$ denotes an arbitrary helicity fermion belonging to any of the
multiplets in table 1. By including electroweak radiative 
corrections in the charged and neutral currents one obtains a very good
description of all known weak interaction processes. In particular, 
all known neutral current phenomena are well described by a single 
parameter, the electroweak mixing angle $\theta_W$.
The success of the LEP experiments in the precise determination 
of the electroweak parameters has been so remarkable that just the 
internal consistency of the various measurements is sufficient to 
provide a very good determination of the mass of the top quark 
$m_t = 180 \pm 14$ GeV, with the error largely due to the lack of 
knowledge of the Higgs boson mass \cite{Martinez}. This is in excellent 
agreement with the direct measurement at Fermilab \cite{top}. 

Similarly, the total $Z$ decay width, as well as its partial widths 
have been precisely measured by the LEP collaborations, leaving little 
room for new physics. Of special interest to us is the measurement of 
the invisible $Z$ width \cite{PDG95}
\beq
\label{NNU}
\Gamma_{inv}^{Z} = 499.9 \pm 2.5 \: \mbox{MeV}
\eeq
which can be translated into a measurement of the effective 
number of Standard Model neutrino generations. This places 
a very stringent constraint on models of neutrino mass where
lepton number is a global symmetry spontaneously broken at low
energies (see section \ref{majorons} below).

Let us now turn to a discussion of the flavour structure of the
charged current weak interactions. First we consider the case of 
quarks. In order to determine the form of the charged current we 
first diagonalize separately the quark mass matrices $M_u$ and $M_d$
\bea
\label{DIAG}
U_{Ru}^\dagger M_u U_{Lu} = diag(m_u,m_c,m_t)\\ \nonumber
U_{Rd}^\dagger M_d U_{Ld} = diag(m_d,m_s,m_b)
\eea
Since these matrices can not in general be made diagonal in the 
same basis, it follows that the charged current quark weak 
interactions, rewritten in terms of the mass-eigenstate 
quarks $u_i$ and $d_i$, take the following form
\beq
\frac{ig}{\sqrt{2}} W_{\mu}^+ \
\bar{u}_{Li} \ \gamma_{\mu} \ V_{ij} \ d_{Lj}
+ H.C.
\label{CKM}
\eeq
where $V = U_{Lu}^\dagger U_{Ld}$. Some of the parameters in the 
unitary matrix $V$ are not physical, since they can be eliminated 
by appropriate quark redefinitions. For $n$ generations $V_{ij}$ 
involves a set of
\beq
n(n-1)/2
\label{ANGLES1}
\eeq
mixing angles $\theta_{ij}$ and
\beq
n(n-1)/2-(n-1)
\label{PHASES1}
\eeq
CP violating phases $\phi_{ij}$. They can be brought to the following
useful form \cite{2227,OLD}
\beq
V = \omega_0 (\alpha) \prod_{i<j}^{n} \omega_{ij} (\eta_{ij})\:
\omega_0^\dagger (\alpha)
\label{CKM2}
\eeq
where $\theta_{ij} = \mid \eta_{ij} \mid$ and $\omega_0 =
diag (e^{i \alpha_i})$ is a diagonal matrix of phases, for example,
\beq
\omega_{12} (\eta_{12}) =
\left(\begin{array}{ccccc}
c_{12} & e^{i \phi_{12}} s_{12} & 0  & 0 & ...\\
-e^{-i \phi_{12}} s_{12} & c_{12} & 0 & 0 & ...\\
0  & 0 & 1 & 0 & ...\\
0  & 0 & 0 & 1 & ...\\
... & ... & ... & ... & ...
\end{array}\right)
\label{CKM3}
\eeq
For the simplest 2 generation case one sees that the $2 \times 2$ 
mixing matrix $V$ is completely determined by one parameter, the 
Cabibbo angle. In this case CP is necessarily conserved. As first 
noted by Kobayashi and Maskawa \cite{KM}, for three generations 
there are 3 mixing angles and a single phase that describes CP 
violation in the Standard Model. The first parametrizations of 
the three generation mixing matrix $V$ were given in ref. 
\cite{KM} and \cite{MAIANI}. A useful truncated form was 
introduced by Wolfenstein \cite{WOL}. So far this scheme has 
been proven sufficient to account for the observed pattern of 
quark mixing and CP violation in the Standard Model \cite{PDG95}. 
However, one lacks a fundamental theory of mass and flavour and,
as a result, the parameters in $V$ are all undetermined and 
must be extracted from experiment. Note that the CP phases in 
\eq{CKM2} are not directly observable: only special combinations 
(as many as given by our counting) which are invariant under quark 
rephasings have physical meaning. For 3 generations this CP violating 
combination is unique.
 
Finally, note that the weak neutral current of the mass-eigenstate 
quarks has a trivial flavour structure, i.e. it is given by the
identity matrix both in the weak and in the mass-eigenstate basis. 
This important property is usually referred to as the 
Glashow-Iliopoulos-Maiani (GIM) mechanism \cite{GIM}. 
It accounts for the absence of flavour changing neutral 
current mediated processes so far observed. Any deviations 
from this rule would indicate the presence of physics beyond 
the Standard Model.
 
\section{Theory of Neutrino Mass}

The first basic concept one needs to develop in order to make a
theoretical discussion of neutrino masses is that of Majorana
and Dirac masses.

\subsection{Dirac and Majorana Masses}
\label{DxM}
 
The most basic kinematical concept necessary to describe the 
physics of massive neutrinos is that of a {\it Majorana fermion},
and how it relates with the more familiar one of a {\it Dirac fermion}.
A massive Majorana fermion has just half of the number of degrees
of freedom of a conventional massive spin $1/2$ Dirac fermion
and corresponds to the lowest representation of the Lorentz group.
The basic Lagrangian describing such particle is \cite{2227,Case}
\beq
L_M=-i\rho^{\dagger} \sigma_{\mu} \partial_{\mu} \rho
-\frac{m}{2}\rho^T \sigma_2 \rho + H.C.
\label{LM}
\eeq
and is expressed in terms of a 2-component spinor $\rho$. In \eq{LM} 
$\sigma_i$ are the usual $2 \times 2$ Pauli matrices and 
$\sigma_4=-i \: I$, with $I$ being the identity matrix.
Pauli's metric convention for Minkowski coordinates is used,
where the dot product of two four vectors is $a.b \equiv a_{\mu}
b_{\mu} \equiv \vec{a} \cdot \vec{b} + a_4 b_4$, where $a_4=ia_0$.
Under a Lorentz transformation, $x \ra \Lambda x$, the spinor
$\rho$ transforms as $\rho \ra S(\Lambda)\rho(\Lambda^{-1}x)$
where $S$ obeys
\beq
S^{\dagger} \sigma_{\mu} S = \Lambda_{\mu \nu}\sigma_{\nu}
\label{HOMO}
\eeq
The kinetic term in \eq{LM} is clearly invariant. Similarly, the mass
term is invariant, as a result of unimodular property $det\:S=1$.
However, since it is $not$ invariant under a phase transformation
\beq
\rho \ra e^{i \alpha} \rho
\label{repha0}
\eeq
the Majorana mass term is only admissible in a theory of {\it electrically
neutral} fermions such as neutrinos, or other more exotic fermions of
the type present in supersymmetric theories, \eg neutral gauginos and
Higgsinos.
 
The equation of motion following from \eq{LM} is
\beq
-i \sigma_{\mu} \partial_{\mu} \rho = m \sigma_2 \rho^*
\label{EQM}
\eeq
As a result of the conjugation properties of the $\sigma$-matrices, 
one can verify that each component of the spinor $\rho$ obeys the 
Klein-Gordon wave-equation.
 
In order to display clearly the relationship between \eq{LM} and 
the usual theory of a massive spin $1/2$ Dirac fermion, defined by 
the familiar Lagrangian
\beq
L_D= - \bar{\Psi} \gamma_{\mu} \partial_{\mu} \Psi -
m\:\bar{\Psi} \Psi,
\label{LD}
\eeq
we construct the solutions to \eq{EQM} in terms of those of \eq{LD}, 
which are well known. For this we can use any representation of the 
Dirac algebra
$\gamma_{\mu} \gamma_{\nu} + \gamma_{\nu} \gamma_{\mu} = 2\:\delta_{\mu \nu}$.
To develop the weak interaction theory, however, it is convenient to use the
{\it chiral} representation, in which $\gamma_5$ is diagonal,
\bea
\gamma_i = \left(\begin{array}{ccccc}
                        0 & -i\sigma_i\\
                        i\sigma_i & 0\\
\end{array}\right)\: &
\gamma_4 = \left(\begin{array}{ccccc}
                        0 & I\\
                        I & 0\\
\end{array}\right)&
\gamma_5 = \left(\begin{array}{ccccc}
                        I & 0\\
                        0 & -I\\
\end{array}\right)
\label{GAMMAS}
\eea
In this representation the charge conjugation matrix $C$ obeying
\bea
C^T = - C\\ \nonumber
C^\dagger = C^{-1}\\ \nonumber
C^{-1}\:\gamma_{\mu}\:C = -\:\gamma_{\mu}^T
\label{conj}
\eea
is simply given by
\beq
C = \left(\begin{array}{ccccc}
                        -\sigma_2 & 0\\
                        0 & \sigma_2 \\
\end{array}\right)
\label{CHCONJ}
\eeq
 
A Dirac spinor may then be written as
\beq
\Psi_D = \left(\begin{array}{ccccc}
                \chi\\
                \sigma_2\:\phi^*\\
                \end{array}\right)
\label{PSID}
\eeq
so that the corresponding charge-conjugate spinor
$\Psi_D^c = C\:\bar{\Psi}_D^T$ is the same as $\Psi_D$ but
exchanging $\phi$ and $\chi$, \ie
\beq
\Psi_D^c = \left(\begin{array}{ccccc}
                \phi\\
                \sigma_2\:\chi^*\\
                \end{array}\right)
\label{PSIDCONJ}
\eeq
A 4-component spinor is said to be Majorana or self-conjugate if
$\Psi = C \bar{\Psi}^T$ which amounts, from \eq{PSID} and
\eq{PSIDCONJ} to setting $\chi = \phi$.
Using \eq{PSID} we can rewrite \eq{LD} as follows
\beq
L_D=-i\sum_{\alpha=1}^2 \rho_{\alpha}^{\dagger} \sigma_{\mu}
\partial_{\mu} \rho_{\alpha}
-\frac{m}{2} \sum_{\alpha=1}^2 \rho_{\alpha}^T \sigma_2 \rho_{\alpha} + H.C.
\label{LD2}
\eeq
where
\bea
\chi = \frac{1}{\sqrt2} (\rho_2 + i\rho_1) \nonumber \\
\phi = \frac{1}{\sqrt2} (\rho_2 - i\rho_1)
\label{DECOMP}
\eea
are the left handed components of $\Psi_D$ and of the charge-conjugate
field $\Psi_D^c$, respectively. This way the Dirac fermion is shown to
be equivalent to two Majorana fermions of equal mass. As a result
of this mass degeneracy, the theory described by \eq{LD} is invariant
under a continuous rotation symmetry between $\rho_1$ and $\rho_2$
\bea
\rho_1 \ra \c \rho_1 + \s \rho_2\nonumber \\
\rho_2 \ra -\s \rho_1 + \c \rho_2
\label{repha1}
\eea
This continuous symmetry is what corresponds to the phase symmetry
$\Psi_D \ra e^{i \alpha} \Psi_D$ associated to {\it fermion number
conservation} in the Dirac theory.
 
The mass term in \eq{LM} vanishes unless $\rho$ and $\rho^*$ are
anti-commuting, so we consider the Majorana fermion, right from the
start, as a second {\sl quantized} field. In order to obtain solutions of
\eq{EQM} we start from the usual Fourier expansion for the Dirac spinor,
\bea
\Psi_D =
(2\pi)^{-3/2} \int d^3k \sum_{r=1}^2 (\frac{m}{E})^{1/2}
[ e^{i k.x} a_r(k) u_r(k) + e^{-i k.x} b_r^{\dagger}(k) v_r(k)]
\label{PSID2}
\eea
where $u=C\:\bar{v}^T$ and $E(k) = (\vec{k}^2 + m^2)^{1/2}$ is the
mass-shell condition. From \eq{DECOMP} we then derive the corresponding
expansion for one of our 2-component Majorana spinors. For example, for
$\Psi_M = \rho_2$ the expansion is
\bea
\Psi_M =
(2\pi)^{-3/2} \int d^3k \sum_{r=1}^2 (\frac{m}{E})^{1/2}
[ e^{i k.x} A_r(k) u_{Lr}(k) + e^{-i k.x} A_r^{\dagger}(k) v_{Lr}(k)]
\label{PSIM}
\eea
with a similar expression for $\rho_1$.
Here $L$ denotes left-handed chiral projection, and the operators $A$
are defined are $A=(a+b)/ \sqrt2$, for each value of $r$ and $k$, from
where it follows that they obey canonical anti-commutation rules. Note that
\eq{PSIM} expresses the massive Majorana field operator in terms of
the chiral projections of the {\it ordinary} massive Dirac wave functions
$u$ and $v$. Note also that the $same$ creation and annihilation operators
appear in \eq{PSIM}, showing explicitly that there are only $half$ the number
of degrees of freedom in the Majorana field. This gives a consistent
Fock-space particle interpretation of the Majorana theory. For example, one
may look at quantum observables such as the energy momentum, written in
terms of the second quantized Majorana field operator given above.
One finds in this case that
\beq
P_{\mu} = \int d^3k \sum_{r=1}^2 k_{\mu} A_r^{\dagger}(k) A_r(k) \:,
\eeq
apart from the zero point energy.
 
Another important concept of the Majorana theory are the propagators
that follow from \eq{LM}. Lorentz invariance implies that there are $2$
different kinds of propagators, i.e.
\bea
<0 \mid \rho(x) \: \rho^*(y) \mid 0>
=i \sigma_{\mu} \partial_{\mu} \Delta_F(x-y;m)\\ 
\label{NORMAL}
<0 \mid \rho(x) \: \rho (y) \mid 0>
= m \: \sigma_2 \: \Delta_F(x-y;m)
\label{LNV}
\eea
where $\Delta_F(x-y;m)$ is the usual Feynman function.
The first one is the "normal" propagator that intervenes in
total lepton number conserving ($\mid \Delta L \mid = 0 $) processes,
while the one in \eq{LNV} describes the virtual propagation of Majorana
neutrinos in $\mid \Delta L \mid = 2 $ processes such as neutrino-less
double-beta decay.
 
It is instructive to consider the massless limit of the Majorana theory.
For this purpose we define helicity eigenstate wave-functions by
\bea
\vec{\sigma} \cdot \vec{k} \: u_L^{\pm}(k) = \pm \mid \vec{k}
\mid u_{L}^{\pm}(k)\\ \nonumber
\vec{\sigma} \cdot \vec{k} \: v_L^{\pm}(k) = \mp \mid \vec{k}
\mid v_{L}^{\pm}(k)
\label{limit}
\eea
The old 2-component massless neutrino theory is recovered from this
by noting that, out of the $4$ linearly independent wave functions
$u_{L}^{\pm}(k)$ and $v_{L}^{\pm}(k)$, {\it  only two} survive as the mass
approaches zero, namely, $u_{L}^{-}(k)$ and $v_{L}^{+}(k)$ \cite{BFD}.
 
In summary, \eq{LM} represents a perfectly consistent Lorentz
invariant quantum field theory \cite{2227}. It can easily be
generalised for a system of an arbitrary number of Majorana neutrinos.
In this case the most general Lagrangian allowed by Lorentz invariance
is of the type
\beq
L_M=-i \sum_{\alpha=1}^{n} \rho_{\alpha}^{\dagger} \sigma_{\mu} \partial_{\mu} \rho_{\alpha}
-\frac{1}{2} \sum_{\alpha,\beta=1}^{n} M_{\alpha\beta}\rho_{\alpha}^T \sigma_2 \rho_{\beta} + H.C.
\label{LM2}
\eeq
where the sum runs over $\alpha$ and $\beta$. By Fermi statistics
the mass coefficients $M_{\alpha \beta}$ must form a symmetric matrix,
in general complex. This matrix can always be diagonalized by
a complex $n \times n$ unitary matrix $U$ \cite{2227}
\beq
M_{diag} = U^T M U \:.
\eeq
When $M$ is taken to be real (CP conserving) its diagonalizing matrix
$U$ may be chosen to be orthogonal and, in general, the mass eigenvalues
can have different signs. These may be assembled as a signature matrix
\beq
\eta = diag(+,+,...,-,-,..)
\eeq
In \eq{LD2} and \eq{DECOMP} I chose to get rid of these signs by 
introducing appropriate factors of $i$ in the wave functions. This is 
perfectly consistent, as long as one only has the free theory given by 
\eq{LM}. In the presence of interactions, such as the charged currents 
of a realistic weak interaction \gau theory, one must bear in mind that
these signs are physical, and
theories characterized by different signature matrices {\it differ in
an essential way}. For example, there are two inequivalent models
containing 2 Majorana neutrinos: one characterised by $\eta = diag(+,+)$
and another by $\eta = diag(+,-)$. A Dirac \neu belongs to the second
class. For example, the condition for CP invariance is different for 
these two cases. These signs play an important role in the discussion 
of neutrino-less double beta decay \cite{CPSIGN}.
 
It should be apparent from the above analysis that there is no reason,
in general, to expect a conserved fermion number symmetry to arise
in a \gau theory where the basic building blocks are 2-component
massive electrically neutral fermions, such as neutrinos or the
supersymmetric $inos$. Unfortunately, the Majorana or Dirac 
nature of \neus can only be distinguished {\sl to the extent
that \neus are massive particles}. So far this has proven to
be a very severe limiting factor.

\subsection{Neutrinos in The Standard Model and Beyond}
\label{models}

In the Standard Model the only fermions which are electrically 
neutral, without right-handed partners, and apparently massless 
are the neutrinos \cite{fae}. It is rather mysterious that they 
seem to be so special. The presence of right handed neutrinos
would make the particle spectrum more symmetric between quarks and 
leptons. In this case neutrinos could get a Dirac mass like the quarks. 
The presence of \21 singlet right-handed neutrinos is required in many 
unified extensions of the Standard Model, such as SO(10), in order to 
realize the higher symmetry. However, having no electric charge, 
such \neus can also have a large Majorana mass term $M_R$.
This term violates total lepton number, or B-L (baryon minus lepton 
number), a symmetry that plays an important role in these extended 
gauge models \cite{LR}. In the presence of the usual Dirac mass 
term this leads to an effective Majorana mass term for the 
left-handed neutrinos via the so-called {\sl seesaw mechanism} 
\cite{GRS}. The masses of the light neutrinos are obtained by 
diagonalizing the following mass matrix
\beq
\begin{array}{c|cccccccc}
& \nu & \nu^c\\
\hline
\nu   & 0 & D \\
\nu^c & D^T & M_R
\end{array}
\label{SS}
\eeq
where $D = h_D v /\sqrt2$ is the Dirac mass matrix and
$M_R = M_R^T$ is the isosinglet Majorana mass. In the seesaw 
approximation, one finds 
\beq
{M_L} = - D M_R^{-1} D^T \:.
\label{SEESAW}
\eeq
This mechanism is able to explain naturally the relative smallness of 
\neu masses \cite{GRS}. Although one expects $M_R$ to be large, its 
magnitude heavily depends on the model. Although the seesaw idea was 
suggested in the context of SO(10) or left-right symmetric extensions
where lepton number is a part of the gauge symmetry \cite{LR}, it may 
be directly 
introduced in the Standard Model and the value of the scale may be as
low as the TeV scale. Although very often one hears about seesaw
"predictions" for light neutrino masses, they often depend 
on specific assumptions. In general, one can not make any generic
prediction for the light neutrino masses that are generated 
through the exchange of the heavy right-handed neutrinos. 
For example, the mass formula in \eq{SEESAW} would suggest
that the light neutrino masses scale with the up-type quark masses 
as \mne $\propto m_u^2$, \mnm $\propto m_c^2$ and \mnt $\propto m_t^2$. 
Unfortunately this is not really true for two reasons: first the existence 
of an induced triplet VEV and, second, due to the dependence upon the
detailed structure not only of the Dirac type entries, but also 
on the possible texture of the large Majorana mass term \cite{Smirnov}.
As a result, the seesaw mechanism provides us only with a general 
scheme, rather than detailed predictions. 

Although attractive, the seesaw mechanism is by no means 
the only way to generate neutrino masses. There is a large 
diversity of possible schemes to generate neutrino masses, 
which do not require any new large mass scale. For example, 
it is possible to start from an extension of the lepton sector 
of the \21 theory by adding a set of $two$ 2-component isosinglet 
neutral fermions, denoted ${\nu^c}_i$ and $S_i$. In this case
there is an exact L symmetry that keeps neutrinos strictly massless, 
as in the Standard Model. The conservation of total lepton number 
leads to the following form for the neutral mass matrix
\beq
\begin{array}{c|cccccccc}
& \nu  & \nu^c & S \\
\hline
\nu & 0 & D & 0 \\
\nu^c & D^T & 0 & M \\
S & 0 & M^T & 0
\end{array}
\label{MAT}
\eeq
This form has also been suggested in various theoretical models
\cite{WYLER}, including many of the superstring inspired models. 
In the latter case the zeros of \eq{MAT} naturally arise 
due to the absence of Higgs fields to provide the usual Majorana 
mass terms, needed in the seesaw model \cite{SST}. The implications
of \eq{MAT} are interesting on their own right, and the model 
represents a conceptually simple and phenomenologically rich
extension of the Standard Model, which brings in the possibility
that a wide range of new phenomena be sizeable. These have to do 
with neutrino mixing, universality, flavour and CP violation in 
the lepton sector \cite{BER,CP,SST}, as well as direct effects 
of associated with the NHL production in $Z$ decays \cite{CERN}. 
Clearly, one can easily introduce non-zero masses in this
model through a $\mu S S$ term that could be proportional
to the VEV of a singlet field $\sigma$ \cite{CON}. In contrast 
to the seesaw scheme, the \neu masses are directly proportional to
$\VEV{\sigma}$, a fact which is very important for the phenomenology
of the Higgs boson sector.

There is also a large variety of possible {\sl radiative} schemes 
to generate \neu masses with explicitly broken lepton number, which 
do not require one to introduce a large mass scale. It is quite 
possible embed such schemes so as to have the spontaneous violation 
of the global lepton number symmetry. The scale at which such a 
symmetry gets broken does not need to be high, as in the original 
proposal \cite{CMP}, but can be rather low, close to the weak 
scale \cite{JoshipuraValle92}. Such models are very attractive 
and lead to a richer phenomenology, as the extra particles required 
have masses at scales that could be accessible to present experiments 
\cite{zee,Babu88}.

The physics of neutrino masses and lepton number violation
could have important implications not only in astrophysics 
and cosmology but also in particle physics.

\subsection{Leptonic Charged Currents}
\label{Lmixing}
 
In analogy with the case of quarks, one can give a mass to neutrinos 
by adding right-handed singlet neutrinos to table 1, denoted 
by $\nu_i^c$, through a new gauge invariant Yukawa interaction, 
\beq
h_{Dia} \nu_i^c \ell_a \tau_2 \phi^*.
\label{MD}
\eeq
This generates a Dirac mass for the neutrinos which preserves
the total lepton number. However, unlike the right-handed charged 
fermions, the right-handed neutrinos, are singlets under the full 
\21 gauge group. As a result, the number $m$ of the $\nu_i^c$
fields, $1 \leq i \leq m$, is completely free and does not need 
to match the number of left-handed neutrinos \cite{OLD}. 
 
As we have seen above, in gauge theories of massive
neutrinos there is in general no reason to have a conserved total
lepton number. Indeed, in many \gau theories it is precisely the
violation of lepton number that generates the masses for the 
neutrinos, either as a result of an enlarged Higgs sector or 
due a seesaw mechanism involving the exchange of some heavy 
neutral leptons. In this case neutrinos are Majorana particles 
and their mass term is of the form given in \eq{LM2}.
 
The form of the leptonic charged current will depend on whether
there are isosinglet neutrinos and how many. It will also be quite 
sensitive to whether neutrinos are of Dirac or Majorana type. 
\newpage 
\begin{itemize}
\item 
{\bf Dirac neutrinos}\\
We first consider the case where the number $m$ of isosinglets
is smaller than that of isodoublet neutrinos, $m  < n$. In this case
$n-m$ \neus remain massless. As a result, the structure of lepton mixing
is $simpler$ than that in \eq{CKM2} because one can perform arbitrary
rotations (not only rephasings) in the degenerate sector of massless
neutrinos and this way eliminate unphysical parameters. The simplest
model of the $(n,m)$ type, with $m \neq 1$ is one with a conserved 
total lepton number and contains \cite{OLD}
\beq
n(n-1)/2 - (n-m)(n-m-1)/2
\label{ANGLES2}
\eeq
mixing angles $\theta_{ij}$ and
\beq
1+ n(m-1)-m(m+1)/2
\label{PHASES2}
\eeq
CP violating phases $\phi_{ij}$. For the case of $3$ generations 
and a single right handed neutrino added ($n=3,\: m=1$) the mixing 
matrix can always be brought to the form \cite{OLD}
\beq
K = R_{23} R_{13} =
\left(\begin{array}{ccccc}
c_{13} & 0 & s_{13} \\
-s_{23} s_{13} & c_{23} & s_{23} c_{13} \\
-s_{13} c_{23} & -s_{23} & c_{13} c_{23}
\end{array}\right)
\label{CKM4}
\eeq
where $R_{ij} = \omega_{ij}(\phi_{ij}=0)$ is a real rotation by an angle
$\theta_{ij}$ in the ${ij}$ plane. \Eq{CKM4} shows explicitly how one of the
$3$ mixing angles has been eliminated, together with the would-be CP violating
phase.
When $n=m$ the structure of lepton mixing of Dirac \neus is identical with
that describing quark mixing, \eq{CKM2}, so the same parametrization
applies.
When the number of isosinglets is larger than that of isodoublets,
$m > n$, new conceptual possibilities arise for the physics of leptons.
For example, if $m=2n$, the masslessness of the observed neutrinos may be
enforced by imposing the conservation of total lepton number, yet allowing
for the remarkable possibility of lepton flavour and CP non-conservation
despite neutrinos being strictly massless \cite{BER,CP,CERN,BER1}. 
\item 
{\bf Majorana neutrinos}\\
In the case of Majorana neutrinos the mass matrix \eq{LM2} is not 
invariant under rephasings of the \neu fields. As a result, there are
new phases present which cause CP to be violated in a theory with just 
two generations of Majorana neutrinos \cite{2227,1666}. The charged 
current of a 2-generation model ($n=2,m=0$) may be parametrised as
\beq
\left(\begin{array}{ccccc}
c_{12} & e^{i \phi_{12}} s_{12} \\
-e^{-i \phi_{12}} s_{12} & c_{12}
\end{array}\right)
\label{MAJCP}
\eeq
where $\phi_{12}$ is the Majorana phase. For $n$ generations these 
Majorana phases make up a total of $(n-1)$ additional phases over 
and above those of \eq{PHASES1}. They are physical parameters, 
and play a role in $ \mid \Delta L \mid \: = 2 $ processes such 
as some class of neutrino oscillations \cite{1666}. Although
conceptually important, the CP violating effects associated to
these phases are too small to be observed, since they are 
helicity suppressed.
An important subtlety arises regarding the conditions for CP 
conservation in theories of massive Majorana neutrinos. Unlike the 
case of Dirac fermions, in the Majorana case the condition for CP 
invariance is \cite{BFD}
\beq
K^* = K \eta
\label{cpeta}
\eeq
where $\eta = diag(+,+,...,-,-,..)$ is the signature matrix describing
the relative signs of the neutrino mass eigenvalues that follow from 
\eq{LM2} obtained by using real diagonalizing matrices \cite{CPSIGN,BFD}.
Only for the trivial case where $\eta$ is trivial, CP invariance requires
the mixing matrix to be real. For example the value $\phi_{12} = \pi/2$ 
(just as $\phi_{12} = 0$) can be CP conserving. 
We now turn to the most general situation where there are $m \neq 0$
two-component $SU(2) \ot U(1)$ singlet leptons (such as RH neutrinos)
present in the theory. In this case there is in general no reason to
forbid a gauge and Lorentz invariant Majorana mass term of the type
\beq
{M_R}_{ij} \nu^c_i \nu^c_j
\label{MRH}
\eeq
which breaks total lepton number symmetry. As a result, the structure of
the weak currents can be substantially more complex.
Since the number $m$ is arbitrary, one may consider models with 
Majorana \neus based on {\sl any value} of $m$ \cite{OLD}. For 
$m \leq n$, $n-m$ \neus will remain massless, while $2m$ \neus 
will acquire Majorana masses.
For example, in a model with $n=3$ and $m=1$ one has one light
(presumably $\nu_{\tau}$) and one heavy Majorana neutrino, in addition
to the two massless \neus ($\nu_e$ and $\nu_{\mu})$. In this case
clearly there will be less parameters than present in a model with
$m=n=3$. The case $m > n$, \eg $m=2n$, may also be interesting because
it allows for an elegant way to avoid constraints related to \neu
masses, and therefore enhance many of the effects possible in these
theories \cite{CON,CERN,SST,ETAU}. I will now analyse some of
the general features of the associated currents. The first is that the
isosinglets, presumably heavy, will now mix with the ordinary isodoublet
neutrinos in the charged current weak interaction. As a result, the
mixing matrix describing the charged leptonic weak interaction is a
$rectangular$ matrix $K$ \cite{2227}, which may be decomposed as
\beq
K = (K_L, K_H)
\label{CC1}
\eeq
where $K_L$ is an $n \times n$ matrix and $K_H$ is an $n \times m$ matrix.
The charged weak interactions of the light (mass-eigenstate) neutrinos
are effectively described by a mixing matrix $K_L$ which is non-unitary.
An explicit parametrization of the weak charged current mixing matrix
$K$ that covers the most general situation present in these $(n,m)$
models has also been given in ref. \cite{2227}. It involves in general
\beq
n(n+2m-1)/2
\label{ANGLES3}
\eeq
mixing angles $\theta_{ij}$ and
\beq
n(n+2m-1)/2
\label{PHASES3}
\eeq
CP violating phases $\phi_{ij}$. This number far exceeds
the corresponding number of parameters describing the charged current
weak interaction of quarks \eq{ANGLES1} and \eq{PHASES1}. The reasons
are that, since \neus are Majorana particles, their mass terms
are not invariant under rephasings \cite{2227}. As a result, there
are less phases that can be eliminated by field redefinitions.
Their possible role in neutrino oscillations was studied in ref.
\cite{1666}. In addition, the isodoublet neutrinos in general
mix with the isosinglets, so CP may also be violated in this mixing,
even in the case where the physical light \neus are massless \cite{CP}.
\end{itemize}

\subsection{Leptonic Neutral Current}
\label{NC0}
 
The neutral current couplings of mass-eigenstate \neus are diagonal,
just as in the case of charged leptons and quarks, in theories where
there are no isosinglet \neus, \ie $m=0$. This occurs irrespective of
whether the mass-eigenstate \neus are Dirac or Majorana particles.
 
An important feature arises in any theory based on $SU(2) \ot U(1)$,
where isosinglet and isodoublet lepton mass terms coexist, regardless
of $L$ number conservation. In this case the \gau currents
mix \neus with isosinglet leptons. As a result, in these theories
there can be non-diagonal couplings of the $Z$ to the
mass-eigenstate neutrinos, even at the tree level \cite{2227}.
The neutral current may be expressed in the following general form
\beq
P = K^\dagger K
\label{NCMAJ1}
\eeq
where the matrix P is a projective hermitian matrix $P^2=P=P^\dagger$,
directly determined in terms of \eq{CC1} as
\beq
P = \left(\begin{array}{ccccc}
K_L^\dagger K_L & K_L^\dagger K_H\\
K_H^\dagger K_L & K_H^\dagger K_H
\end{array}\right)
\label{NCMAJ2}
\eeq
This matrix determines the neutral current couplings of all
mass-eigenstate neutral leptons (denoted $N_{L\alpha}$), both
the light \neus as well as the NHLS, as follows
\beq
\frac{ig'}{2 \sin \theta_W}
Z_{\mu} \
\sum_{\alpha \beta} \bar{N}_{L\alpha} \ \gamma_{\mu} \ P_{\alpha\beta} \
N_{L\beta}
\label{NC2}
\eeq
In summary we see that while the structure of the leptonic weak
interaction is completely trivial within the Standard Model, when
compared to the quark weak interactions, it may be much more complex
in theories beyond the Standard Model. These could give rise to a rich
body of phenomena which may or may not be directly related to the 
magnitude of the neutrino masses, as discussed below.

\subsection{Constraints on neutrino-NHL mixing }

There are important constraints on the leptonic charged current matrix 
elements and, in particular, on the allowed magnitude of neutrino-NHL mixing.
For example, from \eq{CC} and \eq{CC1} we see that the coupling of a given 
light neutrino to the corresponding charged lepton is decreased by a certain
factor. This observed universality of weak interactions constrains the
relative values of these decreases and limits them to be small
\cite{BER1}. In the low mass range the existence of these new leptons
could show up explicitly in {\sl low energy} weak decay processes, 
if such neutrinos can be kinematically produced. Heavier masses
above the few GeV range can be ideally searched for at LEP
experiments at the Z pole. Constraints on the strength of the 
$K_H$ mixing matrix elements follow therefore from low energy weak
decay measurements as well as from LEP experiments. They are shown 
in \fig{nhlimits}, from ref. \cite{lacasta}.

\subsection{Phenomenological Implications.}
\label{impli}

One of the simplest and natural extensions of the lepton sector
is the addition of neutral isosinglet heavy leptons, such as 
right-handed neutrinos. Their existence is required, for example, 
in left-right symmetric models \cite{LR} and leads, via the seesaw
mechanism, to non-vanishing neutrino masses. However, they may be 
added simply at the \21 level and, since they are SU(2)$_L$ singlets, 
their number is totally free, as they carry no anomalies \cite{2227}.
There is a variety of novel phenomena whose existence would be associated 
to these extensions of the lepton sector \cite{fae}. 
After mass matrix diagonalization one finds that these leptons will 
couple in the charged and neutral weak currents. For example, the 
matrix $P$ in \eq{NC2} contains couplings connecting light to heavy 
neutrinos \cite{2227}. Thus, if their mass is below that of the $Z$,
the heavy ones will be singly produced in $Z$ decays, \cite{CERN}
\begin{equation}
Z \rightarrow N_{\tau} + \nu_{\tau}
\end{equation}
Subsequent NHL decays would then give rise to large missing 
momentum events, called zen-events. The attainable rates for 
such processes can be quite large \cite{CERN}, well within 
the sensitivities of the LEP experiments \cite{CERN}. Dedicated 
searches for acoplanar 
jets and lepton pairs from $Z$ decays have provided stringent
constraints on NHL couplings to the $Z$, plotted below \cite{lacasta}
\footnote{There have been also inconclusive hints reported by 
ALEPH \cite{alephmono}.}
\bef
\centerline{\protect\hbox{
\psfig{file=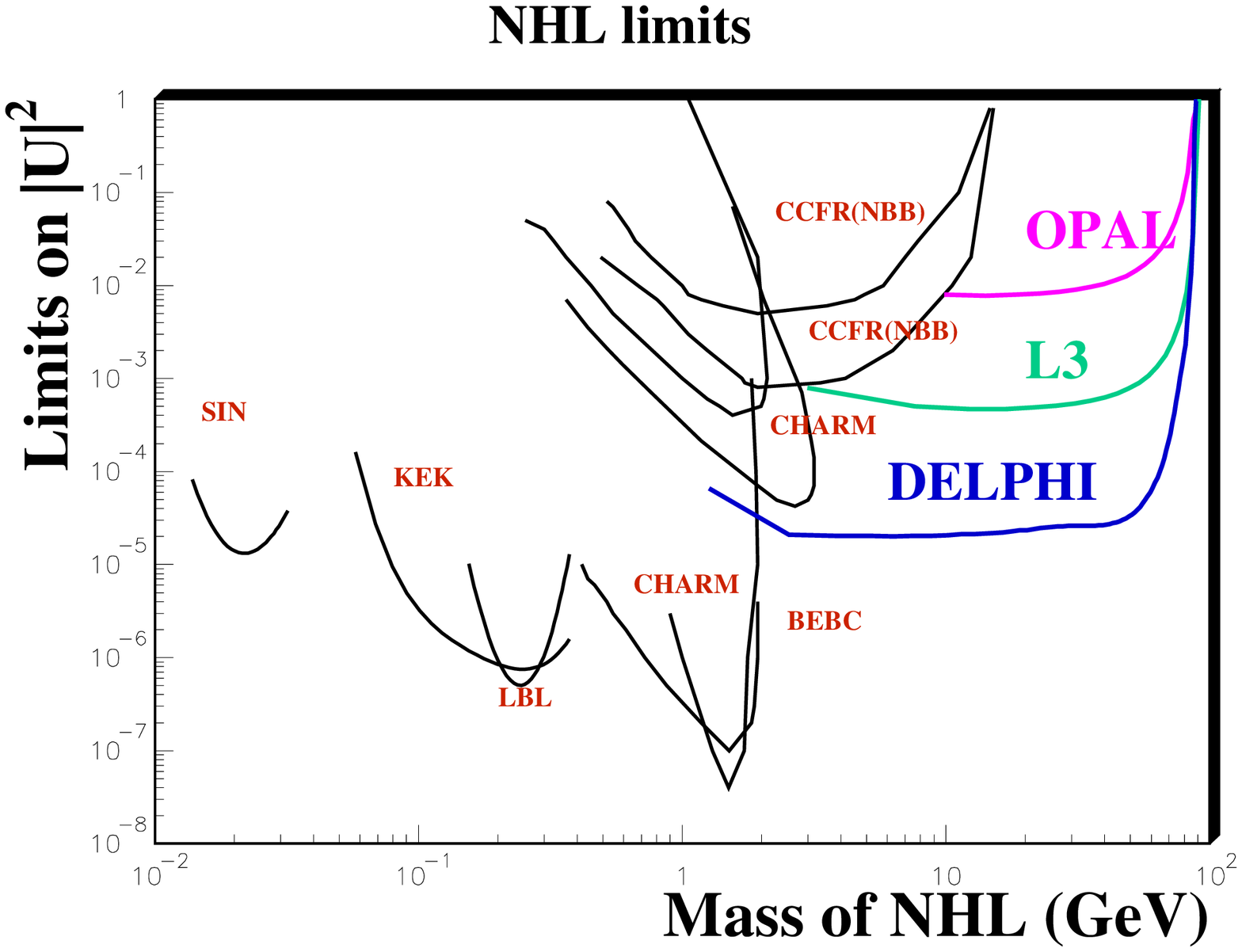,height=9cm}}}
\caption{Limits on NHL mass and couplings.}
\label{nhlimits}
\eef 

One sees that the recent DELPHI constraints supersede by
far the low energy constraints following, e.g. from weak 
universality.

Let us now turn to the case when the isosinglet neutral heavy 
leptons are heavier than the $Z$. Even in this case they can 
produce interesting virtual effects, completely calculable 
in these models, in terms of the NHL masses and electroweak 
charged and neutral current couplings. This way they can
mediate lepton flavour violating (LFV) decays which are 
exactly forbidden in the Standard Model, thus a clear
signature of physics beyond the Standard Model.
In the simplest models of seesaw type where the NHLS are 
Majorana type these decays are expected to be small, due to 
limits on the light \neu masses. However, in other variant
models with Dirac NHLS \cite{SST} this suppression is not 
present \cite{BER,CP} and LFV rates are restricted only
by present constraints on weak universality violation.
These allow for sizeable decay branching ratios, close
to present experimental limits \cite{3E} and within the 
sensitivities of the planned tau and B factories \cite{TTTAU}. 
The situation is summarised in Tables 2 and 3.
The dependence of the attainable LFV $\tau$ and Z decay branching 
ratios upon the NHL mass is illustrated in ref. \cite{Pila}.
\begin{table}
\begin{center}
\caption{Allowed lepton-flavour-violating $\tau$ decay branching ratios. }
\begin{displaymath}
\begin{array}{|c|cr|} 
\hline
\mbox{channel} & \mbox{strength} & \mbox{} \\
\hline
\tau \rightarrow e \gamma ,\mu \gamma &  \lsim 10^{-6} & \\
\tau \rightarrow e \pi^0 ,\mu \pi^0 &  \lsim 10^{-6} & \\
\tau \rightarrow e \eta^0 ,\mu \eta^0 &  \lsim 10^{-6} - 10^{-7} & \\
\tau \rightarrow 3e , 3 \mu , \mu \mu e, \etc &  \lsim 10^{-6} - 10^{-7} & \\
\hline
\end{array}
\end{displaymath}
\end{center}
\label{LFVTAU} 
\end{table}
The study of rare $Z$ decays nicely complements what can 
be learned from LFV muon and tau decays. The stringent limits 
on $\mu \rightarrow e \gamma$ preclude sizeable branching ratios
for the corresponding process $Z \ra e \mu$. However the decays 
$Z \ra e \tau$ and $Z \ra \mu \tau$ can occur at the \O($10^{-6}$) 
level, as illustrated in table 3.
\begin{table}
\label{LFVZ} 
\begin{center}
\caption{Allowed lepton-flavour-violating $Z$ decay branching ratios.}
\begin{displaymath}
\begin{array}{|c|cr|} 
\hline
\mbox{channel} & \mbox{strength} & \mbox{} \\
\hline
Z \rightarrow e \tau &  \lsim 10^{-6} & \\
Z \rightarrow \mu \tau &  \lsim 10^{-7} & \\
\hline
\end{array}
\end{displaymath}
\vglue -2cm
\end{center}
\end{table}
Similar statements can be 
made also for the CP violating Z decay asymmetries in these 
LFV processes \cite{CP}. However, under realistic assumptions, 
it is unlikely that one will be able to see these decays at LEP 
without a high luminosity option \cite{ETAU}. In any case there 
have been dedicated experimental searches which have set good 
limits on LFV Z and $\tau$ decays at LEP \cite{opallfv}. This 
is illustrated in tables 4 and 5.
\begin{table}
\label{DELPHITAU} 
\begin{center}
\caption{Delphi limits on lepton-flavour-violating $\tau$ decays.}
\begin{displaymath}
\begin{array}{|c|cr|} 
\hline
\mbox{channel} & \mbox{90\% C.L. limit} & \mbox{} \\
\hline
\tau \rightarrow e \gamma &  1.1 \times 10^{-4} & \\
\tau \rightarrow \mu \gamma &  6.2 \times 10^{-5} & \\
\hline
\end{array}
\end{displaymath}
\vglue -4cm
\end{center}
\end{table}
\begin{table}
\label{OPALZ} 
\begin{center}
\caption{Opal limits on lepton-flavour-violating Z decays.}
\begin{displaymath}
\begin{array}{|c|cr|} 
\hline
\mbox{channel} & \mbox{95\% C.L. limit} & \mbox{} \\
\hline
Z \rightarrow e \mu &  1.7 \times 10^{-6} & \\
Z \rightarrow e \tau &  9.8 \times 10^{-6} & \\
Z \rightarrow \mu \tau &  17 \times 10^{-6} & \\
\hline
\end{array}
\end{displaymath}
\vglue -2cm
\end{center}
\end{table}

Finally we note that there can also be large rates for lepton flavour 
violating decays in models with radiative mass generation \cite{zee,Babu88}. 
A very interesting observation is that LFV phenomena are expected in 
supersymmetric unified models \cite{Hall}. This has been discussed in 
many papers \cite{SUSYLFV} and the predictions have now been recently
reanalysed in view of the large top quark mass in ref. \cite{SUSYLFV2}.
The expected decay rates may lie within the present experimental 
sensitivities and the situation should improve at PSI or at the 
proposed tau-charm factories.

\subsection{Laboratory Limits on Neutrino Masses}

\begin{itemize}
\item
Of all existing \neu mass limits the most model-independent are
the laboratory limits that follow purely from kinematics. These
can be summarised as \cite{PDG95}
\beq
\label{1}
m_{\nu_e} 	\lsim 5 \: \mbox{eV}, \:\:\:\:\:
m_{\nu_\mu}	\lsim 250 \: \mbox{KeV}, \:\:\:\:\:
m_{\nu_\tau}	\lsim 23 \: \mbox{MeV} 
\eeq
The limit on the \ne mass follows from tritium beta decay studies
\cite{Erice}, while that on the \nt mass has recently been reported
by the ALEPH experiment at CERN \cite{eps95}. This limit may be 
substantially improved at a future tau factory \cite{jj}. 
\item
Neutrino oscillations have been searched in both accelerator and 
reactor experiments \cite{granadaosc}.  So far no effect has been 
conclusively demonstrated, in any of the channels. As a result,
limits have been set on the relevant mixings and mass differences. 
The 90\% confidence level (C.L.) exclusion contours of neutrino 
oscillation parameters in the 2-flavour approximation are given 
in \fig{oscil}, taken from ref. \cite{jjc}. 
\bef
\centerline{\protect\hbox{
\psfig{file=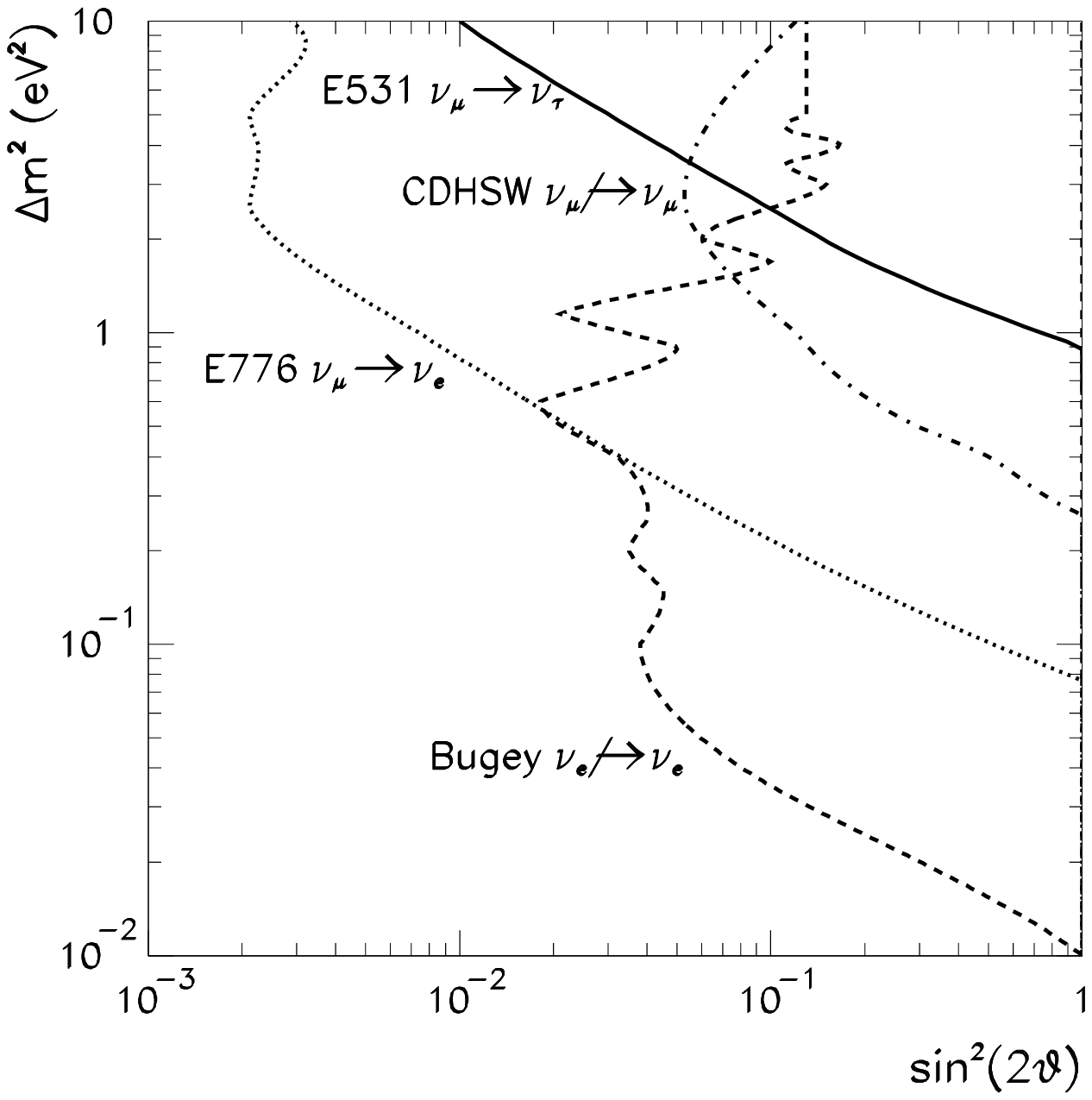,height=9cm}}}
\caption{Status of neutrino oscillation parameters. }
\label{oscil}
\eef
As can be seen, the best limit is for $\nu_{\mu} \ra \nu_e$ 
oscillations. Recently there has been a limit on \ne $\ra$ 
\nm oscillations using data from the LSND experiment at
Los Alamos \cite{Caldwell1}. 
Improvements for the \nm $\ra$ \nt channel are 
expected soon from the ongoing CHORUS and NOMAD experiments 
at CERN, with a similar proposal at Fermilab \cite{chorus}.
For illustration, we give in \fig{Chorus} the region of
sensitivity of Chorus experiment.
\bef
\centerline{\protect\hbox{
\psfig{file=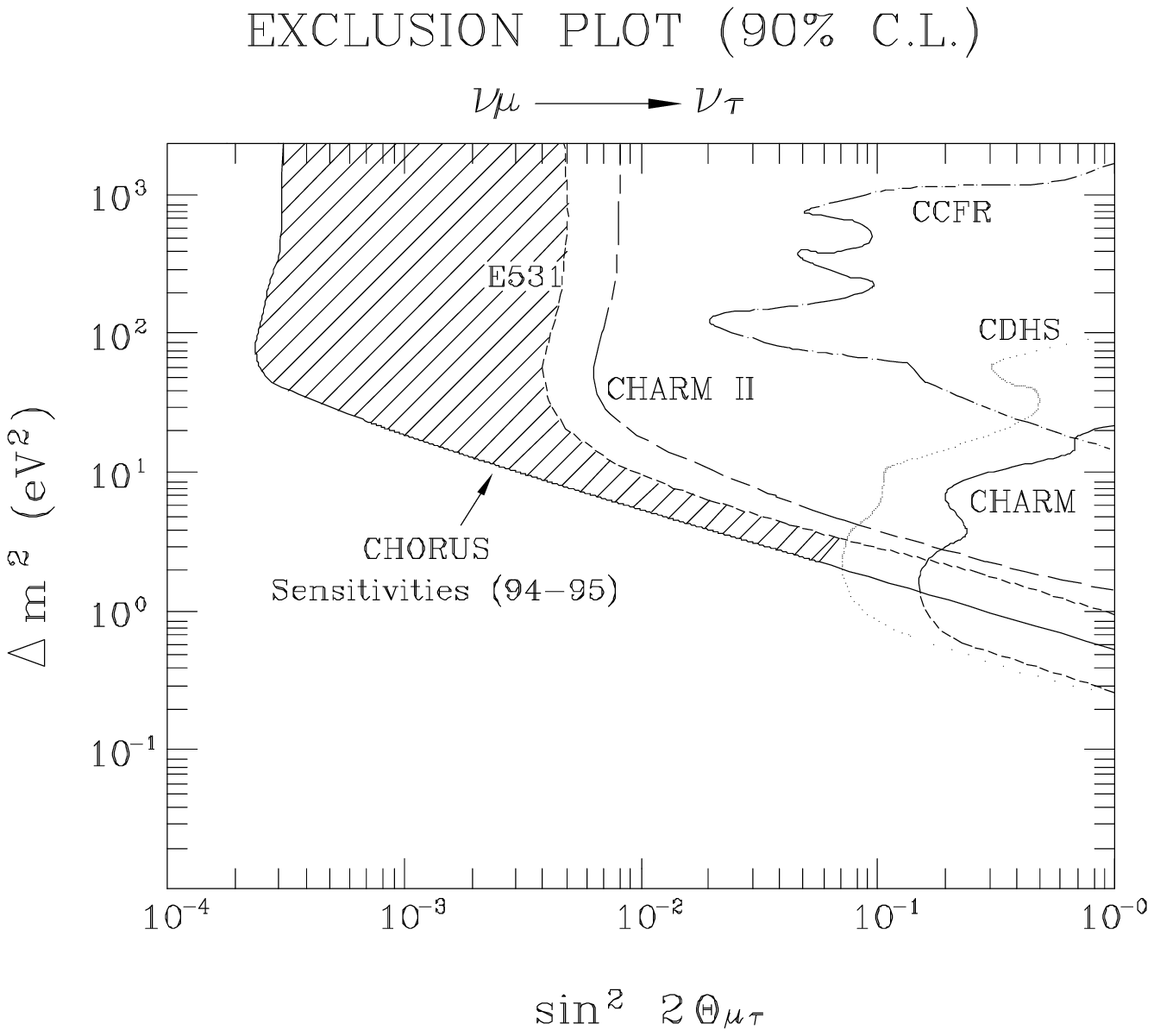,height=10cm}}}
\caption{Sensitivity of Chorus experiment.}
\label{Chorus}
\eef
Finally there are future prospects to probe neutrino oscillation
parameters with much better sensitivity at long baseline experiments
using CERN, Fermilab and KEK neutrino beams in conjunction with 
Gran Sasso, Soudan and SuperKamiokande underground installations,
respectively. There are also long baseline reactor experiments 
such as Chooz and San Onofre, with very good sensitivity to
neutrino oscillation parameters.
\item
Another important limit follows from the non-observation of 
neutrino-less double beta decay - ${\beta \beta}_{0\nu}$ - i.e. 
the process by which an $(A,Z-2)$ nucleus decays to $(A,Z) + 2 \ e^-$. 
This process would arise from the virtual exchange of a Majorana 
neutrino from an ordinary double beta decay process. Unlike the 
latter, the neutrino-less process is lepton number violating and
its existence would signal the Majorana nature of neutrinos.
Because of the phase space advantage, this process is a very 
sensitive tool to probe into the nature of neutrinos.
In fact, as shown in ref. \cite{BOX}, a non-vanishing 
${\beta \beta}_{0\nu}$ decay rate requires \neus to be Majorana 
particles, {\sl irrespective of which mechanism induces it}. 
This establishes a very deep connection which, in some special 
models, may be translated into a lower limit on the \neu masses.
The negative searches for ${\beta \beta}_{0\nu}$ in $^{76} \rm{Ge}$ 
and other nuclei leads to a limit of about two eV \cite{Avignone} on
the weighted average \neu mass parameter $\VEV{m}$
\beq
\label{bb}
\VEV{m} \lsim 1 - 2 \ \mbox{eV}
\eeq
depending to some extent on the relevant nuclear matrix elements 
characterising this process \cite{haxtongranada}. Improved sensitivity 
is expected from the upcoming enriched germanium experiments. 
Although rather stringent, this limit in \eq{bb} may allow relatively 
large \neu masses, as there may be strong cancellations between the 
contributions of different neutrino types. This happens automatically
in the case of a Dirac \neu as a result of the lepton number symmetry 
\cite{QDN}. 
\end{itemize}

\subsection{The Cosmological Density Limit }

In addition to laboratory limits, there is a cosmological bound that 
follows from avoiding the overabundance of relic neutrinos \cite{KT}
\beq
\label{RHO1}
\sum_i m_{\nu_i} \lsim 92 \: \Omega_{\nu} h^2 \: \mbox{eV}\:,
\eeq
where $\Omega_{\nu} h^2 \leq 1$ and the sum runs over all isodoublet 
neutrino species with mass less than \O(1) MeV. Here 
$\Omega_{\nu}=\rho_{\nu}/\rho_c$, where $\rho_{\nu}$ is the neutrino
contribution to the total density and $\rho_c$ is the critical density.
The factor $h^2$ measures the uncertainty in the determination of the
present value of the Hubble parameter, $0.4 \leq h \leq 1$. 
The factor $\Omega_{\nu} h^2$ is known to be smaller than 1.

For the $\nu_{\mu}$ and $\nu_{\tau}$ this bound is much more 
stringent than the corresponding laboratory limits in (\ref{1}). 

Recently there has been a lot of work on the possibility of
an MeV tau neutrino \cite{ma1,SF}. Such range seems to be an 
interesting one from the point of view of structure formation 
\cite{ma1,SF}. Moreover, it is theoretically viable as the
constraint in \eq{RHO1} holds only if \neus are stable on the 
relevant cosmological time scales. 
\begin{figure*}
\centerline{\protect\hbox{
\psfig{file=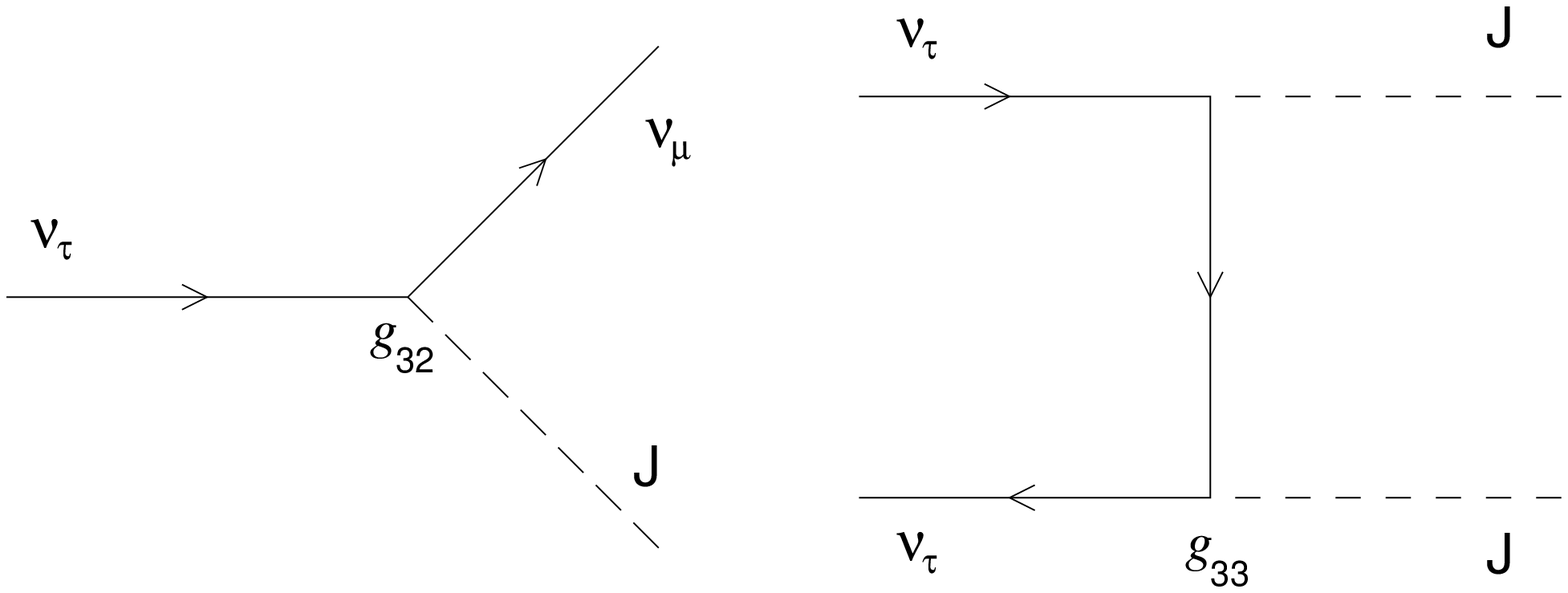,height=5cm,width=12cm}}}
\caption{Majorana tau neutrino decays and annihilations to Majorons.}
\label{couplings}
\end{figure*}
As shown in \fig{couplings}, in models with spontaneous violation of 
total lepton number \cite{CMP} neutrinos may decay into a lighter \neu 
plus a Majoron, for example \cite{fae},
\beq
\label{NUJ}
\nu^{\prime} \ra \nu + J \:\: .
\eeq
or have sizeable annihilations (there is also a crossed diagram)
\beq
\label{BBNJJ}
\nu^{\prime} \nu^{\prime} \ra J + J \:\: .
\eeq
{couplings}
The possible existence of fast decay and/or annihilation channels 
could eliminate relic neutrinos and therefore allow them to be heavier
than \eq{RHO1}. The cosmological density constraint on neutrino decay 
lifetime (for neutrinos lighter than an MeV or so) may be written as
\beq
\tau_{\nu^{\prime}} \ltap 1.5 \times 10^7 (\mbox{KeV}/m_{\nu^{\prime}})^{2} 
\mbox{yr} \:,
\label{RHO2}
\eeq
and follows from demanding an adequate red-shift of the heavy neutrino
decay products. For  neutrinos heavier than $\sim 1 \: MeV$, such as
possible for the case of $\nu_{\tau}$, the cosmological limit on the
lifetime is less stringent than given in \eq{RHO2}.
 
As we already mentioned the possible existence of non-standard 
interactions of neutrinos due to their couplings to the Majoron
brings in the possibility of fast invisible neutrino decays with 
Majoron emission \cite{fae}. These 2-body decays can be
much faster than the visible decays, such as radiative decays of 
the type $\nu' \ra \nu + \gamma$. As a result the Majoron decays
are almost unconstrained by astrophysics and cosmology. For
a more detailed discussion see ref. \cite{KT}. 

A general method to determine the Majoron emission decay rates 
of neutrinos was first given in ref. \cite{774}. The resulting 
decay rates are rather subtle \cite{774} and model dependent
and will not be discussed here. The reader may consult ref.
\cite{V,CON,fae}. The conclusion is that there are many ways
to make neutrinos sufficiently short-lived that all mass values
consistent with laboratory experiments are cosmologically acceptable.
For neutrino decay lifetime estimates see ref. \cite{fae,V,CON,RPMSW}.

\subsection{The Cosmological Nucleosynthesis Limit}

Recently there has been extensive discussion about cosmological 
nucleosynthesis \cite{bbncrisis,sarkar}. The observation of the 
primordial light element abundances can be used to place
important constraints on neutrino properties, such as masses,
decay lifetimes and annihilation cross sections. Here we focus
on the case of the tau neutrino since, from (\ref{1}), it is the 
only neutrino that could have mass in the relevant MeV region. 
If massive $\nu_\tau$'s are stable during nucleosynthesis 
($\nu_\tau$ lifetime longer than $\sim 100$ sec), one can constrain
their contribution to the total energy density from the observed 
amount of primordial helium. This bound can be expressed through 
an effective number of massless neutrino species ($N_\nu$). Using
$N_\nu < 3.4-3.6$, the following range of $\nu_\tau$ mass has been 
ruled out \cite{KTCS91,DI93}
\begin{equation}
0.5 \mbox{MeV} < m_{\nu_\tau} < 35 \mbox{MeV}
\label{cons1}
\end{equation}
If the nucleosynthesis limit is taken less stringent the limit
loosens somewhat. However it has recently been argued that 
non-equilibrium effects from the light neutrinos arising from
the annihilations of the heavy \nt's make the constraint stronger
and forbids all $\nu_\tau$ masses on the few MeV range. 

One can show that if the \nt is unstable during nucleosynthesis 
\cite{unstable} the bound on its mass is substantially weakened,
depending on the assumed neutrino lifetime.

\begin{figure}
\centerline{\protect\hbox{
\psfig{file=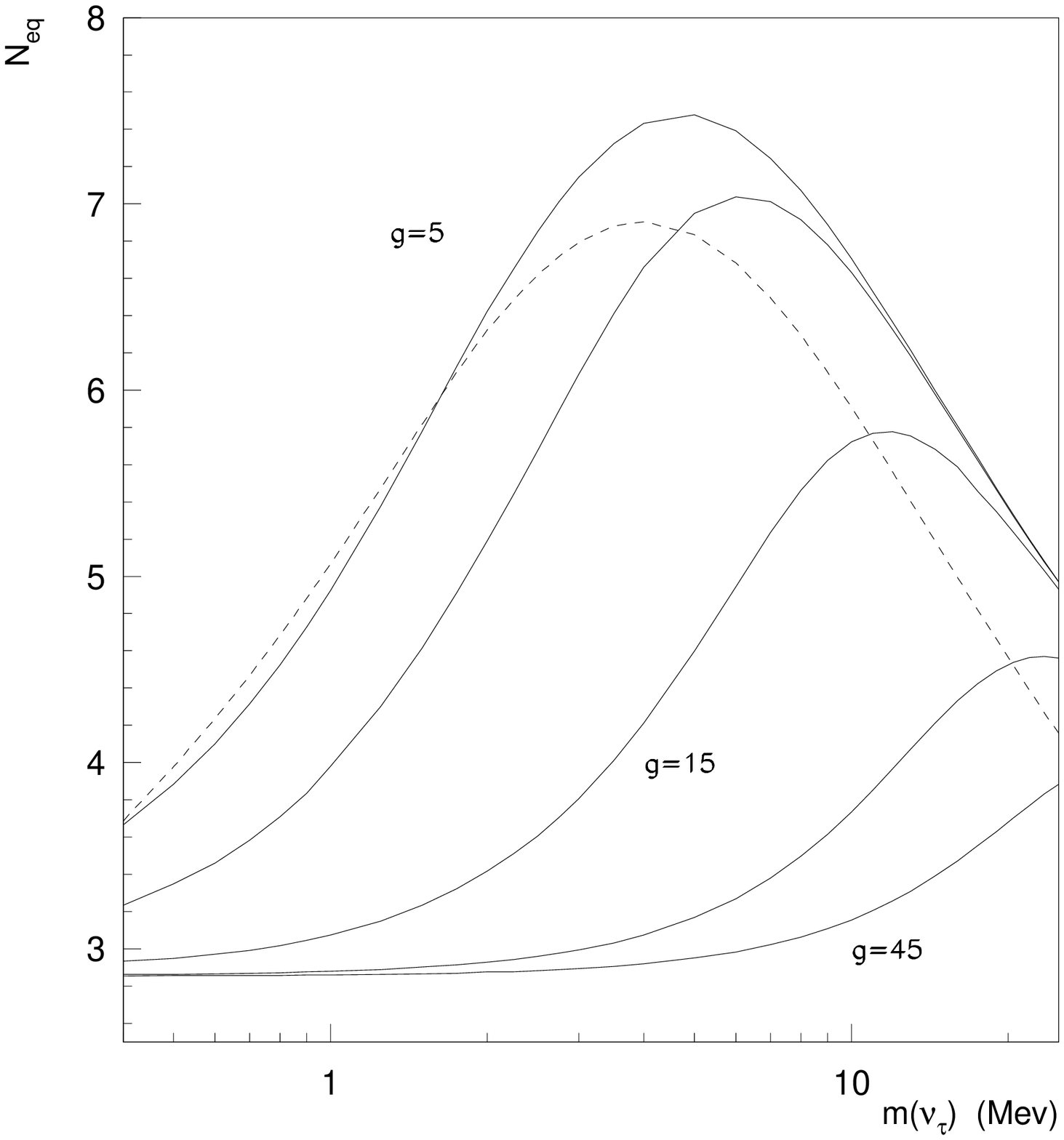,height=9cm}}}
\caption{Effective number of massless neutrinos equivalent to
the contribution of massive neutrinos. }
\label{neq}
\end{figure}

Even more drastic is the effect of \neu annihilations \cite{DPRV}.
The results are shown in \fig{neq}. The solid line gives the 
effective number of massless neutrinos equivalent to the 
contribution of the massive \nt for different $g$ values 
expressed in units $10^{-5}$, while the dashed line 
corresponds to the Standard Model case when $g=0$ and 
no Majorons are present. One sees that no \nt masses below 23 MeV 
can be ruled out as long as the coupling between $\nu_\tau$'s and $J$'s 
exceeds a few times $10^{-4}$ or so. Such values are reasonable for 
many Majoron models \cite{Jmodels}. For more details see ref. \cite{DPRV}.
In short one sees that the constraints on the mass of a Majorana
$\nu_\tau$ from primordial nucleosynthesis can be substantially 
relaxed if annihilations $\nu_\tau \bar{\nu}_\tau \leftrightarrow JJ$ 
are present.

As a result of the considerations in section 2
one concludes that it is worthwhile to continue the efforts to 
improve present laboratory \neu mass limits in the laboratory. 
One method sensitive to large masses is to search for distortions 
in the energy spectra of leptons coming from $\pi, K$
weak decays such as $\pi K \ra e \nu$, $\pi, K \ra \mu \nu$, 
as well as kinks in nuclear $\beta$ decays.

\subsection{Hints for Neutrino Masses}

The only hint for neutrino masses following from accelerator
experiments is the controversial claim for neutrino oscillations 
reported by the LSND experiment. I will not describe it here,
but the reader is encouraged to read the original paper, 
now published \cite{Caldwell2}.

Thus all positive indications in favour of nonzero neutrino masses
follow from astrophysics and cosmology. We now turn to these.

\subsubsection{Dark Matter}

With the COBE detection of fluctuations in the Cosmic Microwave
Background (CMB) radiation by the COBE satellite \cite{cobe} it 
is now possible to accurately normalize fluctuations on the 
largest observable scales \cite{cobe2}. By combining these 
observations on large scales performed with cluster-cluster 
correlation data e.g. from IRAS \cite{iras} one finds that 
it is not possible to fit well the data on all scales within 
the framework of the popular cold dark matter (CDM) model. 
The situation is illustrated in \fig{cobe_iras}. This figure
shows the measured power spectrum of density perturbations
and how it compares with the predictions of various models 
of structure formation.
\bef
\centerline{\protect\hbox{
\psfig{file=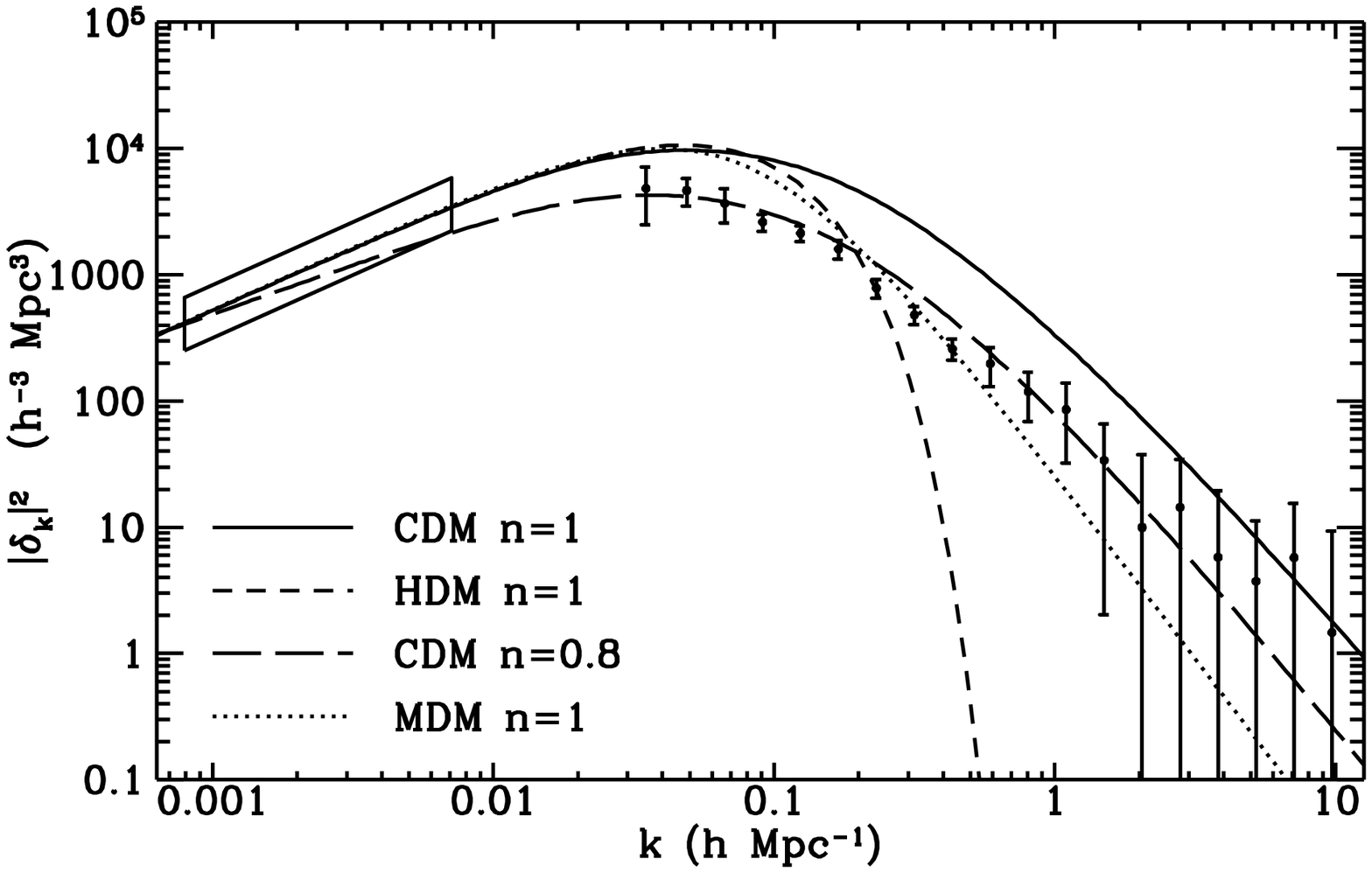,height=9cm}}}
\vglue -.5cm
\caption{Power spectrum of density perturbations}
\label{cobe_iras}
\eef
A good fit is obtained for an otherwise {\sl ad hoc} Mixed 
Dark Matter (MDM) universe, consisting of about 75\% CDM with 
about 25\% {\sl hot dark matter} (HDM), and a very small amount 
in baryons \cite{cobe2} and with the standard Harrison-Zeldovich 
n=1 spectrum predicted by inflation.

One way to make up for the hot dark matter component is 
through a massive neutrino in the few eV mass range. It has been 
argued that this could be the tau neutrino, in which case one might 
expect the existence of \ne $\ra$ \nt or \nm $\ra$ \nt oscillations. 
These are now being searched by the CHORUS and NOMAD experiments at 
CERN. These is also a similar proposal at Fermilab \cite{chorus}.

\subsubsection{Solar Neutrino Experiments}

So far the averaged data collected by the chlorine \cite{cl}, 
Kamiokande \cite{k}, as well as by the low-energy data on pp 
neutrinos from the GALLEX and SAGE experiments \cite{ga,sa} 
still pose a persisting puzzle. The most recent data can be
summarised as:
\beq
\label{data}
R_{Cl}^{exp}= (2.55 \pm 0.25) \mbox{SNU}, \,\,\,\,
R_{Ga}^{exp}= (74 \pm 8) \mbox{SNU}, \,\,\,\
R_{Ka}^{exp}= (0.44 \pm 0.06) R_{Ka}^{BP95} 
\eeq 
where  $R_{Ka}^{BP95}$ is the Bahcall-Pinsonneault (BP95) SSM 
prediction of ref. \cite{SSM}.
For the gallium result we have taken the average of the GALLEX 
$R^{exp}_{Ga}= (77\pm8\pm5)$SNU \cite{ga} and the SAGE measurements
$R^{exp}_{Ga}= (69\pm 11\pm 6)$SNU \cite{sa}. 

Comparing the data of gallium experiments with the Kamiokande data
one sees the need for a reduction of the $^7$Be flux relative to 
Standard Solar Model \cite{SSM} expectations. Inclusion of the 
Homestake data only aggravates the discrepancy, suggesting that the 
solar \neu problem is indeed a real problem. The totality of the 
data strongly suggests that the simplest astrophysical solutions 
are ruled out, and that new physics is needed \cite{CF}. The most 
attractive possibility is to assume the existence of \neu 
conversions involving very small \neu masses. 

The detection rates in the chlorine and gallium experiments are given as 
\beq
\label{rate1}
R_{Cl,Ga} = \int dE \sigma(E) \cal{P}(E)\sum_i \phi_i (E) ,
\eeq
where $\cal{P}$ is the neutrino survival probability and
the sum is over the relevant neutrino sources ($i= ^7$Be, $^8$B...) 
and $\sigma(E)$ are the corresponding neutrino cross sections. For the 
Kamiokande experiment, the detection rate is 
\beq
\label{rate2}
R_k = \int_{Th} dE \Bigl[ \sigma_{\nu_e}(E) \cal{P}(E)
+ \sigma_{\nu_x}(E)(1- \cal{P}(E))\Bigr] \phi_B (E) ,
\eeq
where $\sigma_{\nu_e}(E)$ and $\sigma_{\nu_x}(E)$ ($x=\mu,\tau$)  
are the $\nu_e\!-\!e$ and $\nu_x\!-\!e$ elastic scattering cross 
sections, respectively, and 'Th' stands for the detection energy 
threshold. In the case of sterile conversion   $\sigma_{\nu_x}=0$.    

The quantity $\cal{P}$ depends on the particle physics mechanism 
used in order to suppress the solar neutrino fluxes. 

One of the simplest mechanisms suggested in order to solve the 
solar neutrino problem is the long-wavelength or just-so oscillation.
The 68 \% C.L. allowed region of oscillation parameters
is displayed in \fig{rossi}, taken from ref. \cite{Rossi}. 
\bef
\centerline{\protect\hbox{
\psfig{file=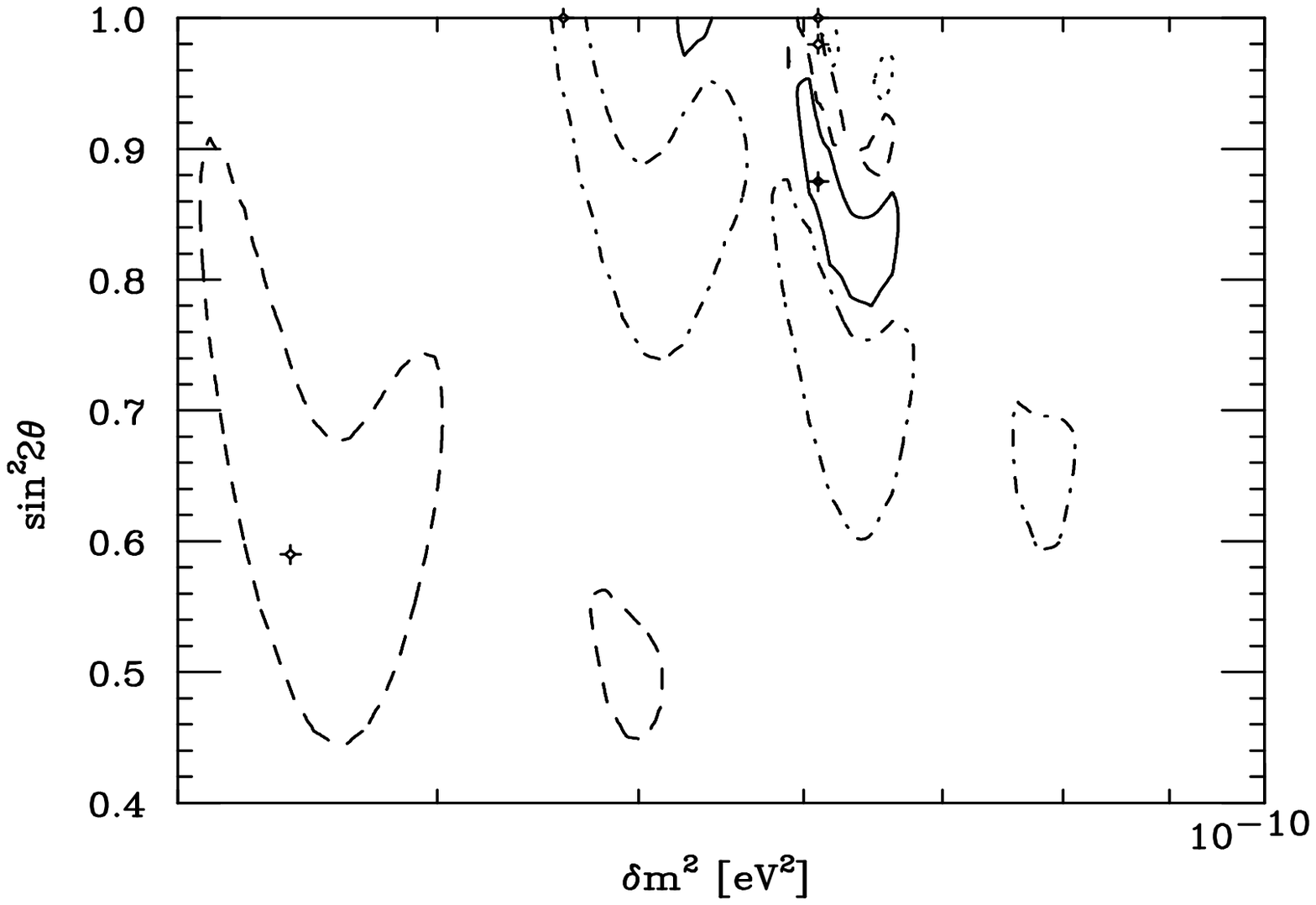,height=8.5cm}}}
\vskip 1.3cm
\centerline{\protect\hbox{
\psfig{file=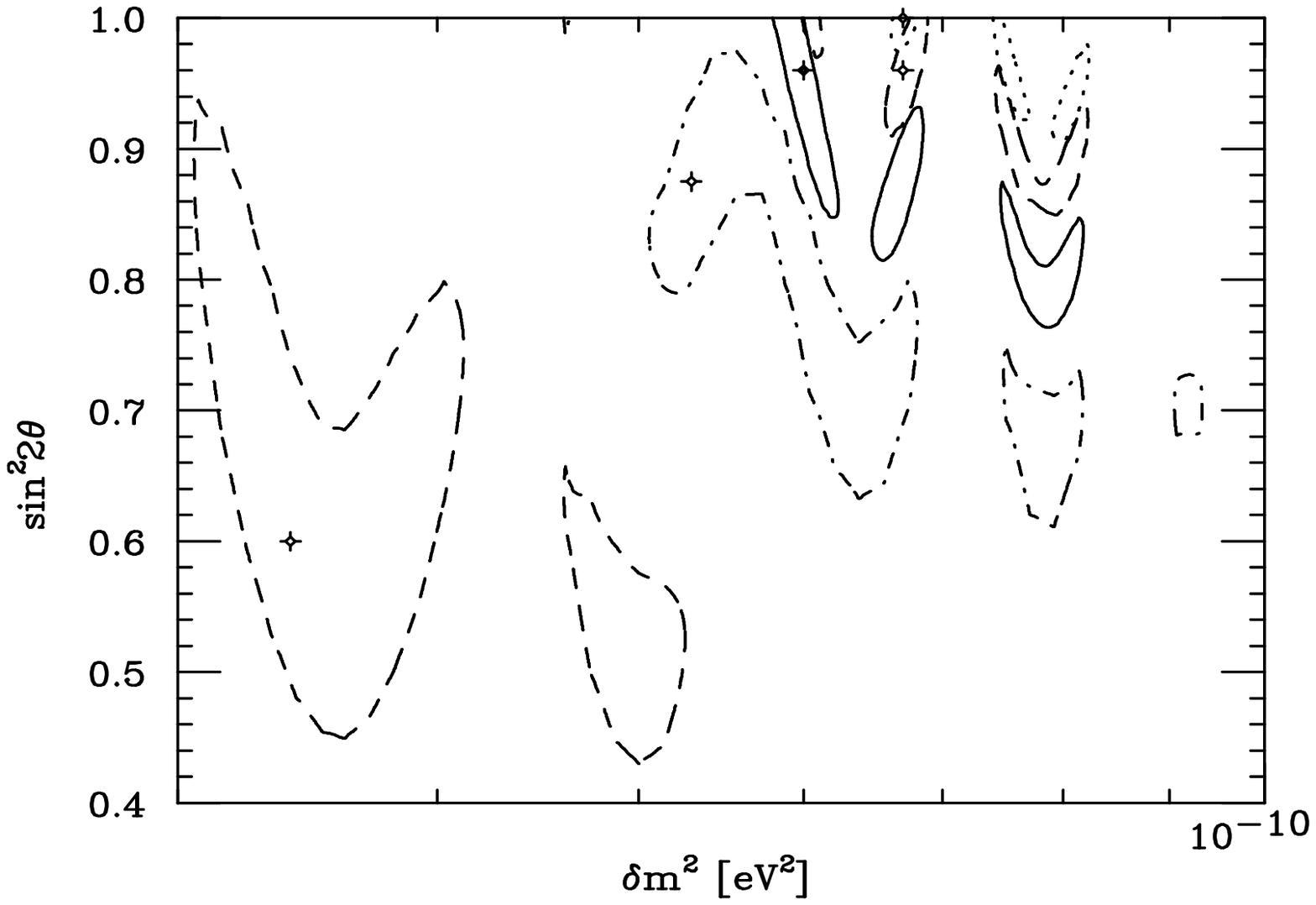,height=8.5cm}}}
\caption{Allowed parameters for active (upper figure) and 
sterile (lower figure) just-so solar neutrino oscillations.}
\label{rossi}
\eef

We now turn to the most elegant and popular solution to
the solar neutrino deficit, namely the MSW effect \cite{MSW}.
This takes into account the fact that solar neutrinos, produced
mostly close to the solar centre, interact with solar matter 
before they escape the Sun. In this case Mikheyev and Smirnov
discovered in 1985 a beautiful and very simple mechanism of 
resonantly enhancing the neutrino conversion probabilities.
In order to make detailed solar neutrino predictions in the 
framework of the MSW effect and to determine the required neutrino
parameters one needs to integrate the system of neutrino evolution 
equations describing oscillations in matter of varying density. 
The required set of solar neutrino parameters $\Delta m^2$ and 
$\sin^2 2\theta$ are determined through a $\chi^2$ fit of the 
experimental data 
\footnote{For simplicity here we neglect theoretical uncertainties,
as well as the details of the neutrino production region and the 
earth effects.}. 

In \fig{mswactive}, taken from ref. \cite{noise}, one shows the 90\% 
C.L. areas for the BP95 model for the case of active neutrino 
conversions. The fit favours the small mixing solution over the 
large mixing one, due mostly to the larger reduction of the 
$^7$Be flux found in the former.
\bef
\centerline{\protect\hbox{
\psfig{file=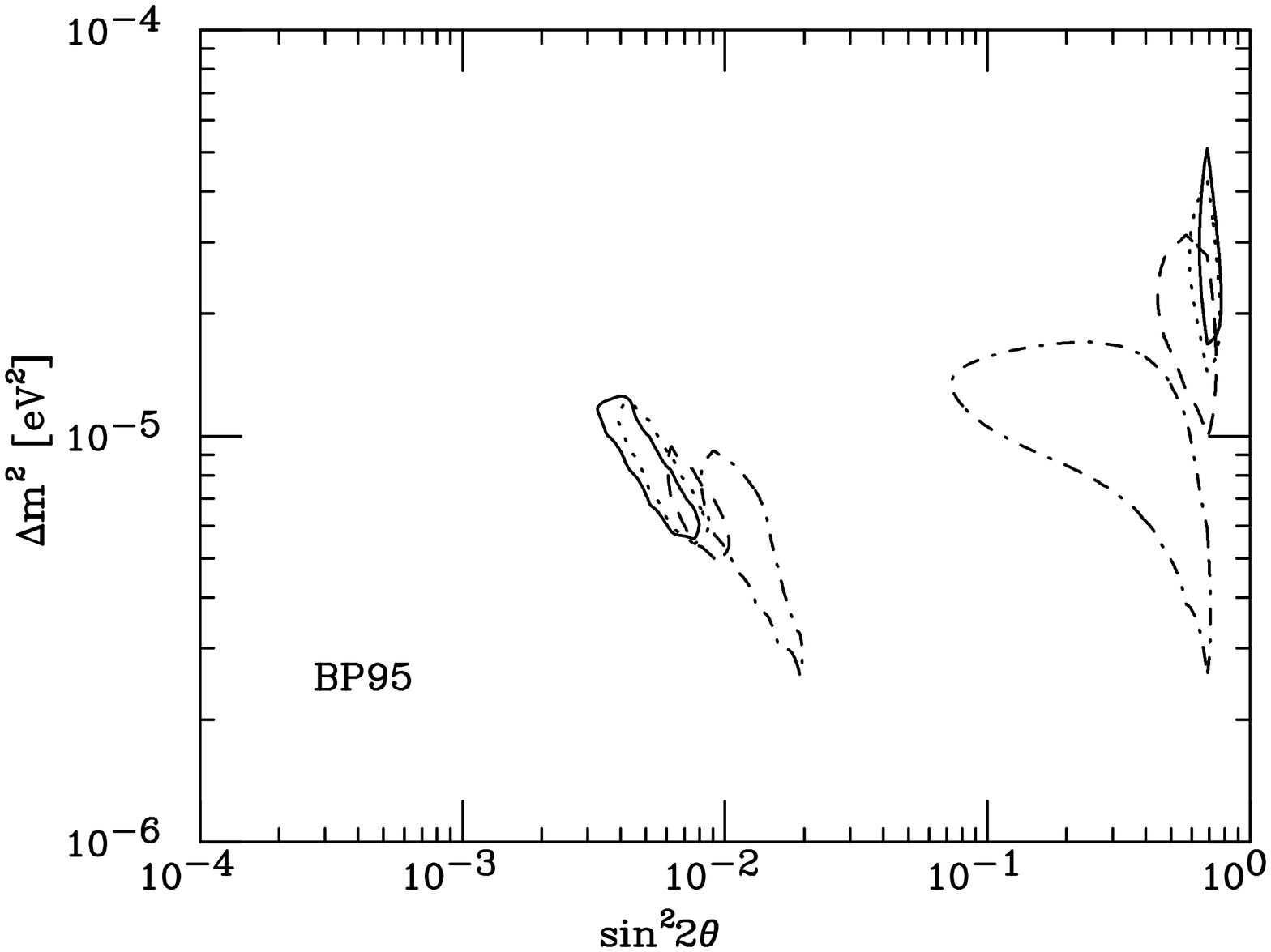,width=9cm,height=8cm,angle=90}}}
\caption{Allowed solar neutrino oscillation parameters for
active neutrino conversions.}
\label{mswactive}
\eef
Here $\xi$ denotes the assumed level of noise fluctuations in 
the solar matter density \cite{BalantekinLoreti}, not excluded by 
the SSM nor by present helioseismology studies. The solid curves 
are for the standard $\xi=0$ assumption corresponding to a smooth Sun. 
The regions inside the other curves correspond to the case
where matter density fluctuations are assumed. Noise causes
a slight shift of $\Delta m^2$ towards lower values and a 
larger shift of $\sin ^2 2 \theta$ towards larger values. 
The corresponding allowed $\Delta m^2$ range is 
$2.5 \times 10^{-6} <\Delta m^2< 9 \times 10^{-6}$ eV$^2$
instead of 
$5 \times 10^{-6} <\Delta m^2< 1.2 \times 10^{-5}$ eV$^2$
in the noiseless case.
The large mixing area is less stable, exhibiting a tendency to shift 
towards smaller $\Delta m^2$ and $\sin^2 2 \theta$. For example, if
we take $\xi=8\%$, for the sake of argument, we find that the small 
mixing region is much more stable than the large mixing one, even 
for such a relatively large value of the noise. For details, such
as some discussion of other solar models see ref. \cite{noise}. 

The results for the case of sterile solar neutrino conversions 
are given, also for the BP95 model, in Fig. 9
\bef
\centerline{\protect\hbox{
\psfig{file=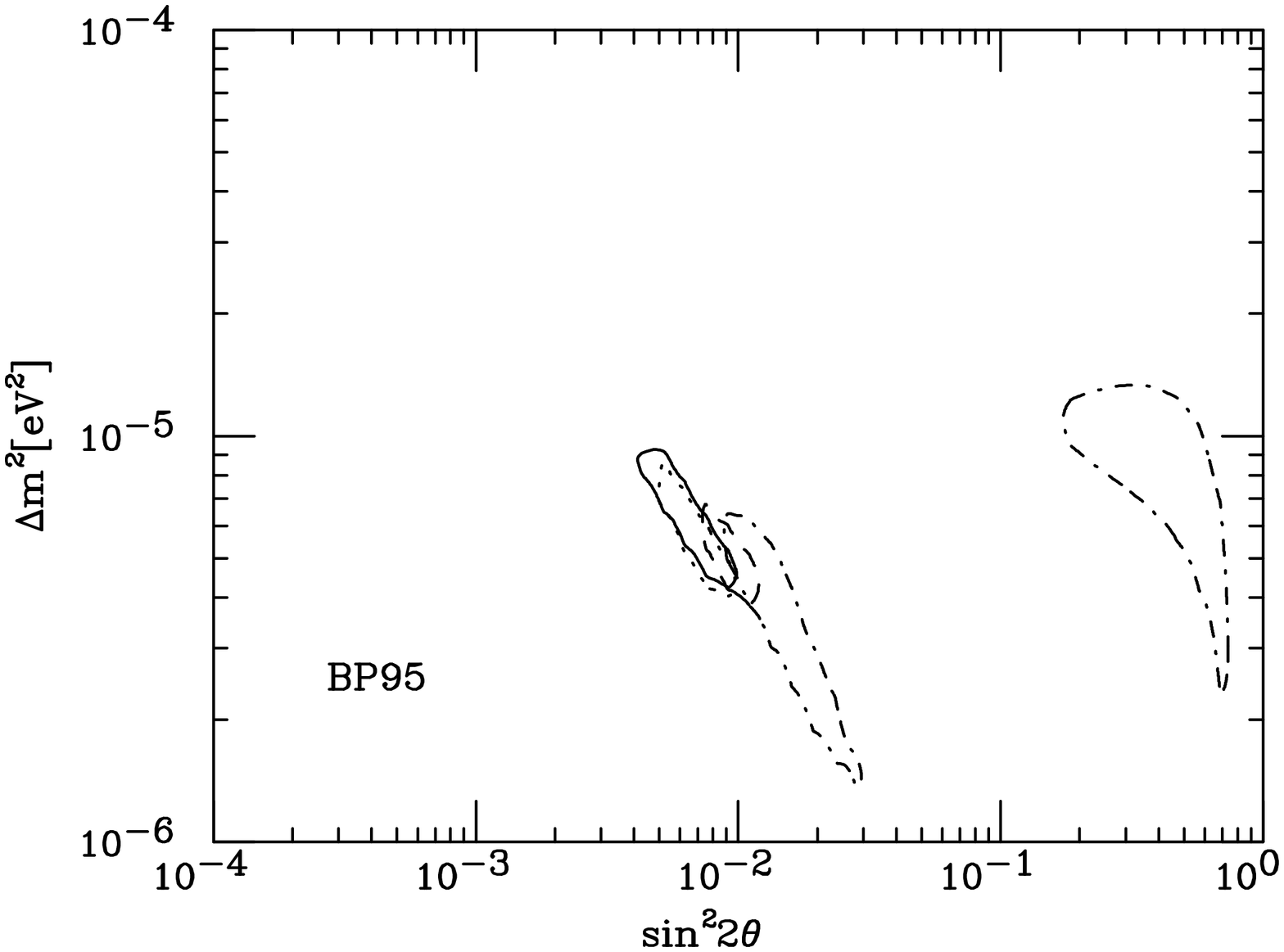,width=9cm,height=8cm,angle=90}}}
\label{mswsterile}
\caption{Allowed solar neutrino oscillation parameters for
sterile neutrino conversions.}.
\eef
The fit is now worse than for the active case and excludes, even 
at 95\% C.L., the large mixing region (in the noiseless case). 
The presence of matter density noise may restore this region 
and avoid a conflict with the primordial helium abundance 
constraints \cite{sbbn}.

As we have seen the $^7$Be neutrinos are the component of the
solar neutrino spectrum which is most affected by the presence 
of matter noise. Therefore the future Borexino experiment, aimed 
to detect the $^7$Be neutrino flux \cite{borex} through the elastic 
$\nu-e$ scattering should be an ideal tool for studying the solar 
matter fluctuations. Its potential in "testing" the level of matter 
density fluctuations in the solar interior through the measurement 
of the $^7$Be neutrino flux is illustrated in \fig{borexactive}.
The solid lines in this figure show the iso-signal contours for the 
predicted over expected (in the BP95 model) $^7$Be signal and the 
corresponding 90$\%$ allowed regions lie inside the dashed lines, 
with the best fit points denoted by a cross. The top figure is for
the smooth Sun, while the other corresponds to noise at the 4\% level.
\bef
\centerline{\protect\hbox{
\psfig{file=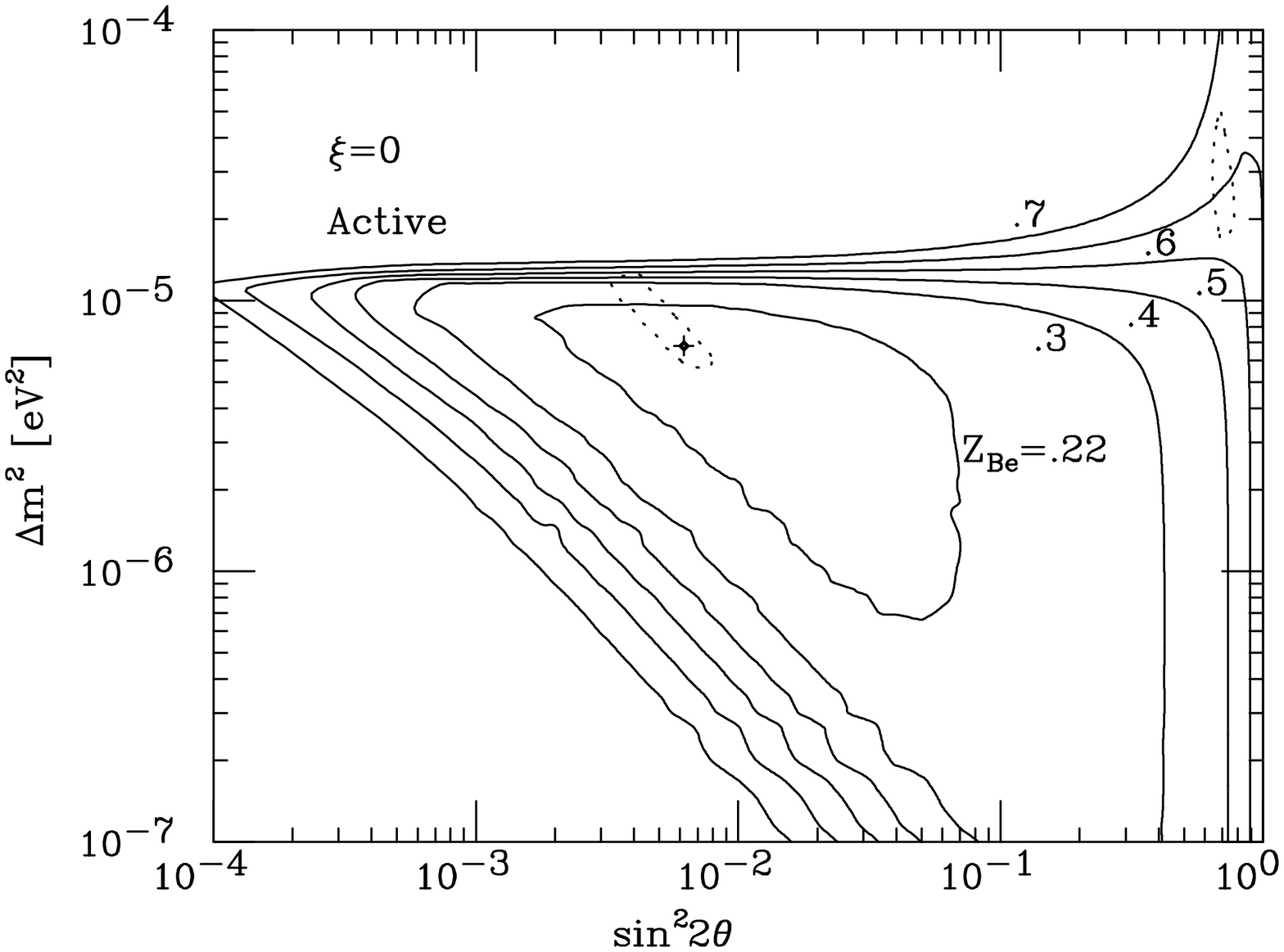,height=8cm,width=12cm,angle=90}}}
\centerline{\protect\hbox{
\psfig{file=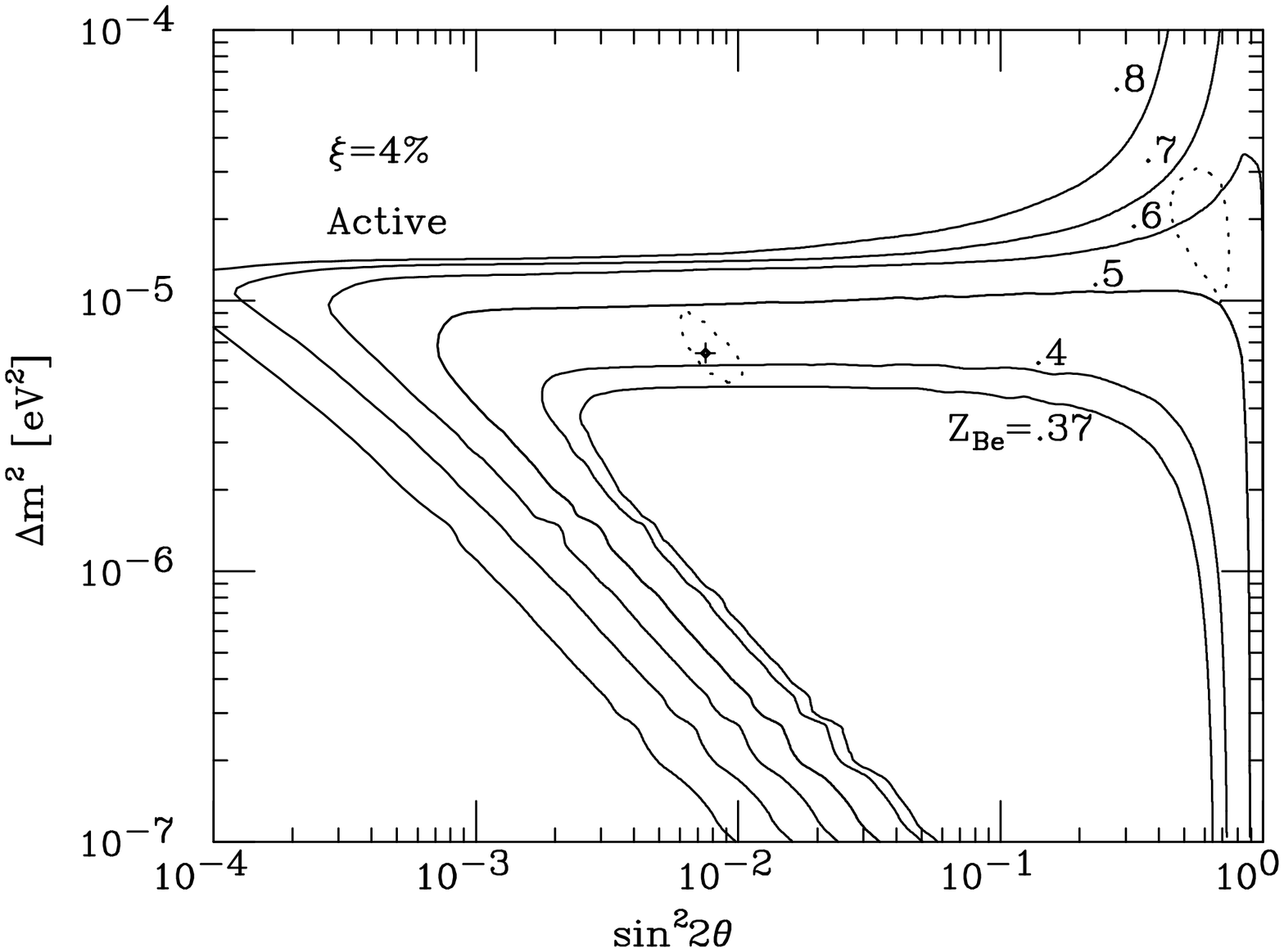,height=8cm,width=12cm,angle=90}}}
\label{borexactive}
\caption{Potential of Borexino to test solar density noise fluctuations.}
\eef
One sees that, with sufficiently small statistical errors, the Borexino 
experiment would have great potential in testing the level of density
fluctuations in the solar interior. See ref. \cite{noise} for more 
details.

\subsubsection{Atmospheric Neutrinos}

Two underground experiments, Kamiokande and IMB, and possibly 
also Soudan2, have indications which support an apparent deficit 
in the expected flux of atmospheric $\nu_\mu$'s relative to that 
of $\nu_e$'s that would be produced from conventional decays of 
$\pi$'s, $K$'s as well as secondary muon decays. Although the 
predicted absolute fluxes of \neus produced by cosmic-ray 
interactions in the atmosphere are uncertain at the 20\% level, 
their ratios are expected to be accurate to within 5\%. While 
some of the experiments, such as Frejus and NUSEX, have not 
found a firm evidence, it has been argued that there may be 
a strong hint for an atmospheric neutrino deficit that 
could be ascribed to \neu oscillations. It is not our
purpose here to give a detailed discussion, specially in
view of the controversy between water Cerenkov and iron
calorimetry experiments, for that we refer the reader to 
the experimental reviews in ref. \cite{Barish}. We will
stress, however, that Kamiokande results on higher energy 
\neus strengthen the case for an atmospheric \neu problem
due to the observed zenith-angle dependence. The relevant 
oscillation parameters are shown in ref. \cite{atm}.

\subsection{Models Reconciling Present Hints.}

Can we reconcile the present hints from astrophysics and 
cosmology in the framework of a consistent elementary 
particle physics theory? The above observations suggest 
an interesting theoretical puzzle whose possible 
resolutions will now be discussed.

\subsubsection{Three Almost Degenerate Neutrinos}

It is difficult to reconcile these three observations 
simultaneously in the framework of the simplest seesaw model 
with just the three known \neus. The only possibility to fit
these observation in a world with just the three neutrinos
of the Standard Model is if all of them have nearly the
same mass $\sim$ 2 eV \cite{caldwell}.

It is known that the general seesaw models have 
two independent terms giving rise to the light neutrino masses. 
The first is an effective triplet vacuum expectation value 
\cite{2227} which is expected to be small in left-right
symmetric models \cite{LR}. Based on this fact one can 
in fact construct extended seesaw models where the main 
contribution to the light \neu masses ($\sim$ 2 eV) is universal,
due to a suitable horizontal symmetry, while the splittings 
between \ne and \nm explain the solar \neu deficit and that 
between \nm and \nt explain the atmospheric \neu anomaly \cite{DEG}.

\subsubsection{Three Active plus One Sterile Neutrino}

The alternative way to fit all the data is to add a 
fourth \neu species which, from the LEP data on the 
invisible Z width, we know must be of the sterile type,
call it \ns. The first scheme of this type gives mass
to only one of the three neutrinos at the tree level,
keeping the other two massless \cite{OLD}. 
In a seesaw scheme with broken lepton number, radiative 
corrections involving gauge boson exchanges will give 
small masses to the other two neutrinos \ne and \nm
\cite{Choudhury}. However, since the singlet \neu is 
super-heavy in this case, there is no room to account 
for the three hints discussed above.

Two basic schemes have been suggested to keep the sterile
neutrino light due to a special symmetry. In addition to the
sterile \neu \ns, they invoke additional Higgs bosons beyond 
that of the Standard Model, in order to generate radiatively
the scales required for the solar and atmospheric \neu conversions. 
In these models the \ns either lies at the dark matter scale 
\cite{DARK92} or, alternatively, at the solar \neu scale 
\cite{DARK92B}. 
In the first case the atmospheric \neu puzzle is explained by 
\nm $\ra$ \ns oscillations, while in the second it is explained by 
\nm $\ra$ \nt oscillations. Correspondingly, the deficit of solar 
\neus is explained in the first case by \ne $\ra$ \nt conversions,
while in the second it is explained by \ne $\ra$ \ns oscillations. 
In both cases it is possible to fit all observations together. 
However, in the first case there may be a clash with the bounds 
from big-bang nucleosynthesis. In the latter case the \ns is very
light, at the solar neutrino scale, so that nucleosynthesis limits 
are satisfied and, if nucleosynthesis limits are taken at face value 
they single out the small mixing MSW solution uniquely. Moreover,
the mixing angle characterising the \nm $\ra$ \nt oscillations is 
nearly maximal, in nice agreement with ref. \cite{atm}.
Moreover, it can naturally fit the recent preliminary hints of neutrino 
oscillations of the LSND experiment \cite{Caldwell2}. Finally, 
there is another theoretical possibility is that all active \neus 
are very light, while the sterile \neu \ns is the single \neu 
responsible for the dark matter \cite{DARK92D}.

\subsubsection{MeV Tau Neutrino}

A tau neutrino with a mass in the MeV range is an 
interesting possibility to consider for two different 
reasons. On experimental side such a neutrino is within 
the range of the detectability, for example at a tau-charm
factory \cite{jj,tcf}. On the other hand, if such neutrino 
decays  before the matter dominance epoch, its decay
products could then add energy to the radiation thereby 
delaying the time at which the matter and radiation 
contributions to the energy density of the universe 
become equal. Such delay would allow one to reduce
the density fluctuations at the smaller scales \cite{latedecay} 
purely within the standard cold dark matter scenario \cite{cdm},
and could reconcile the large scale fluctuations observed by
COBE \cite{cobe} with the earlier observations such as those 
of IRAS \cite{iras} on the fluctuations at smaller scales.
An MeV  $\nu_{\tau}$ may, however, conflict with the 
big-bang nucleosynthesis picture \cite{bbncrisis}.

We present a model where an unstable MeV Majorana tau
\neu can naturally reconcile the cold dark matter model
(CDM) with cosmological observations of large and small 
scale density fluctuations and, simultaneously, with data 
on solar and atmospheric neutrinos. The solar \neu deficit 
is explained through long wavelength, so-called 
{\sl just-so} oscillations involving conversions of \ne 
into both \nm and a sterile species \ns, while atmospheric
\neu data are explained through \nm to \ne conversions.
Future long baseline \neu oscillation experiments, as
well as some reactor experiments should test this hypothesis. 
The model is based on the spontaneous violation
of a global lepton number symmetry at the weak scale. 
This symmetry plays a key role in generating the 
cosmologically required decay of the \nt with lifetime
$\tau_{\nu_\tau} \sim 10^2 - 10^4$ seconds, as well 
as the masses and oscillations of the three light 
\neus \ne, \nm and \ns required in order to account for solar
and atmospheric \neu data. It also leads to the
invisibly decaying Higgs signature that can be
searched at LEP and future particle colliders.

\section{ Electroweak Symmetry Breaking: The Higgs Sector}

In the Standard Model the spontaneous breaking of the \gau 
symmetry follows from energetics, namely the minimum of the
so-called Higgs potential favours a non-zero value for
the scalar field VEV in \eq{WEAKSCALE}. This eliminates three of 
the four degrees of freedom present in the complex scalar 
doublet $\phi$ in favour of the longitudinal degrees of freedom
of the $W^\pm$ and the $Z$. 
The surviving electrically neutral Higgs scalar, the so-called 
Standard Model Higgs boson, has a mass given by
\beq
m_h \propto \sqrt{\lambda} \VEV{\phi}
\eeq
where $\lambda$ is the quartic coupling in the Higgs potential.
A great effort has been devoted in designing a search for the 
Standard Model Higgs boson. Clearly this is one of the main open 
questions of the Standard Model \cite{electroweakefforts}.

\subsection{Standard Model Higgs}

Unfortunately both the mass and self-coupling strengths of the
Higgs boson are undetermined by the theory. There is, however,
an upper bound on the Higgs boson mass that follows from requiring 
that our simple theoretical picture be valid up to a given 
cutoff scale $\Lambda$, as otherwise a Landau pole would
develop in the renormalization group equation describing the
evolution of the quartic coupling $\lambda$ in the Higgs 
potential. This upper bound depends on the top quark mass through
the renormalization group equation, as illustrated in \fig{sher}.
\bef
\centerline{\protect\hbox{
\psfig{file=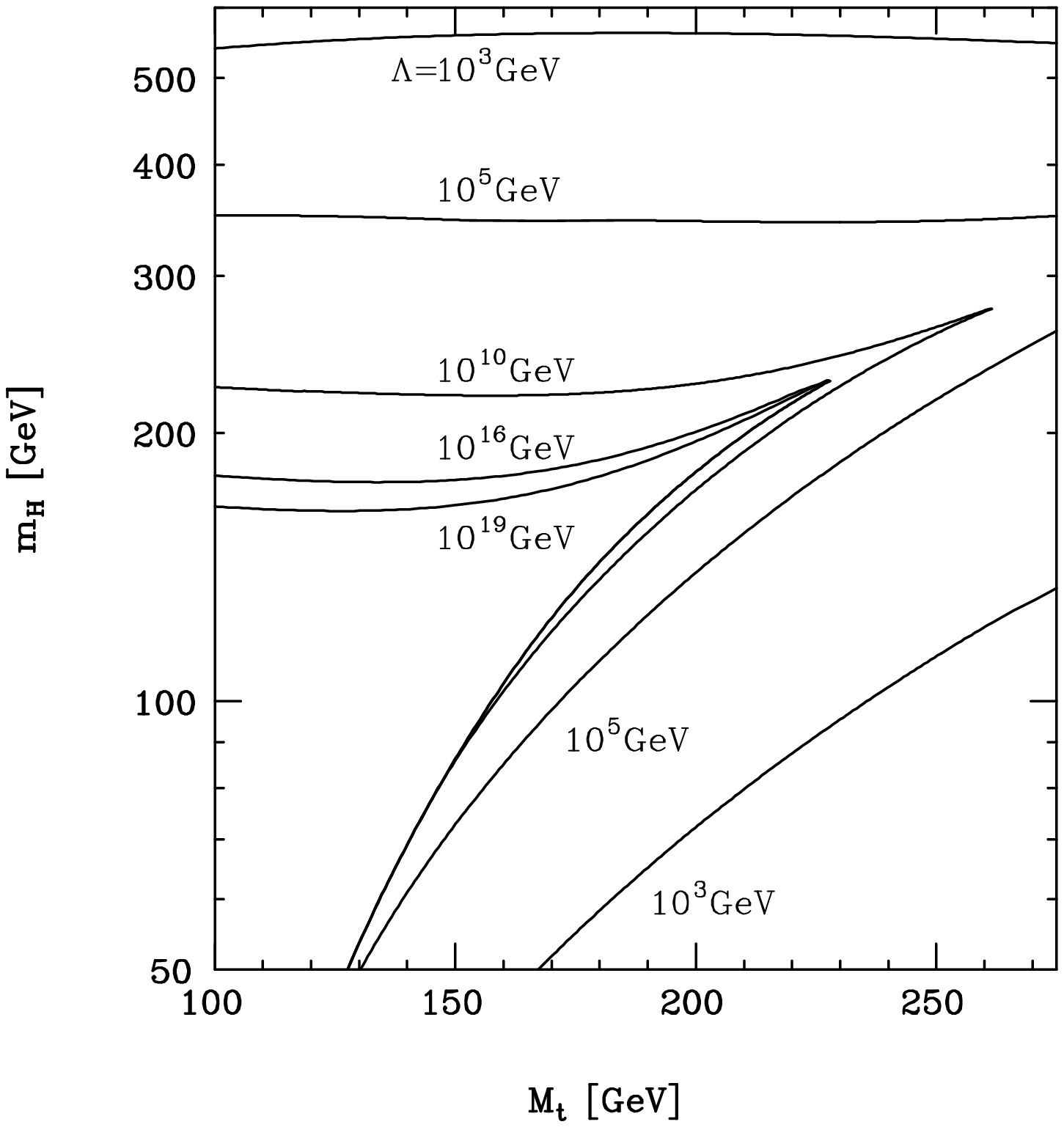,height=8.5cm,angle=90}}}
\caption{Theoretical Bounds on the Standard Model Higgs Boson.}
\label{sher}
\eef 
The limit varies from $m_H \ltap 600 \: GeV$ down to
$m_H \ltap 200 \: GeV$ if one also assumes that there is no new 
physics below $M_{Planck}$ \cite{Sher}. 

There is also a theoretical lower bound on the Higgs boson mass 
obtained by requiring vacuum stability. Since the contribution 
of the top quark Yukawa coupling to the beta function of the scalar 
self-coupling, $\lambda$, is negative, the large top quark mass 
drives $\lambda$ to a negative value, thus destabilizing the 
Standard Model vacuum. This instability can be avoided by 
requiring that the Higgs mass be sufficiently large \cite{Sher}.
Indeed a lower bound on the Higgs mass can be obtained by requiring 
that the Standard Model vacuum be the only stable minimum up to that 
scale.  Requiring of vacuum stability up to the Planck scale gives, 
for a top quark mass of 175 GeV, a lower bound of 130 GeV on the 
Higgs boson mass \cite{hinstability}
\footnote{One may refine this limit by requiring only a meta-stable 
potential with a lifetime longer than the age of the universe.}.
If the Higgs mass is lighter 
than this bound, then the Standard Model must break down at some
scale $\Lambda$. At energy scales below $\Lambda$, the 
physics beyond the Standard Model generally decouples, leaving 
a low-energy effective theory almost identical to the Standard Model.  
However, the Higgs boson allows us to probe $\Lambda$ through
the above stability constraint on the Higgs mass, and due to 
the large top quark mass.  

By combining these two arguments one obtains for the allowed region 
for Higgs boson and top quark masses the one delimited by the solid 
lines in \fig{sher}.

This means that if a smaller Higgs boson mass of 100 GeV is 
discovered then the Standard Model Higgs boson is ruled out
if there is no physics below $\Lambda=M_{Pl}$. If, on the
other extreme, $\Lambda$ is rather close to the TeV scale, 
one would expect the lightest Higgs boson to retain all the
properties of the so-called Standard Model Higgs boson.

\begin{figure*}
\centerline{\protect\hbox{
\psfig{file=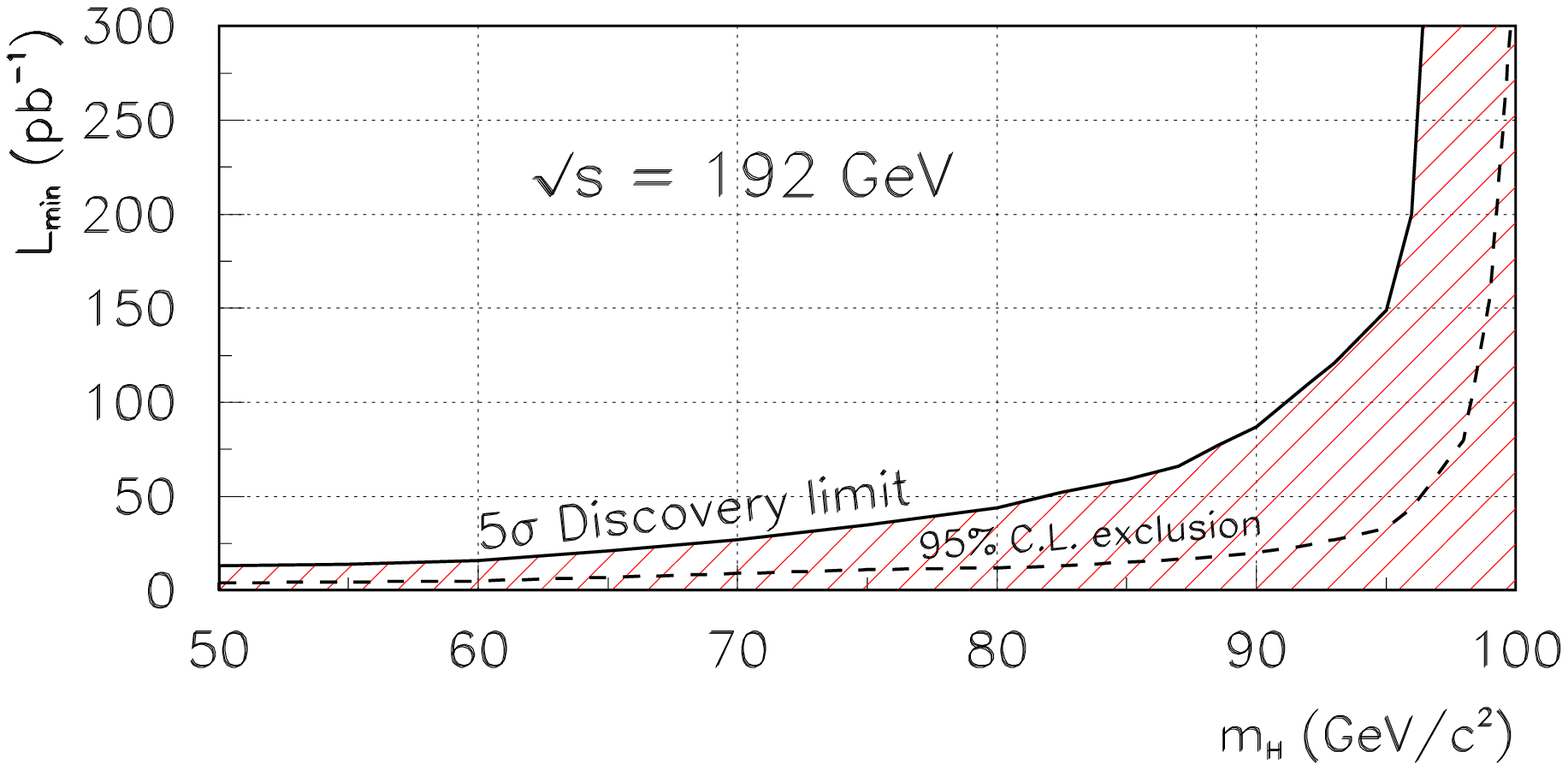,width=13.5cm}}}
\caption{Standard Model Higgs boson search potential at LEP200.}
\label{lumin2}
\end{figure*}

Now we turn to the experimental limits on the Higgs boson mass.
In order to determine this one needs to know the Higgs boson 
decays which follow from its interactions with the fermions 
as well as gauge bosons. The interaction of the physical Higgs 
with the mass eigenstate fermions is clearly diagonal when 
written in the same basis that diagonalizes the fermion masses
\eq{DIAG}. As a result there are no flavour changing neutral 
currents mediated by Higgs boson exchange in the Standard Model.
Moreover, the couplings of the physical Higgs boson to both 
gauge boson and fermions is proportional to 
their mass. This leads to a well defined Higgs boson decay pattern,
which is of fundamental importance in designing the strategies for
Higgs searches at accelerators. Indeed, the Higgs boson will decay 
mostly to the heaviest kinematically accessible particle, as
illustrated in \fig{higgsbr}.
\bef
\centerline{\protect\hbox{
\psfig{file=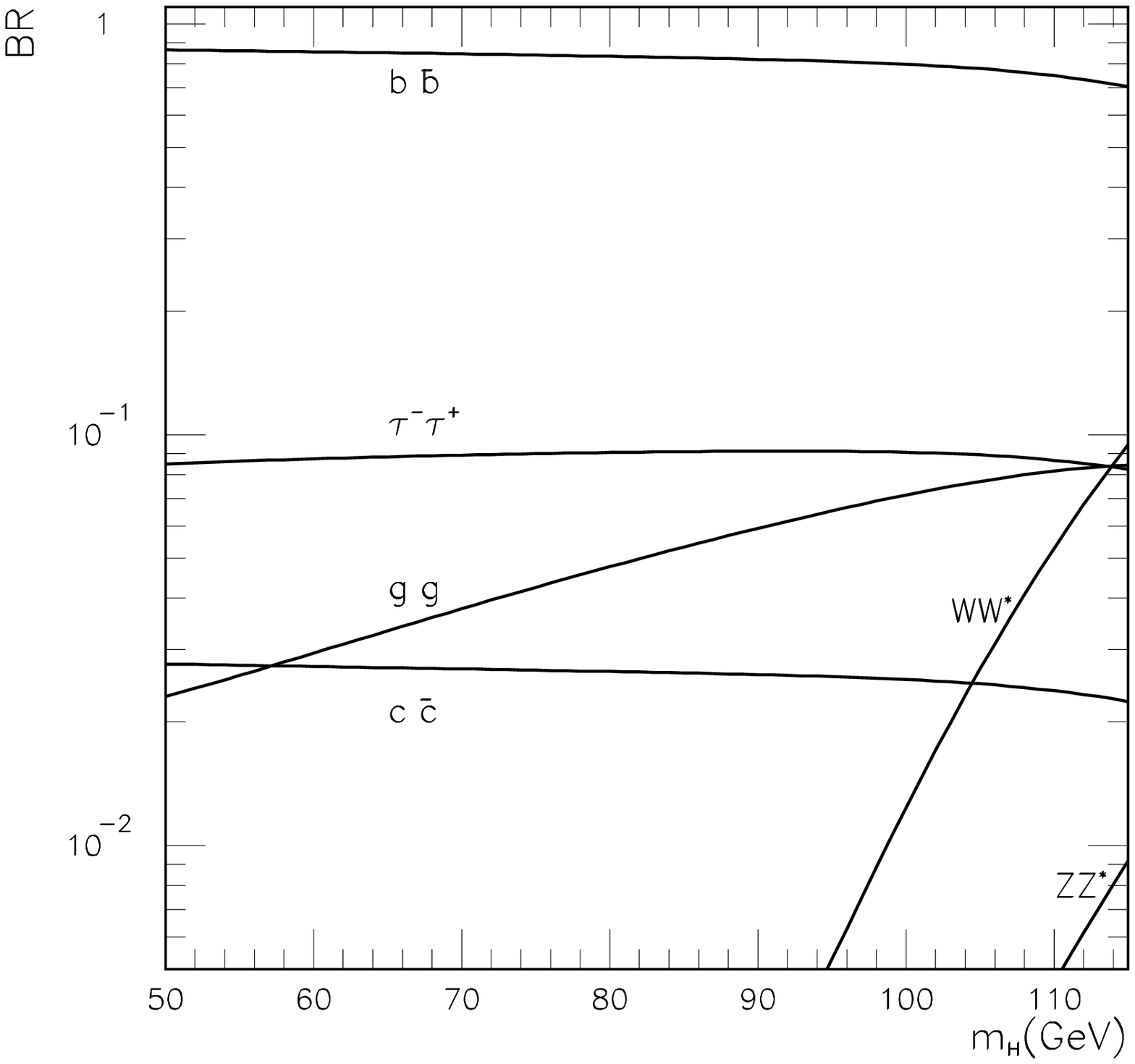,width=7.0cm,height=8cm}}}
\vglue -1cm
\caption{Standard Model Higgs decay branching ratios 
in mass range accessible at LEP.}
\label{higgsbr}
\eef 
From the analysis of the data collected at LEP one can place
the following lower limit on the Standard Model Higgs boson mass 
\cite{LEPSEARCH}
\beq
\label{hlimit}
m_H \gtap 65 \: GeV
\eeq
As illustrated in \fig{lumin2}, if lighter than $\sim 100$ GeV,
the Standard Model Higgs boson should be found at LEP 200 
\cite{Treille}. The minimum required luminosity per experiment, 
in pb$^{-1}$, for a 5$\sigma$ Higgs boson discovery is displayed 
in the solid line of  \fig{lumin2}, while the corresponding 95$\%$ C.L. 
exclusion limit is shown as dashed. Results for other centre-of-mass 
energies can be found in refs. \cite{Treille,CarenaZerwas}.
Higher Higgs boson masses can be probed at higher centre-of-mass 
energies, such as expected in the next linear collider (NLC), or
or at the LHC. Unfortunately the prospects for finding the Higgs 
boson in the intermediate mass range between $m_Z$ and $2m_Z$ at
the LHC are not too optimistic \cite{Denegri}. Above this mass 
the detection would be very easy, through the 4-lepton
signal, as illustrated in Fig. 9, 10, 11 and 12 of 
ref. \cite{Denegri}. 

From the above vacuum stability argument it follows that if the 
Higgs has a mass just above its current experimental limit, then 
the Standard Model must break down at a scale of roughly a TeV.  
Since, by definition, the Standard Model assumes that there is no 
new physics until a scale of several TeV (at least), it follows 
that the discovery of a Higgs boson at LEP 200 could, depending 
on the precise top quark mass, rule out the Standard Model \cite{Sher2}.

Let us now move back to experiment and note that one 
can get indirect experimental information on the Higgs boson mass 
just on the basis of precision tests of the electroweak theory.
In addition to testing the Standard Model, one has the possibility 
of constraining the value of the Higgs mass, which enters through 
the radiative corrections to the $Z$ and $W$ boson self-energies.  
Combining the most recent LEP and SLC electroweak results 
\cite{precision95} with the recent top-quark mass measurement
at the Tevatron, \cite{top} a weak preference is found
for a light Higgs boson mass of order $m_Z$ \cite{precision95}.
One can illustrate the situation by showing in \fig{SMFIT} a 
typical chi-square Standard Model fit constraining the 
Standard Model Higgs mass.  
\bef
\centerline{\protect\hbox{
\psfig{file=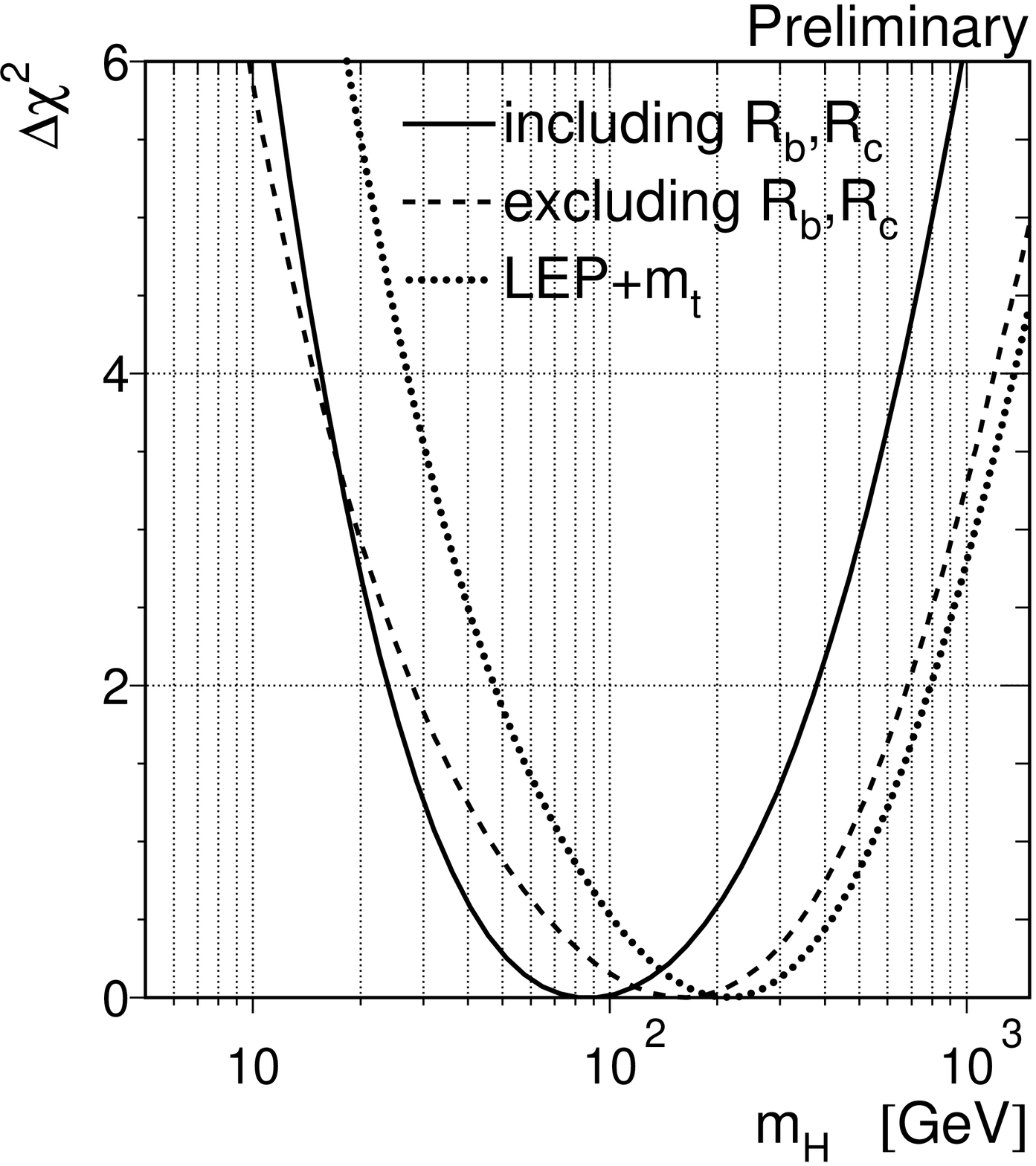,width=8cm,height=10cm}}}
\vglue -1cm
\caption{Standard Model Higgs mass determination from precision data.}
\label{SMFIT}
\eef 
The solid line includes all LEP, SLD, $p \bar{p}$ and deep
inelastic neutrino data, while the dashed one excludes the 
measurements of the Z width into $b\bar{b}$ and $c\bar{c}$.
The dotted line corresponds to the LEP data including $R_b$
and $R_c$. In all cases one includes the direct top mass determination
from the Tevatron.

\subsection{Majoron Models}
\label{majorons} 

Many extensions of the lepton sector seek to give masses to 
\neus involve the spontaneous violation of an ungauged U(1) 
lepton number symmetry. Although the original Majoron proposal 
was made in the framework of the minimal seesaw model, and 
required the introduction of a relatively high energy scale 
associated to the mass of the right-handed \neus \cite{CMP}, 
there are many attractive theoretical alternatives where 
lepton number is violated spontaneously at the weak scale or 
lower. Here we are interested precisely in these models, of
which there are several examples. In some of these models the
neutrinos acquire mass at the tree level \cite{CON}, while
in others it may appear either at one-loop or two-loop level 
in radiative corrections \cite{JoshipuraValle92,Zurab}.
The main common feature of these models is that the associated 
neutrino mass vanishes as the scale of spontaneous lepton number 
violation goes to zero, in sharp contrast to the standard seesaw 
idea which exhibits an inverse relationship. 

All these models imply the existence of a physical Goldstone 
boson, generically called Majoron. The existence of such Majoron 
is consistent with the LEP measurements of the invisible $Z$ decay 
width if the Majoron is (mostly) a singlet under the \21 \gau symmetry.
Although the Majoron has very tiny couplings to matter and \gau bosons, 
it can have significant couplings to the Higgs bosons, leading to 
the possibility that the Higgs boson may decay with a substantial 
branching ratio into the channel \cite{JoshipuraValle92}
\begin{equation}
h \rightarrow J\;+\;J
\label{JJ}
\end{equation}
Since the Majoron $J$ is weakly coupled to the rest of the 
particles, once produced in the accelerator, it will escape
detection, leading to a missing momentum signal. Since the 
strategies to search for the Higgs boson depend heavily on 
its expected decay pattern, the presence of such an invisible 
decay signal affects them in a very remarkable way.

\subsubsection{Simplest Model}
\label{hf}

Models in this class contain one doublet and one singlet Higgs
multiplet. They include both models with tree level generation \cite{CON},
as well as radiative \cite{Babu88}. They are characterized by a very 
simple form of the scalar potential given by:
\beq
\label{V1}
V_{N_1} = \mu_{\phi}^2\phi^{\dagger}\phi
+\mu_{\sigma}^2\sigma^{\dagger}\sigma
+ \lambda_{1}(\phi^{\dagger}\phi)^2 + 
\lambda_2
(\sigma^{\dagger}\sigma)^2
+\delta (\phi^{\dagger}\phi)(\sigma^{\dagger}\sigma)
\eeq
Terms like $\sigma^2$ are omitted above in view of the imposed
$U(1)$ invariance under which we require $\sigma$ to transform
non-trivially and $\phi$ to be trivial. Let $\sigma \equiv
\frac{w}{\sqrt 2}+\frac{R_2+iI_2}{\sqrt{2}}$, $\phi^0\equiv
\frac{v}{\sqrt 2}+\frac{R_1+iI_1}{\sqrt {2}}$, where we have set
$\VEV{\sigma} =\frac{w}{\sqrt{2}}$ and $\VEV{\phi^0}=\frac{v}{\sqrt{2}}$.
The above potential then leads to a physical massless Goldstone
boson, namely the Majoron $J \equiv {\rm Im}\; \sigma$ and two
massive neutral scalars $H_i$ ($i$= 1,2)
\beq
H_i={\hat O}_{ij}\;R_j
\eeq
The mixing ${\hat O}$ can be parametrized as
\begin{equation}
{\hat O}=\left( \begin{array}{cc}
cos\theta&sin\theta\\
-sin\theta&cos\theta\\
\end{array} \right)
\end{equation}
mixing angle $\theta$ as well as the Higgs masses $M_i^2$
are related to the parameters of the potential in the
following way:
\begin{eqnarray}
\label{teta}
2\delta v w &=& (M^2_2-M^2_1) \sin 2 \theta \nonumber\\
2 \lambda _1 v^2&=&M^2_1 \cos ^2\theta+M^2_2 \sin ^2 \theta \nonumber\\
2 \lambda _2 w^2&=&M^2_2 \cos ^2\theta+M^2_1 \sin ^2 \theta. \nonumber\\
tan2\theta&=&-\frac{\delta v \omega}{\lambda_1v^2 - \lambda_2\omega^2}
\end{eqnarray}
The  masses $M_{1,2}^2$, the mixing angle $\theta$, and the ratio
of two vacuum expectation values $\tan  \beta =\frac{v}{w}$
can be taken as independent parameters in terms of which all
couplings can be fixed. There are no physical charged Higgs
bosons in this case.

The potential in \eq{V1} generates the
following coupling of $H_i$ to the Majoron $J$:
\begin{equation}
\label{J1}
{\cal L}_J=
\frac{(\sqrt 2 G_F)^{1/2}}{2}\tan  \beta[M_2^2 cos \theta H_2-M_1^2 sin
\theta H_1]J^2
\end{equation}
The Higgs decay to two Majorons follows immediately from this equation.

{\bf Invisibly Decaying Higgs boson searches in the 
$e^+ e^- \ra H \:Z$ channel}

The production and subsequent decay of a Higgs boson which may decay 
visibly or invisibly via the process $e^+ e^- \ra H \:Z$ production
involves three independent parameters: its mass $M_H$, its coupling 
strength to the Z boson, normalized by that of the Standard Model,
$\epsilon^2$, and its invisible decay branching ratio. 
The LEP searches for various exotic channels can be used in order to 
determine the regions in parameter space that are already ruled out, 
as described in ref. \cite{alfonso}. The exclusion contour in the plane 
$\epsilon^2$ vs. $M_H$, can be found in ref. \cite{alfonso}.

The invisible decay of the Higgs boson may also affect the 
strategies for searches at higher energies. For example, the 
ranges of parameters that can be covered by LEP200 searches 
for various integrated luminosities and centre-of-mass energies 
have been investigated \cite{ebolepp2}, and the results are 
illustrated in \fig{bjlep2}.
\bef
\centerline{\protect\hbox{
\psfig{file=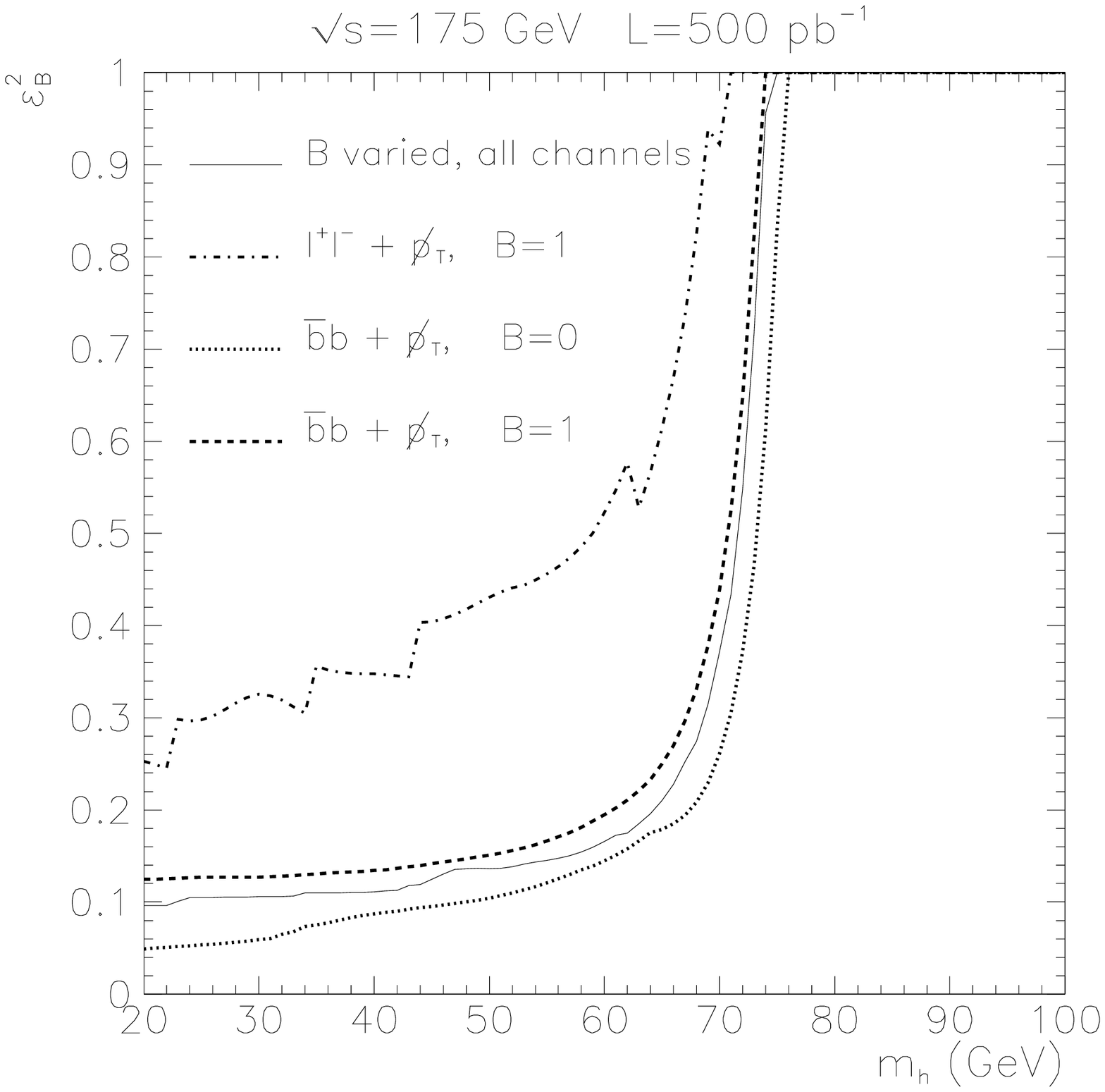,height=11cm}}}
\caption{Higgs mass and coupling that can be explored at 
LEP200 in $e^+ e^- \ra H \:Z$ production.}
\label{bjlep2}
\eef
Similar analysis can be made for the case of a high energy linear 
$e^+ e^-$ collider (NLC) \cite{EE500}, as well as the LHC 
\cite{granada}. In the latter case the invisible decay has 
an advantage for searches in the intermediate mass region,
namely, that the invisible decay branching ratio can be of order 1, 
while the standard $H \ra \gamma \gamma$ decay branching ratio 
in either the SM or the MSSM is rather small, \O$(10^{-3}$).
Although it can lead to sizeable signals, the invisible decay has the 
disadvantage that the Higgs mass can not be reconstructed at a 
hadron collider. In any case, Higgs boson masses in this range 
can be probed in less than a year LHC running.
However, the NLC would be a cleaner machine for invisibly
decaying Higgs boson searches beyond the LEP200 reach.

\subsubsection{Two Scalar Higgs Doublet Extensions}
\label{hhf}

The models of this class include the Majoron embedding 
\cite{JoshipuraValle92} of the original Zee model of 
radiative neutrino mass generation \cite{zee}. In this 
model a second doublet of scalar bosons is required in order
to close the loop diagram yielding the neutrino masses
\footnote{Another type of model of this type is the 	
model discussed in section 4. In that model neutrinos get
mass at the tree level. The presence of two doublets of 
scalar bosons, is  required by supersymmetry, leading to
the existence of a massive CP-odd scalar Higgs boson.}.

The part of the scalar potential containing the 
neutral Higgs fields is given in this case by
\begin{eqnarray}
\label{N2}
V_{N2}&=&\mu_{i}^2\phi^{\dagger}_i\phi_i+\mu_{\sigma}^2\sigma^{\dagger}
\sigma
+ \lambda_{i}(\phi^{\dagger}_i\phi_i)^2+
   \lambda_{\sigma}
(\sigma^{\dagger}\sigma)^2+\nonumber\\
& &\lambda_{12}(\phi^{\dagger}_1\phi_1)(\phi^{\dagger}_2\phi_2)
+\lambda_{13}(\phi^{\dagger}_1\phi_1)(\sigma^{\dagger}\sigma)+
\lambda_{23}(\phi^{\dagger}_2\phi_2)(\sigma^{\dagger}\sigma)\nonumber\\
& & +\delta(\phi^{\dagger}_1\phi_2)(\phi^{\dagger}_2\phi_1)+
    \frac{1}{2}\beta[(\phi^{\dagger}_1\phi_2)^2+\;\;h.\;c.]
\end{eqnarray}
where a sum over repeated indices $i$=1,2 is assumed.
Here $\phi_{1,2}$ are the doublet fields and $\sigma$
corresponds to the singlet carrying nonzero lepton number.

In writing down the above equation, we have imposed a discrete
symmetry $\phi_2\rightarrow-\phi_2$ needed to obtain natural
flavour conservation in the presence of more than one Higgs
doublets. For simplicity, we assume all couplings and VEVS
to be real. Then the conditions for the minimization of the
above potential are easy to work out and are given by
\begin{equation}
\mu_1^2+v_1^2\lambda_1+\frac{1}{2}(\lambda_{12}+\delta)v_2^2+\frac{1}{2}
\lambda_{13}v_3^2+\frac{1}{2}\beta v_2^2=0 \nonumber
\end{equation}
\begin{equation}
\mu_2^2+v_2^2\lambda_2+\frac{1}{2}(\lambda_{12}+\delta)v_1^2+\frac{1}{2}
\lambda_{23}v_3^2+\frac{1}{2}\beta v_1^2=0
\end{equation}
\begin{equation}
\mu_{3}^2+v_3^2\lambda_3+\frac{1}{2}\lambda_{13}v_1^2 +
                         \frac{1}{2} \lambda_{23}v_2^2 = 0 \nonumber
\end{equation}

These conditions can be used to work out the mass matrix for
the Higgs fields. To this end we shift the fields as ($i$=1,2)
\begin{equation}
\phi_i=\frac{v_i}{\sqrt{2}}+\frac{R_i+iI_i}{\sqrt{2}} \nonumber
\end{equation}
\begin{equation}
\sigma=\frac{\omega}{\sqrt2}+\frac{R_3+iI_3}{\sqrt{2}}
\end{equation}
The masses of the CP-even fields $R_a$ ($a$=1...3) are obtained from
\begin{equation}
{\cal L}_{mass}= - \frac{1}{2}R^T\;M_R^2\;R
\end{equation}
with
\begin{equation}
\label{MR}
M_R^2=\left (
\begin{array}{ccc}
2\lambda_1v_1^2&(\beta+\lambda_{12}+\delta)v_1v_2&\lambda_{13}v_1v_3\\
(\beta+\lambda_{12}+\delta)v_1v_2&2\lambda_2v_2^2&\lambda_{23}v_2v_3\\
\lambda_{13}v_1v_3&\lambda_{23}v_2v_3&2\lambda_3v_3^2\\
\end{array}  \right)
\end{equation}
The physical mass eigenstates $H_a$ are related to
the corresponding weak eigenstates as
\begin{equation}
\label{me}
H_a=O_{ab}\;R_b
\end{equation}
where,  $O$ is a 3$\times$3 matrix diagonalizing $M_R^2$
\begin{equation}
\label{diag}
O\;M_R^2\;O^T=diag \: (M_1^2,M_2^2,M_3^2)
\end{equation}
The  Majoron is  given in  this  case by  $J=I_3$.  The  coupling of  the
physical Higgses to $J$ follows from  \eq{N2}. As in the previous case,
it is possible to express this coupling entirely in terms of the masses
$M_a^2$ and the mixing angles characterising the matrix $O$
\begin{eqnarray}
\label{J2}
{\cal L}_J&=&\frac{1}{2}J^2(2\lambda_3v_3R_3+\lambda_{13}v_1R_1+
           \lambda_{23}v_2v_3R_2)\\
         &=&\frac{J^2}{2v_3}(M_R^2)_{3a}R_a\\
       &=&\frac{1}{2}(\sqrt{2}G_F)^{1/2}tan\gamma (O^T)_{3a}M_a^2H_a J^2
\end{eqnarray}
tan$\gamma\equiv\frac{V}{v_3}$; $ V=(v_1^2+v_2^2)^{1/2}$.
We have made use of \eq{me} and \eq{diag} in writing the last line.

Unlike in the previous case, now there exists also a massive
CP-odd state A, related to the doublet fields as follows
\begin{equation}
A=\frac{1}{V}(v_2I_1-v_1I_2)
\end{equation}
Its mass is given by
\begin{equation}
M_A^2=-\beta V^2
\end{equation}
When $\beta \to 0$ this pseudoscalar boson becomes massless,
as the potential acquires a new symmetry.

{\bf Invisibly Decaying Higgs boson searches in the 
$e^+ e^- \ra H \:A$ channel}

Due to the existence of two \21 doublets of scalar bosons,
this class of extended Majoron models of neutrino mass predict
another mode of production of invisibly decaying Higgs bosons,
namely one in which a CP-even Higgs boson is produced in association 
with a massive CP-odd scalar. 

Present LEP1 limits on the corresponding coupling strength parameter 
were given in ref. \cite{HA}.  The region of parameters that can be
explored at LEP200 is shown in \fig{halep2}, as a function of 
the A and H masses, for the case of a visibly decaying A boson and 
an invisibly decaying H boson. This figure is taken from ref. 
\cite{ebolepp2} which contains extensive discussion of various 
integrated luminosities and centre-of-mass energy assumptions. 
\bef
\centerline{\protect\hbox{
\psfig{file=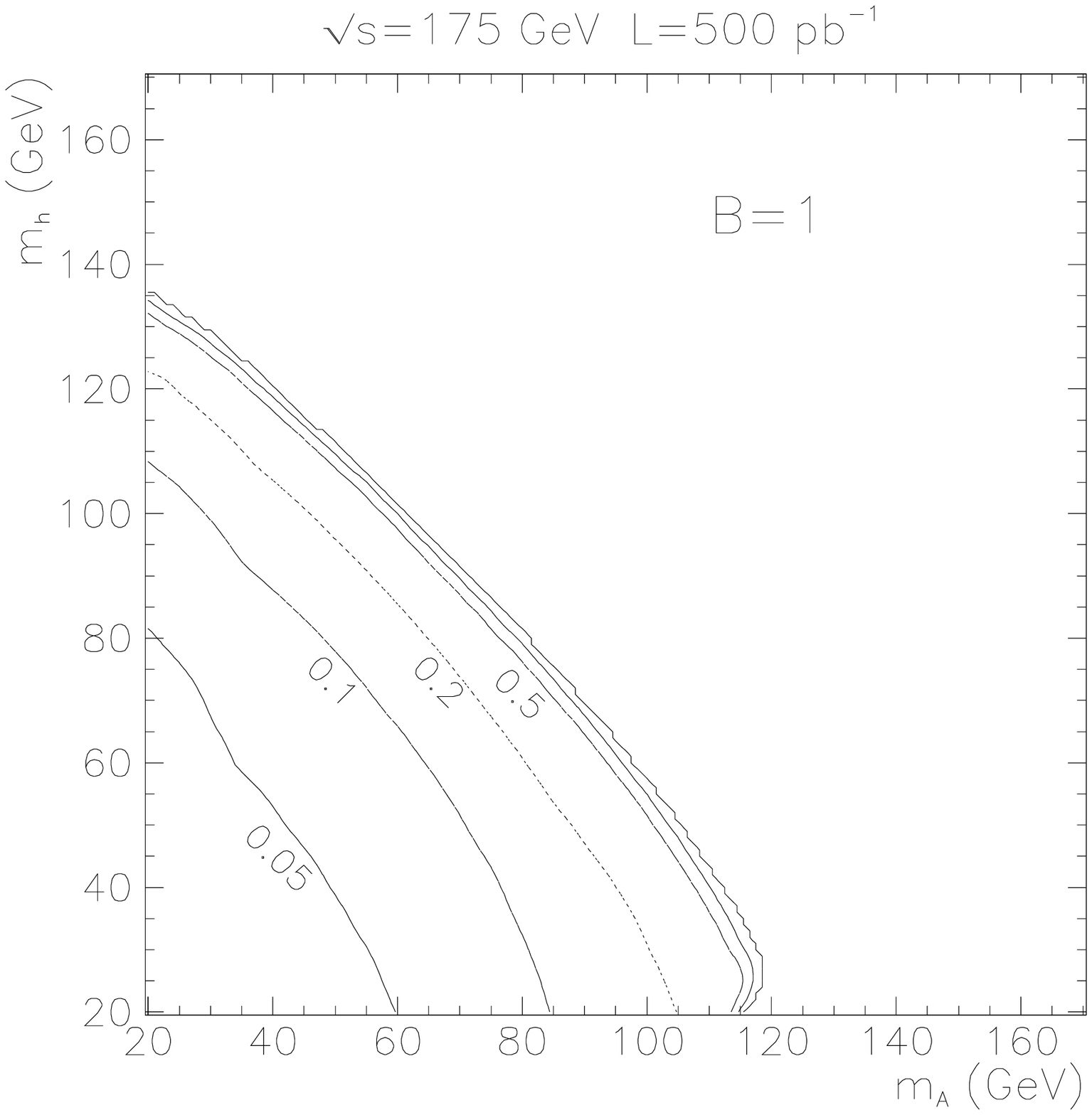,width=10cm,height=8cm}}}
\vglue -0.5cm
\caption{Higgs masses and coupling that can be probed at LEP200 
in $e^+ e^- \ra H \:A$ production.}
\label{halep2}
\eef

\section{Electroweak Symmetry Breaking: Supersymmetry}

The physics associated to the electroweak breaking sector
plays a central role in particle physics. One of the most
important physics motivations in favour of supersymmetry 
is the fact that it is the only symmetry one know which 
can stabilize the elementary Higgs boson mass with respect 
to otherwise uncontrollable radiative corrections. These
would be expected in any fundamental unified theory
including gravity, or simply encompassing the electroweak 
and strong interactions. Either way one has a very large 
mass scale - the Planck scale or the grand unification 
scale - which can mix through loops and destabilize the
electroweak scale in \eq{WEAKSCALE}. This so-called {\sl 
hierarchy problem} can be solved, at least technically, 
through supersymmetry \cite{HIER}, to the extent that it holds at
TeV energies and helps to cancel the offending loops.

Supersymmetry is also theoretically attractive as it is the most 
general symmetry consistent with the basic principles of 
field theory \cite{Zumino}. Unlike most symmetries discussed 
in particle physics, that relate particles of the same spin, 
SUSY relates bosons to fermions, and vice-versa (see table 6).

Finally, the experimental determination of gauge couplings at 
low energies shows a circumstantial evidence in favour of the 
existence of SUSY particles
\footnote{For definiteness, one assumes here those present in 
the so-called MSSM.} at the TeV scale. This hint is provided by the 
joining of these gauge couplings at high energies of order of 
the unification scale 10$^{16}$ GeV \cite{Amaldi91} as illustrated 
in \fig{gaugeunif}, taken from ref. \cite{Nir}.
\bef
\centerline{\protect\hbox{
\psfig{file=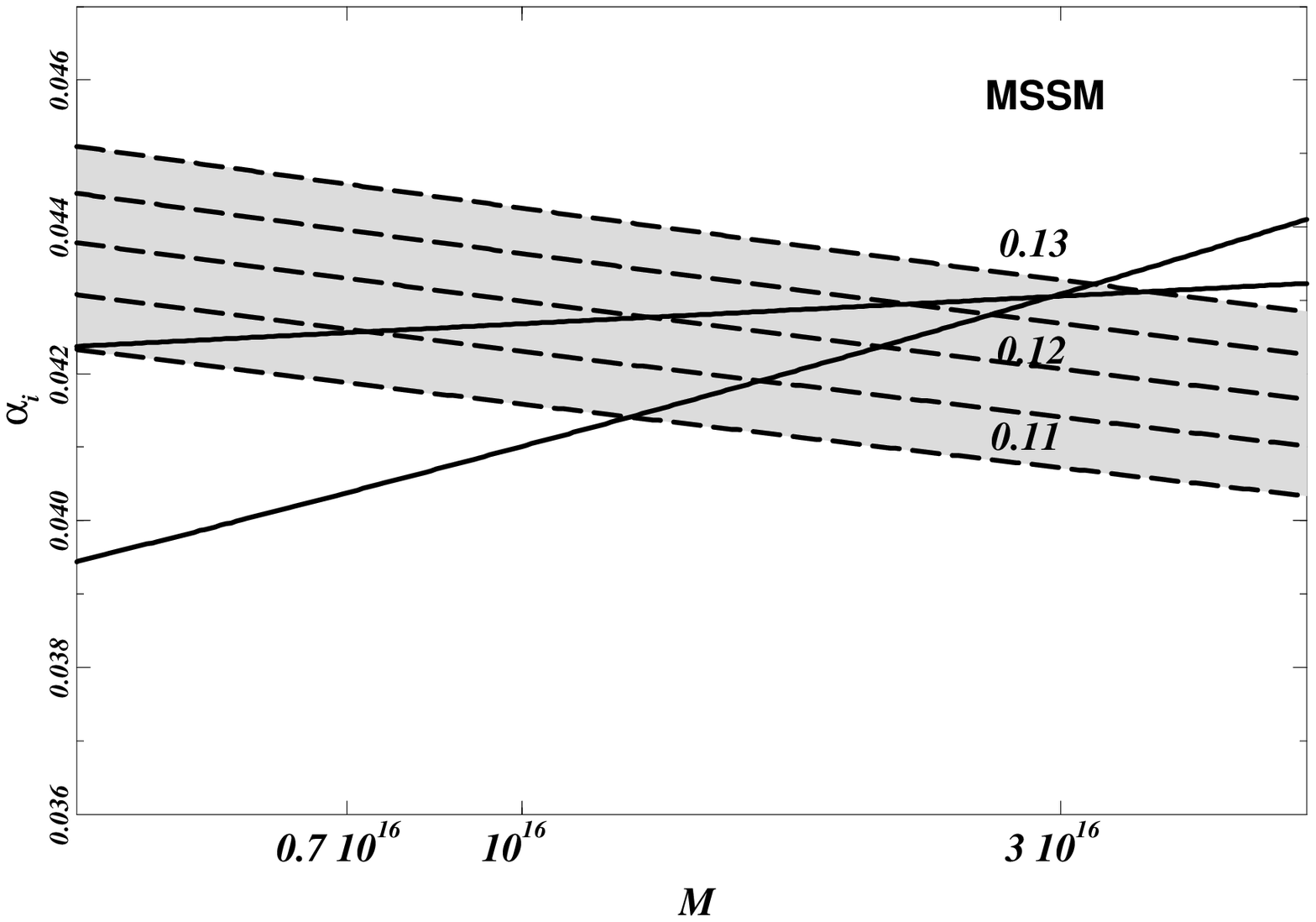,width=10cm,height=7.5cm}}}
\vglue -0.5cm
\caption{Gauge coupling unification in the MSSM.}
\label{gaugeunif}
\eef 
For all these reasons the study of supersymmetric extensions of the 
Standard Model has attracted a lot of research effort, including both
the theoretical understanding of supersymmetric models and their possible 
connections with unification schemes, such as provided by string theory
as well as the simulation of the expected signals at present and future 
particle colliders.

\subsection{ The MSSM}

The simplest supersymmetric model is the so-called Minimal 
Supersymmetric Standard Model (MSSM) \cite{mssm}, defined by 
the particle content given in table 6.
\noindent
\begin{table}
\label{mssmcontent}
\begin{center}
\begin{tabular}{||lc|cl||}\hline
Vector & Supermultiplet & Chiral & Supermultiplet \\ \hline
$J=1 $ & $J=1/2$ & $J=1/2$ & $J=0$ \\ \hline
$g$ & $\tilde{g}$ & $Q_L$,$U^c_L$,$D^c_L$ & $\tilde{Q_L}$, $\tilde{U^c_L}$,
$\tilde{D^c_L}$ \\
$W^{\pm}$, $W^0$ & $\tilde{W^{\pm}}$, $\tilde{W^0}$ & $L_L$, $E^c_L$ &
$\tilde{L_L}$, $\tilde{E^c_L}$ \\
$B$ & $\tilde{B}$ & $\tilde{H}_d$, $\tilde{H}_u$ & $H_d$, $H_u$ \\ \hline
\end{tabular}
\end{center}
\caption{MSSM multiplet content.}
\end{table}
and supplemented by the {\sl ad hoc} hypothesis that the basic 
interactions conserve a discrete R parity ($R_p$) symmetry, 
under which all Standard Model particles are even while their 
partners are odd. The presence of two doublets of Higgs
superfields is required by supersymmetry, anomaly cancellation,
and in order to give masses to both up and down-type charged fermions.

With this assumption the MSSM is characterised by the following 
superpotential, 
\beq 
\label{PMSSM}
W_0 = \varepsilon_{ab}\left [h_{ij}\hat{L}_i^a \hat{H}_1^b
\hat{E}_j^c + h'_{ij}\hat{Q}_i^a \hat{H}_1^b \hat{D}_j^c + h''_{ij}
\hat{Q}_i^a \hat{H}_2^b \hat{U}_j^c + \mu \hat{H}_1^a \hat{H}_2^b \right]
\eeq
For our subsequent discussion we need the chargino and neutralino 
mass matrices. The form of the chargino mass matrix is given by
\beq
\begin{array}{c|cccccccc}
& \tilde{H^+_u} & -i \tilde{W^+}\\
\hline
\tilde{H^-_d} & \mu & \sqrt{2} g_2 v_d\\
-i \tilde{W^-} & \sqrt{2} g_2 v_u & M_2
\end{array}
\label{MSSMchino}
\eeq
Two matrices U and V are needed to diagonalize the $2 \times 2$ 
(non-symmetric) chargino mass matrix
\bea
{\chi}_i^+ = V_{ij} {\psi}_j^+\\
{\chi}_i^- = U_{ij} {\psi}_j^-
\label{MSSM_UV}
\eea
where $\psi_j^+ = (\tilde{H^+_u}, -i \tilde{W^+}$)
and $\psi_j^- = (\tilde{H^-_d}, -i \tilde{W^-}$).

On the other hand the neutralino mass matrix is 
$4\times 4$ and has the following form 
\beq
\begin{array}{c|cccccccc}
& \tilde{H}_u & \tilde{H}_d & -i \tilde{W}_3 & -i \tilde{B}\\
\hline
\tilde{H}_u & 0 & - \mu & -g_2 v_u & g_1 v_u\\
\tilde{H}_d & - \mu & 0 & g_2 v_d & -g_1 v_d\\
-i \tilde{W}_3 & -g_2 v_u & g_2 v_d & M_2 & 0\\
-i \tilde{B} & g_1 v_u & -g_1 v_d & 0 & M_1
\end{array}
\label{MSSMnino}
\eeq
This matrix is diagonalized by a $4 \times 4$ unitary matrix N,
\beq
{\chi}_i^0 = N_{ij} {\psi}_j^0
\eeq
where $\psi_j^0 = (\tilde{H}_u,\tilde{H}_d,-i \tilde{W}_3,-i \tilde{B}$),
(the indices $i$ and $j$ run from $1$ to $4$).

In the above two equations $M_{1,2}$ denote the supersymmetry 
breaking gaugino mass parameters and $g_{1,2}$ are the \21 gauge
couplings divided by $\sqrt{2}$. We assume the canonical relation 
$M_1/M_2 = \frac{5}{3} tan^2{\theta_W}$.
Typical values for the SUSY parameters $\mu$, $M_2$
and $\tan \beta$ lie in the range given by 
\beq
\label{param0}
-1000\leq\frac{\displaystyle\mu}{\mbox{GeV}}\leq 1000 \: \:;\:\:
20\leq \frac{\displaystyle M}{\mbox{GeV}}\leq 1000 \: \:; \:\:
1 \lsim \tan \beta \lsim 40
\eeq
The bilinear term may be replaced by a cubic term of 
the form $h_0 H_u H_d \Phi$. In this case the effective Higgsino 
mixing parameter $\mu$ is given as $\mu = h_0 \VEV \Phi$, where 
$\VEV \Phi$ is the VEV of the appropriate singlet scalar.

Adding the soft supersymmetry breaking scalar mass terms to
the supersymmetric gauge interactions dictated by table 6
(D terms) and the supersymmetric Yukawa interactions following
from \eq{PMSSM} one can write the scalar potential characterising 
the MSSM. Its general form may be written schematically as 
\beq
V_{MSSM}  = 
	\sum_i \abs { \frac{\partial W}{\partial z_i}}^2 
+ \tilde{m}_0 \left[ A W_3 + B W_2 + h.c. \right]
	+ \sum_{i} \tilde{m}_i^2 \abs{z_i}^2 
+ \alpha ( \abs{H_u}^2 - \abs{H_d}^2 - \abs{\tilde{\nu}}^2)^2
\label{VMSSM}
\eeq
where $W_3$ and $W_2$ denote the cubic and quadratic parts 
of the superpotential, $\alpha \equiv \frac{g^2 + {g'}^2}{8}$
and $z_i$ denotes any neutral scalar field in the theory. The 
parameter A is the cubic soft breaking parameter and B=A-1 is 
the corresponding quadratic one \cite{mssm}.

One can show that supersymmetry brings in an attractive
possibility to spontaneously break the electroweak symmetry 
radiatively \cite{Ibanez}, through renormalization effects from 
the unification scale down to low energies. Alternatively,
assuming colour and electric charge conservation, one can show
that the presence of a linear superpotential term in addition to
the cubic $h_0 H_u H_d \Phi$ term allows for the possibility of 
breaking the electroweak symmetry at the tree level \cite{BFS}.

\subsection{The MSSM Higgs Sector}

A complete discussion of the MSSM Higgs sector is totally outside 
the scope of these lectures. For extensive discussions see ref.
\cite{Hunters}. Here we will confine ourselves to a very brief 
summary of the situation, with emphasis on the prospects for 
Higgs searches at future colliders. For this we will borrow
many results from ref. \cite{CarenaZerwas}.

Due to the necessary presence of two Higgs boson doublets in the 
MSSM implies that there are two physical CP-even neutral Higgs 
scalars (h, H), a CP-odd neutral scalar particle, A, and a physical 
electrically charged scalar boson $H^\pm$. At the tree level 
the mass of the lightest CP-even neutral Higgs boson $h$ can be 
calculated in terms of two parameters, which may be chosen as
$m_{A}$ and the ratio of Higgs VEVS $\tan\beta$ \cite{Hunters}. 

Due to the special structure of the MSSM Higgs potential, there is
an upper bound on the lightest CP even Higgs boson mass. At the
tree level, this bound is exactly the Z mass. However, it is 
sensitive to radiative corrections, which are depend on the
soft supersymmetry breaking parameters \cite{susyhiggs.radcor}. 

The full one-loop radiatively corrected $h$ mass is given in refs. 
\cite{honeloop} and \cite{carena}. A simple procedure for accurately 
approximating $m_{h}$ was described by Haber \cite{epshaber}. 
The dominant radiative corrections to $m_{h}$ arise from an incomplete 
cancellation of virtual top-quark and top-squark loops. The two top-squark 
masses ($M_{\widetilde t_1}$ and $M_{\widetilde t_2}$) are obtained by 
diagonalizing the corresponding $2\times 2$ top-squark squared-mass 
matrix.

We assume that the ratio of Higgs VEVS lies in the range 
$1 \lsim \tan\beta \lsim \frac{m_t}{m_b}$ and that the scale 
characterizing supersymmetry breaking $M_S$ is less than 2 TeV. 
This scale can be roughly regarded as a common supersymmetric 
scalar mass. A large $M_S$ value takes into account the possibility 
of large radiative corrections to the lightest CP even Higgs boson 
mass. We used a top quark mass in the range $m_t = 175 \pm 35$ GeV 
which generously covers the region indicated by the recent experimental 
data from the Tevatron.
\begin{figure*}
\centerline{
\epsfig{figure=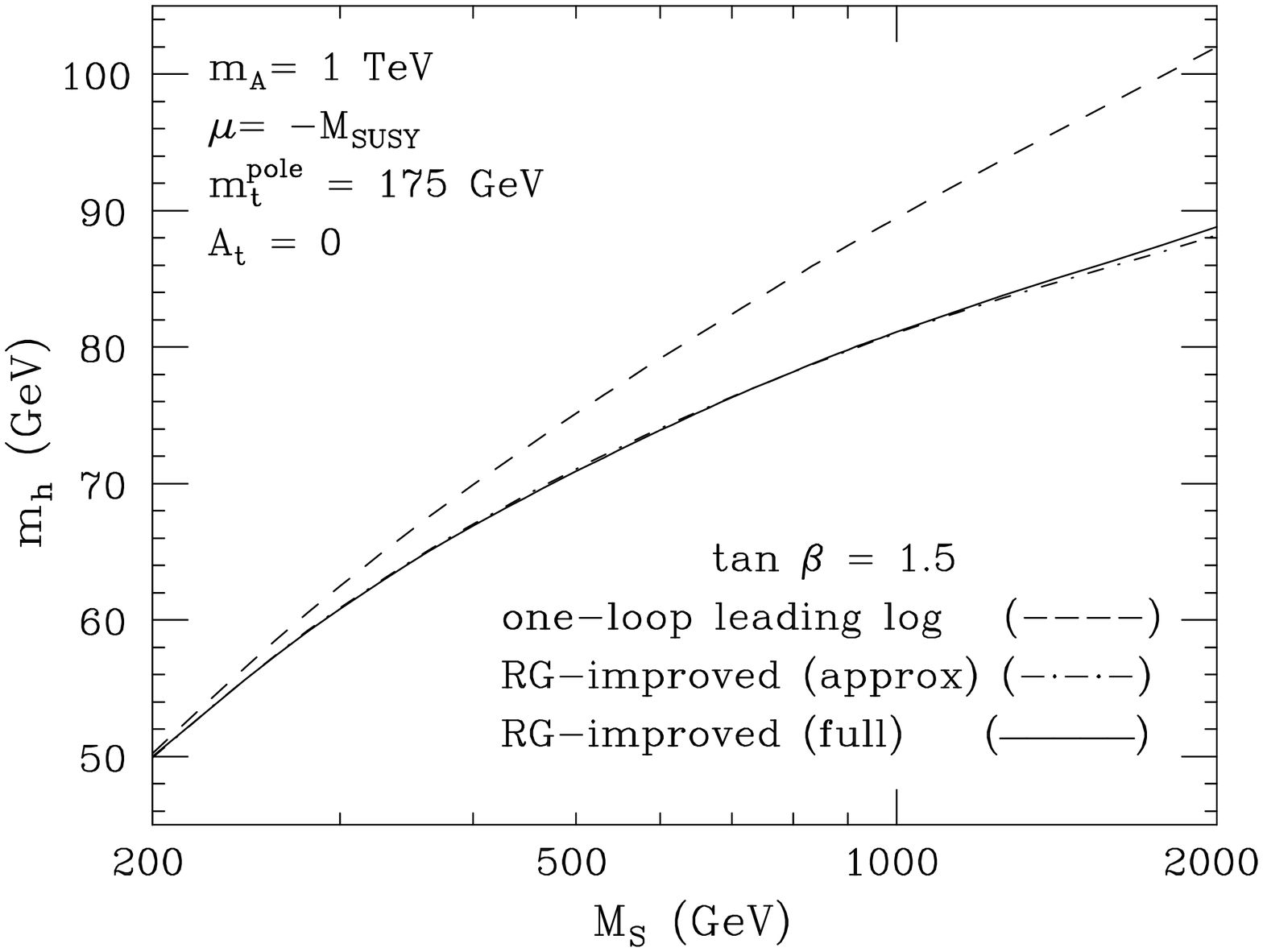,height=6.5cm,angle=90}
}
\caption{Radiatively corrected MSSM lightest Higgs mass versus SUSY scale.}
\label{fig2}
\end{figure*}
In \fig{fig2} we illustrate the dependence of the radiatively corrected 
lightest CP-even Higgs mass with respect to $M_S$ for $\tan\beta=1.5$.  
The one-loop leading logarithmic computation is compared with the 
RG-improved result which was obtained by numerical analysis and by 
using the simple analytic result of ref. \cite{epshaber}. 

\begin{figure*}
\centerline{
\epsfig{figure=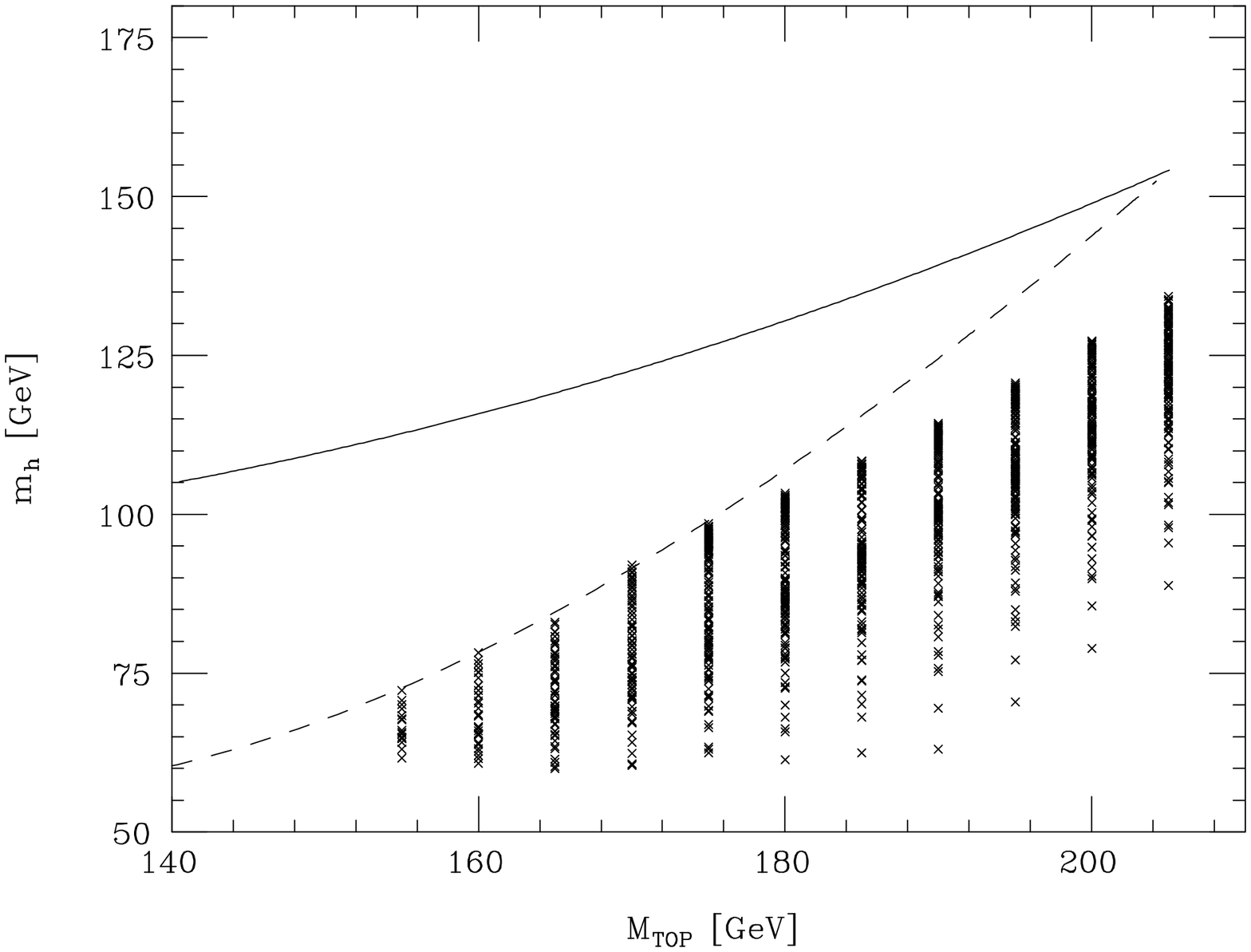,height=7cm,angle=90}
}
\caption{Radiatively corrected MSSM lightest Higgs mass versus $m_t$.}
\label{fig3}
\end{figure*}

\begin{figure*}
\centerline{
\epsfig{figure=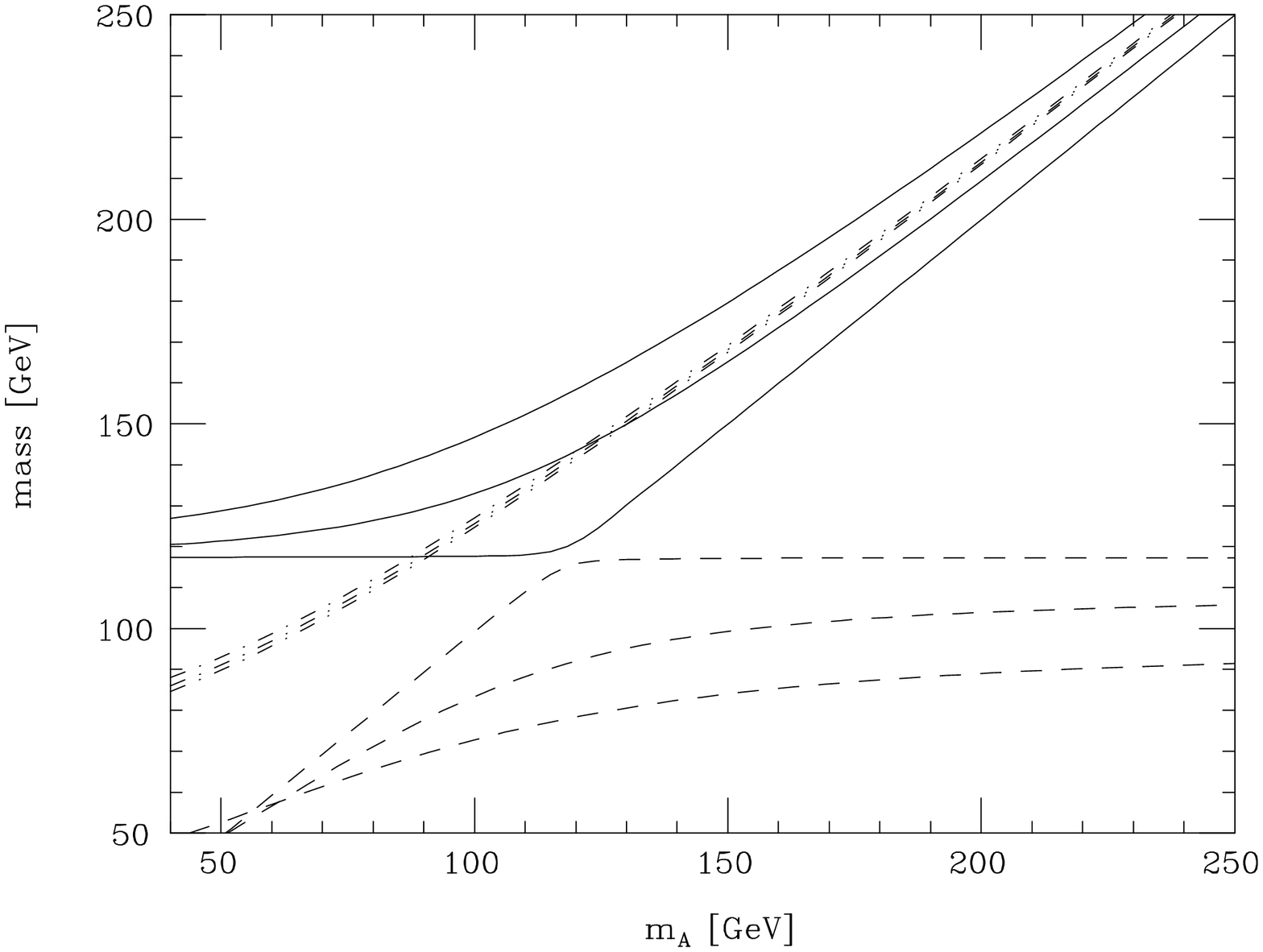,height=8cm,angle=90}
}
\caption{Radiatively corrected MSSM Higgs boson masses.}
\label{fig4}
\end{figure*}

The dependence of the upper bound on the lightest CP-even Higgs 
boson mass in the MSSM with respect to the top quark mass is given 
by the solid line in \fig{fig3}. The dashed line shows the
corresponding result for the special case of b-$\tau$ unification
under several assumptions, explained in ref. \cite{carena}.
The complete spectrum of MSSM scalar boson masses, including the
h, H, A and $H^\pm$ masses is shown in \fig{fig4} from ref. 
\cite{carena}. The dashed, solid and dot-dashed lines refer to 
h, H and $H^\pm$ masses respectively. 
Here one has assumed $m_t = 175$ GeV, $M_S = 1$ TeV, $A = - \mu = M_S$
and three typical $\tan\beta$ values, 1.6, 3 and 30 (for h from bottom 
to top and for H the other way around).
The region of interest is above 40 GeV, which is roughly the
lower limit on the A mass accessible at LEP1. On the other
hand, one sees that for $m_A$ above 200 GeV or so there is
a very slow variation in $m_h$. 

\begin{figure*}
\centerline{
\epsfig{figure=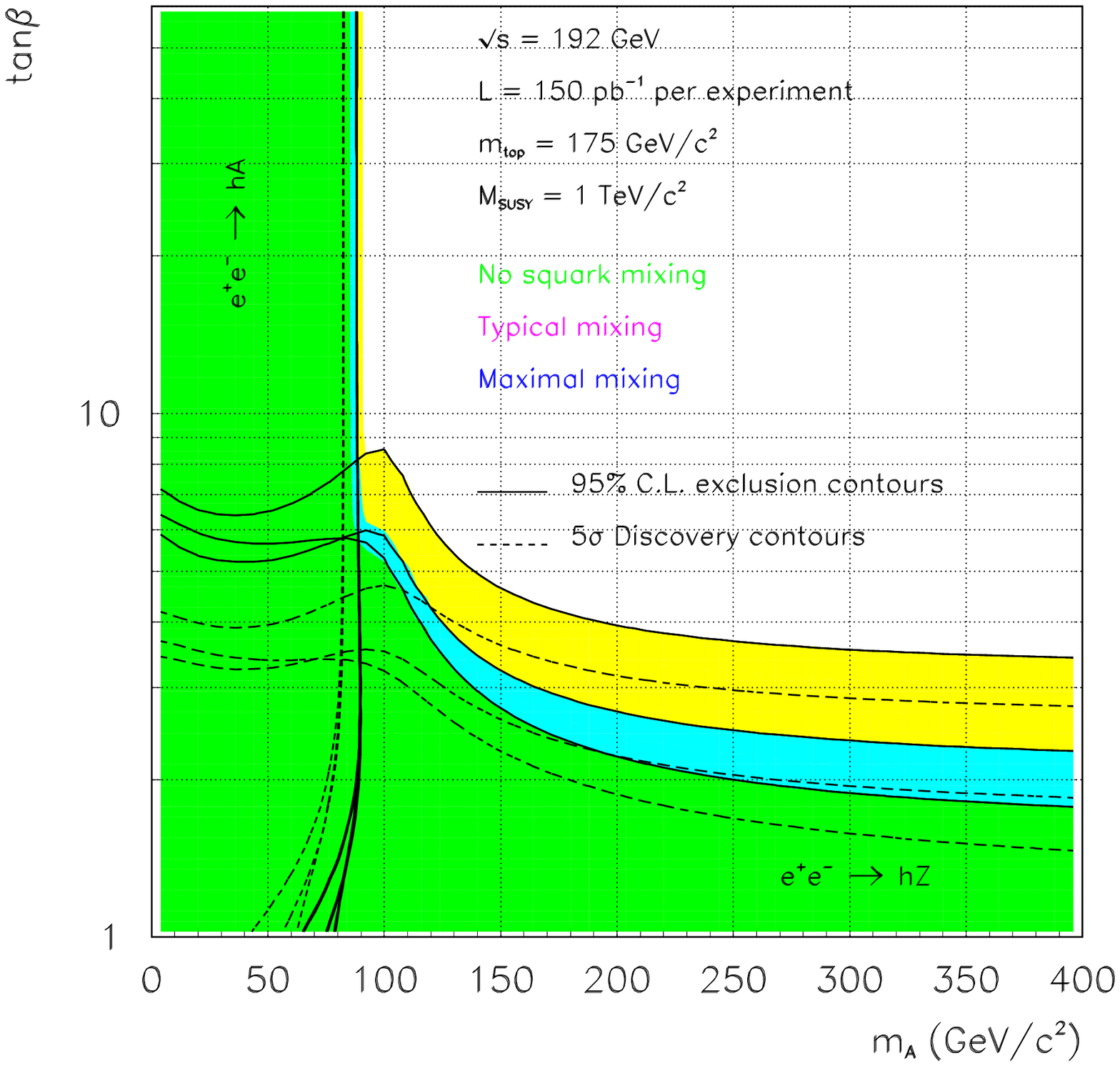,height=9.5cm}
}
\caption{MSSM Higgs boson discovery contours at LEP200.}
\label{fig5}
\end{figure*}

The MSSM Higgs boson discovery contours at LEP200 are illustrated
in \fig{fig5}, from ref. \cite{CarenaZerwas}. This plot corresponds to
centre-of-mass energies 192 GeV, substantially better for Higgs
bosons searches at LEP than 175 GeV, and for three stop quark
mixing assumptions $A_t = 0$ and $|\mu| \ll M_S$ (no mixing),
$A_t = M_S$ and $\mu = - M_S$  (typical mixing), and
$A_t = \sqrt6 M_S$  and $|\mu| \ll M_S$ (maximal mixing), with $M_S = 1$ TeV.

\subsection{Limits on SUSY Particles }
\label{susylim}
 
As a result of the assumption that R parity is conserved,
the interactions of the MSSM are such that all SUSY particles 
must be only produced in pairs, with the lightest of them (LSP) 
being absolutely stable. We take this as a defining feature
of the MSSM
\footnote{The LSP has been suggested as a natural candidate 
for the cold dark matter (CDM) of the universe. Thus, in
addition to searches at accelerators, several methods of 
detection at underground installations have been suggested.
However, as we already saw, the combined COBE and IRAS 
data do not favour a pure CDM model of structure formation.
}.

So far all searches for supersymmetric particles have been 
negative. The best existing search site for the weakly interacting 
SUSY particles is the LEP accelerator. The most recent results 
follow from searches performed at 130 and 136 GeV centre-of-mass 
energies and supersede some of the previous LEP1 results. 

The basic theoretical considerations involved in the analysis are:
\begin{itemize}
\item
The {\sl chargino} production cross section in $e^+ e^-$ collisions
has both s-channel Z-mediated as well as t-channel sneutrino-
mediated contributions, and may be in the few picobarn range
in the 130 GeV energy region. 
\item
Neutralino production also receives a selectron-mediated
t-channel contribution. Due to the R parity conservation hypothesis
the $\chi\chi$ production channel does not lead to a visible
signal, and one looks for events caused by $\chi\chi^\prime$,
where $\chi^\prime$ decays to $\chi f \bar{f}$, f being any 
kinematically accessible fermion. This process will lead to 
acoplanar jets and lepton events with a substantial amount of 
missing momentum. 
\item
Above the Z peak the highest charged slepton production cross 
section is for the selectron case, when the LSP is gaugino-like.
For the smuons and staus there is no t-channel contribution to 
the cross section. The decay $\tilde e^\pm \ra e^\pm + \chi$ 
will give rise to dilepton + missing momentum events.
\end{itemize}

From the non-observation of acoplanar lepton pairs, hadronic 
events with isolated leptons, hadronic events with missing energy, 
and acoplanar jet topologies, the Aleph collaboration has recently 
placed the following limits \cite{LEPSEARCH}:
\begin{itemize}
\item
The new LEP run can be used
in order to improve the limit on the lightest chargino mass. 
Here we quote the result of the Aleph collaboration:
\beq
m_{\chi^{\pm}} \gsim 65 \: GeV
\label{chino2}
\eeq
If the chargino is mostly gaugino this assumes that the
sneutrino mass exceeds 200 GeV and, when it is mostly
Higgsino, it assumes that the chargino-neutralino mass
difference exceeds 10 GeV. 
\item
The searches for neutralinos at Aleph lead to the hatched excluded 
region displayed in \fig{chi_selectron}, for the case $\mu = 1$ TeV 
and $\tan \beta =2$. Note, however, that there is no model 
independent limit on the lightest neutralino. It heavily 
depends on the assumption of universal soft-breaking gaugino 
masses and on the value of the selectron mass. 
\bef
\centerline{\protect\hbox{
\psfig{file=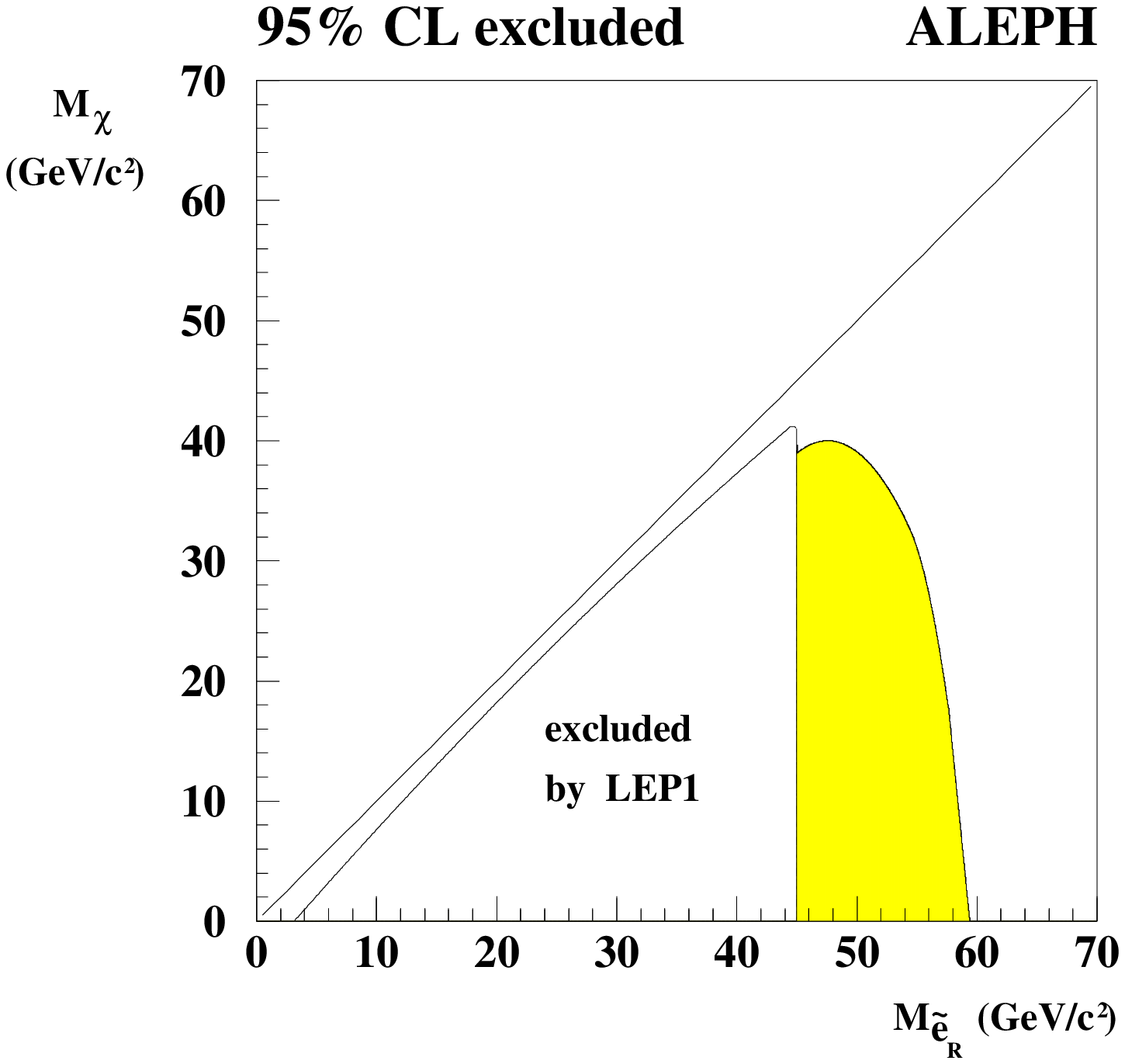,height=9cm}}}
\caption{Limits on neutralino and selectron masses in the MSSM.}
\label{chi_selectron}
\eef
The limits also substantially depend on the assumed decay 
modes of the heavier neutralino. 
\item
Searches for dilepton + missing momentum events have been 
performed by the LEP collaborations. The recent Aleph data 
give the following limit \cite{LEPSEARCH}
\beq
m_{\tilde{\ell^{\pm}}} \gsim 60 \: GeV
\eeq
which has improved over the previous LEP1 results, but is
still worse than previous Tristan limits from single $\gamma$
searches. For the smuons and staus there is no improvement over 
the 45 GeV LEP1 limit. For sneutrinos, the limit is worse than
for charged sleptons.
\end{itemize}

The limits on squark and gluino masses come mostly from hadron 
collisions \cite{D0}. These limits are correlated. For a very heavy
gluino, one has 
$m_{\tilde{q}} \gtap 100 \:$ GeV
for the lower limit on the squarks, with a weaker limit on the top 
squark. On the other hand, in the limit of very heavy squarks one gets
$m_{\tilde{g}} \gtap 140 \: $ GeV
as the corresponding limit on the gluino mass. The allowed region in 
squark-gluino masses is illustrated in \fig{d0}, taken from ref. \cite{D0}. 
\bef
\centerline{\protect\hbox{
\psfig{file=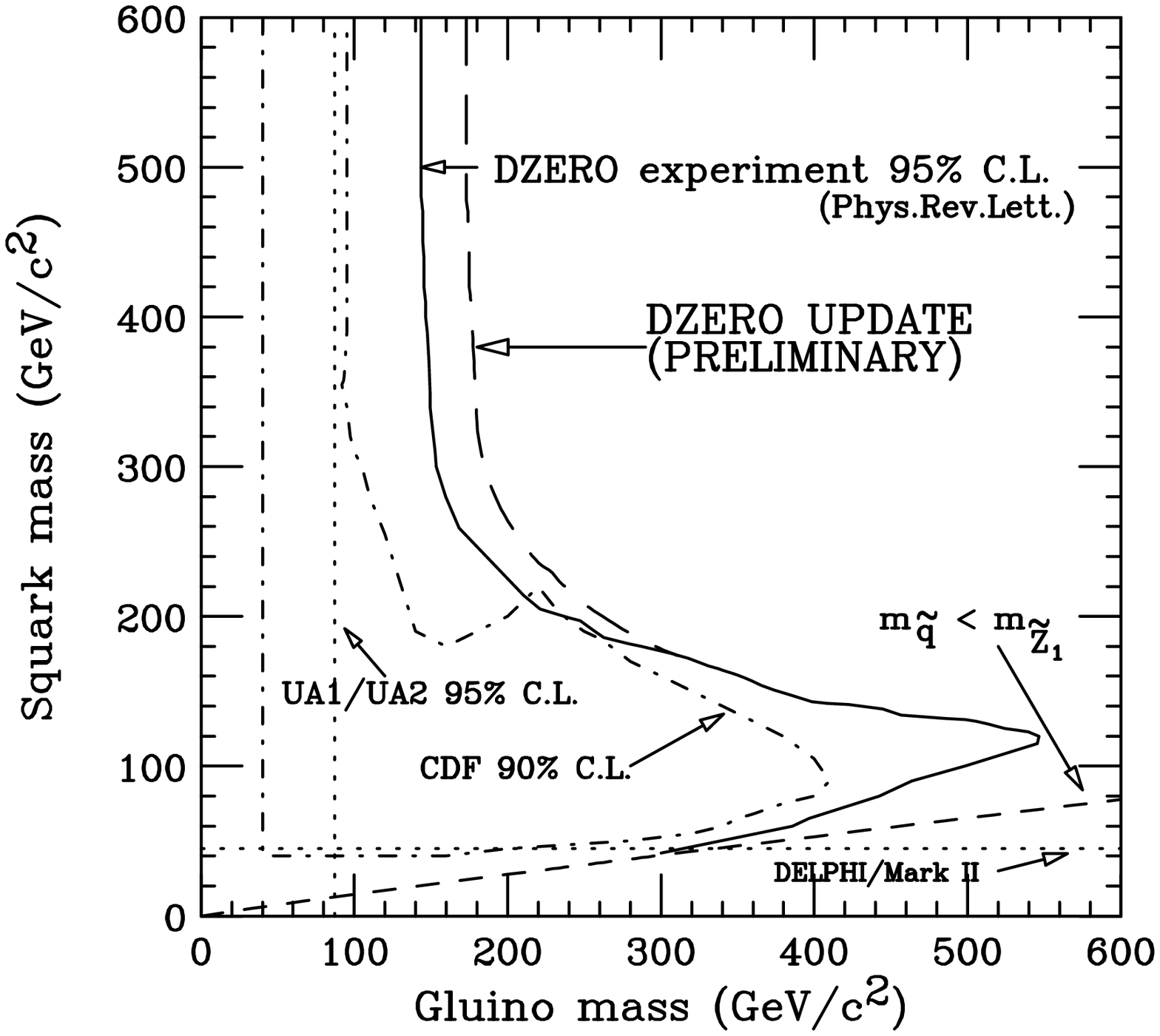,height=9cm}}}
\caption{Tevatron limits on squark and gluino masses in the MSSM.}
\label{d0}
\eef
The limits given depend on simplifying assumptions, and some of them 
may become stronger if one adopts specific parameter choices in the MSSM. 
On the other hand, they may get weaker in extended models. 

The limits for SUSY fermion searches may be combined 
in order to determine the shape of the corresponding allowed 
region of region of SUSY parameters $\mu$ and $M_2$, for given 
choices of the ratio of Higgs doublet VEVS $\tan \beta$, as 
shown in \fig{mum2}. 
\bef
\centerline{\protect\hbox{
\psfig{file=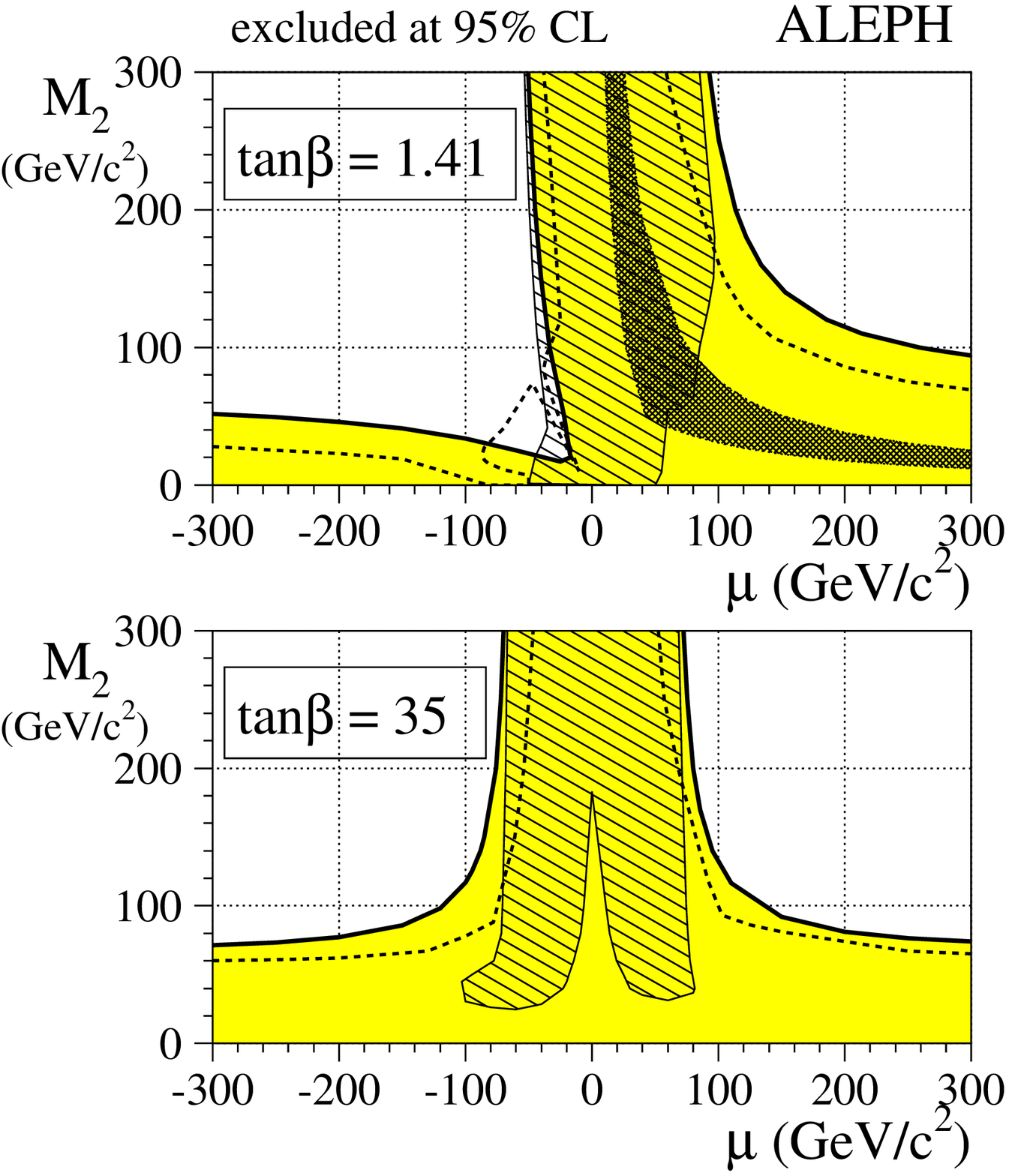,height=12cm}}}
\vglue -0.7cm
\caption{Presently allowed region of MSSM parameters.}
\label{mum2}
\eef
The region excluded by the chargino search is the shaded region, 
while the dashed line indicates the previous LEP1 region. The slepton 
masses are assumed to be 500 GeV. The dark area corresponds to the 
(unlikely) case that the chargino is lighter than all neutralinos.
The searches for neutralinos at Aleph lead to the hatched excluded 
region displayed in \fig{chi_selectron}, for the case $\mu = 1$ TeV 
and $\tan \beta =2$. 
Additional region of parameter space will become accessible 
to further searches at higher LEP energies and are eagerly
awaited for.

In short, one concludes that there is still a very large domain 
of parameters where SUSY would be a meaningful symmetry, in terms 
of being a solution of the hierarchy problem \cite{Barbieri}, 
and where its effects could still have been missed. From this 
point of view it is of great interest to look for its possible 
effects at higher energies, such as will be accessible at the
large hadron collider (LHC) and other future elementary particle 
accelerators such as the next linear collider (NLC).

An important assumption underlying all SUSY searches conducted so far,
is that of R parity conservation. This assumption dictates that all 
SUSY particles must be produced in pairs, the lightest of these (LSP),
typically a neutralino, being absolutely stable. Thus the signal 
associated to the LSP is missing momentum. These properties have 
been taken as the basis of all searches of SUSY particles. 

Unfortunately there is no clue as to how SUSY is realized. Nobody 
knows the origin of the R parity symmetry and whether it is indeed 
a necessary requirement to impose on supersymmetric extensions of 
the Standard Model. Therefore there is no firm theoretical basis 
for the MSSM - it is no more than an $ansatz$, which is the most 
popular mainly because of its simplicity. Whether or not R parity 
is conserved is an important dynamical issue. However, for all we 
know so far, R parity conservation may very well break down at some level. 

Present SUSY particle search strategies are inadequate for the analysis 
of extended models where SUSY is realized with broken R parity. For 
example, if R parity is broken, it would be possible to probe SUSY 
even at the LEP1 energies through genuinely new signatures, such as single 
SUSY particle production \cite{ROMA}! Therefore one needs to re-analyse 
the existing data in order to place limits on these models. We now turn 
to a discussion of explicit and spontaneous breaking of R parity.

\subsection{Explicit R Parity Violation}
\label{ERPB}
 
The minimal supersymmetric extension of the \21 theory 
{\sl in general violates lepton and baryon number conservation}. 
Indeed, \21 \gau invariance and SUSY are consistent with adding
to the basic superpotential, \eq{PMSSM}, many Yukawa terms that 
violate lepton number conservation, such as
\beq 
\label{wr}
W_R=\varepsilon_{ab}\left[\lambda_{ijk}\hat{L}_i^a \hat{L}_j^b
\hat{E}_k^C + \lambda'_{ijk} \hat{L}_i^a \hat{Q}_j^b \hat{D}_k^C 
+ \epsilon_i \hat{L}_i^a \hat{H}_u^b \right]                                
\end{equation} 
Here i,j,k denote flavour indices and $\lambda$ is anti-symmetric 
in i,j. Similarly, one could add terms such as $\hat{U^c} \hat{D}^c 
\hat{D}^c$, as they are consistent with all symmetries of the 
Standard Model, plus supersymmetry. The presence of such terms,
along with those in \eq{wr}, will lead to baryon number violating 
processes such as proton decay.
Such terms may arise as residuals from many extended models, 
such as supersymmetric GUTS \cite{expl}. In their presence R parity 
symmetry is broken explicitly, as can easily be checked. The bilinear
in $W_R$ parametrizes effectively, for many purposes, the main 
effects of theories with spontaneous breaking. Moreover, as we
will see below, it plays a very important role in the physics of 
the Higgs sector.

There are several constraints on these couplings, some of
which are quite stringent \cite{Barger}. Recently the Aleph 
collaboration has placed limits on explicitly broken R-parity 
models by considering the pair production of the lightest 
neutralino followed by its decay as would be induced under
the very restrictive assumption that a single of the 
$\hat{L} \hat{L} \hat{E^c} $ coupling is present \cite{aleph95}.
Other couplings are very much constrained. For example, the presence 
of $\hat{U^c} \hat{D}^c \hat{D}^c$ terms, along with those in 
\eq{wr}, will lead to unacceptably fast proton decay. They may also 
be constrained by the requirement of non-erasure of a primordial 
baryon asymmetry \cite{giudice}. Therefore one normally forbids 
these terms by hand, invoking R parity conservation. It is 
possible, however, that R parity is explicitly broken only 
by a subset of these terms, at a sizeable level, yet fully
consistent with observation. The missing terms could arise 
by imposing some global and/or discrete symmetry. Moreover, 
explicit $R_p$ violating interactions could be tolerated in 
the presence of a mechanism that could generate a nonzero 
baryon asymmetry at low energy, as suggested in ref. \cite{late}.

What is least pleasant of models with explicitly broken R parity 
is that they involve too large a number of arbitrary parameters 
in the Yukawa sector, which limits considerably their predictive
power and the ability to map out systematically their possible
effects at accelerators. For this reason, we will focus our 
discussion mostly on the case of Spontaneous R Parity Violation.

\subsection{Spontaneous R Parity Violation}
\label{sbrp}

In Spontaneous R Parity Violation scenarios the breaking of 
R-parity is driven by right-handed {\sl isosinglet} sneutrino 
vacuum expectation values (VEVS) \cite{MASIpot3,MASI}, so that 
the associated Goldstone boson (Majoron) is mostly singlet.
As a result the $Z$ does not decay by Majoron emission, in 
agreement with LEP observations \cite{Martinez}.

Here we focus on what is the conceptually simplest model for
Spontaneous R Parity Violation, in which two \21 singlet leptons,
instead of one, are added in each family \cite{MASIpot3}
\footnote{One may add just a single pair of singlet lepton
superfields, instead of three. }.
The conceptual simplicity of the model follows from the fact 
that the magnitude of all R Parity violating effects is strictly 
correlated to the mass of the tau neutrino.
Although most of the subsequent discussion applies equally well 
to models where the Majoron is absent, due to an enlarged gauge 
structure \cite{ZR,RPCHI}, or to other \21 models with spontaneously 
broken R parity \cite{MASI}, for definiteness we will focus on the
simplest model and start by recalling its main ingredients.
Indeed, many of the phenomenological features relevant for the 
accelerator studies already emerge in an effective model where 
the spontaneous violation of R parity is reproduced through a the
addition of the explicit bilinear superpotential term in
\eq{wr} \cite{epsi}.

The superpotential is given by
\beq
h_u  Q H_u U^c + h_d  H_d Q D^c + h_e \ell H_d E^c +
(h_0 H_u H_d - \epsilon^2) \Phi +
h_{\nu} \ell H_u \nu^c  + h \Phi  S \nu^c + h.c.
\label{P}
\eeq
where we have omitted the hats in the superfields, as well as
generation space indices in the coupling matrices $h_u,h_d,h_e,h_{\nu},h$.
This superpotential conserves $total$ lepton number and R parity.
The superfields $(\Phi ,{\nu^c}_i,S_i)$ are singlets under \21 and
carry a conserved lepton number assigned as (0,-1,1), respectively. 
These additional singlets $\nu^c, S$ \cite{SST} and $\Phi$ \cite{BFS} 
may drive the spontaneous violation of R parity in the model \cite{MASIpot3}. 
This leads to the existence of a Majoron given by the imaginary part of 
\beq
\frac{v_L^2}{Vv^2} (v_u H_u - v_d H_d) +
              \frac{v_L}{V} \tilde{\nu_{\tau}} -
              \frac{v_R}{V} \tilde{\nu^c}_{\tau} +
              \frac{v_S}{V} \tilde{S_{\tau}}
\label{maj}
\eeq
where the isosinglet VEVS
\beq
\begin{array}{lr}
v_R = \VEV {\tilde{\nu}_{R\tau}}\:, &
v_S = \VEV {\tilde{S_{\tau}}}
\end{array} 
\eeq
with $V = \sqrt{v_R^2 + v_S^2}$ characterize R-parity or lepton 
number breaking and the isodoublet VEVS
\beq
\begin{array}{lr}
v_u = \VEV {H_u} \:, &
v_d = \VEV {H_d} 
\end{array} 
\eeq
drive electroweak breaking and the fermion masses. The combination 
$v^2 = v_u^2 + v_d^2 + v_L^2$ is fixed by the W,Z masses. Finally, there 
is a small seed of R parity breaking in the doublet sector, i.e.
\beq
v_L = \VEV {\tilde{\nu}_{L\tau}} 
\eeq
whose magnitude is now related to the Yukawa coupling
$h_{\nu}$. Since this vanishes as $h_{\nu} \ra 0$, 
we can naturally obey the limits from stellar energy 
loss \cite{KIM}. 

For our subsequent discussion we need the chargino and neutralino 
mass matrices. The form of the chargino mass matrix is common to 
a wide class of \21 SUSY models with spontaneously broken R parity. 
It is given by
\beq
\begin{array}{c|cccccccc}
& e^+_j & \tilde{H^+_u} & -i \tilde{W^+}\\
\hline
e_i & h_{e ij} v_d & - h_{\nu ij} v_{Rj} & \sqrt{2} g_2 v_{Li} \\
\tilde{H^-_d} & - h_{e ij} v_{Li} & \mu & \sqrt{2} g_2 v_d\\
-i \tilde{W^-} & 0 & \sqrt{2} g_2 v_u & M_2
\end{array}
\label{chino}
\eeq
Two matrices U and V are needed to diagonalize the $5 \times 5$ 
(non-symmetric) chargino mass matrix
\bea
{\chi}_i^+ = V_{ij} {\psi}_j^+\\
{\chi}_i^- = U_{ij} {\psi}_j^-
\label{INO}
\eea
where the indices $i$ and $j$ run from $1$ to $5$ and
$\psi_j^+ = (e_1^+, e_2^+ , e_3^+ ,\tilde{H^+_u}, -i \tilde{W^+}$)
and $\psi_j^- = (e_1^-, e_2^- , e_3^-, \tilde{H^-_d}, -i \tilde{W^-}$).

Under reasonable approximations, we can truncate the neutralino 
mass matrix so as to obtain an effective $7\times 7$ matrix of 
the following form \cite{MASIpot3}
\beq
\begin{array}{c|cccccccc}
& {\nu}_i & \tilde{H}_u & \tilde{H}_d & -i \tilde{W}_3 & -i \tilde{B}\\
\hline
{\nu}_i & 0 & h_{\nu ij} v_{Rj} & 0 & g_2 v_{Li} & -g_1 v_{Li}\\
\tilde{H}_u & h_{\nu ij} v_{Rj} & 0 & - \mu & -g_2 v_u & g_1 v_u\\
\tilde{H}_d & 0 & - \mu & 0 & g_2 v_d & -g_1 v_d\\
-i \tilde{W}_3 & g_2 v_{Li} & -g_2 v_u & g_2 v_d & M_2 & 0\\
-i \tilde{B} & -g_1 v_{Li} & g_1 v_u & -g_1 v_d & 0 & M_1
\end{array}
\label{nino}
\eeq
This matrix is diagonalized by a $7 \times 7$ unitary matrix N,
\beq
{\chi}_i^0 = N_{ij} {\psi}_j^0
\eeq
where 
$\psi_j^0 = ({\nu}_i,\tilde{H}_u,\tilde{H}_d,-i \tilde{W}_3,-i \tilde{B}$),
with $\nu_i$ denoting weak-eigenstate neutrinos (the indices $i$ and $j$ 
run from $1$ to $7$).

Here we make the same parameter assumptions and conventions as used 
in the MSSM. Typical values for the SUSY parameters $\mu$ and $M_2$ 
are as before. The parameters $h_{\nu i,3}$ lie in the range given by 
\beq
\label{param1}
\begin{array}{llll}
10^{-10}\leq h_{\nu 13}, h_{\nu 23} \leq 10^{-1} & & & 
10^{-5}\leq h_{\nu 33} \leq 10^{-1}\\
\end{array}
\eeq
while the expectation values are chosen as
\beq
\label{param2}
\begin{array}{llll}
v_L=v_{L3}=100\:\: \mbox{MeV} & & & v_{L1}=v_{L2}=0 \\
v_R=v_{R3}=1000\:\: \mbox{GeV} & & & v_{R1}=v_{R2}=0\\
v_S=1000\:\: \mbox{GeV}   & & &  \mbox{$1 \lsim \tan\beta = \frac{v_u}{v_d}
\lsim \frac{m_t}{m_b}$}
\end{array}
\eeq 
The diagonalization of \eq{nino} gives rise to the mixing of the 
neutralinos with the neutrinos, leading to R-parity violating
gauge couplings and to neutrino masses, mainly the \nt mass.
Although the \nt can be quite massive, it is perfectly consistent 
with cosmology \cite{KT}, including primordial nucleosynthesis, as
it can both decay through Majoron emission $\nt \ra \nm + J$
\cite{V,RPMSW} due to flavour non-diagonal couplings such as
${h_{\nu}}_{23}$, as well as annihilate to a Majoron pair due
to the diagonal coupling ${h_{\nu}}_{33}$ \cite{DPRV}. Both
processes can be quite efficient cosmologically in order to
bypass the required restrictions, leaving the \nt mass free
to attain its maximum valued allowed by laboratory experiments.
On the other hand, the tiny \ne and \nm masses may be chosen 
to lie in the range where resonant \ne $\ra$ \nm conversions 
provide an explanation of solar neutrino deficit. Due to this 
peculiar hierarchical pattern, one can regard the associated
R parity violating processes as a tool to probe the physics 
underlying the solar \neu conversions in this model \cite{RPMSW}. 
Indeed, the rates for such rare decays can be used to discriminate 
between large and small mixing angle MSW solutions to the solar 
\neu problem \cite{MSW}. Typically, in the small mixing region 
can have larger rare decay branching ratios than in the large 
mixing region, as seen in Figure 5 of ref. \cite{RPMSW}.

As already mentioned, the \nt mass shows a direct correlation with the 
magnitude of R-parity violating phenomena, making this model a especially
useful way to parametrize the resulting physics. 

Using the above diagonalizing matrices U, V and N one can write the 
electroweak currents of the mass-eigenstate fermions. For example, 
the charged current Lagrangian describing the weak interaction between 
charged lepton/chargino and neutrino/neutralinos may be written as
\beq
\frac{g}{\sqrt2} W_\mu \bar{\chi}_i^- \gamma^\mu 
(K_{Lik} P_L + K_{Rik} P_R) {\chi}_k^0 + H.C.
\label{CC2}
\eeq
where $P_{L,R}$ are the two chiral projectors and
the $5\times 7$ coupling matrices $K_{L,R}$ may
be written as
\bea
K_{Lik} = \eta_i (-\sqrt2 U_{i5} N_{k6} - U_{i4}
N_{k5} - \sum_{m=1}^{3} U_{im}N_{km})\label{KL}\\
K_{Rik} = \epsilon_k (-\sqrt2 V_{i5} N_{k6} + V_{i4} N_{k4})
\label{KR}
\eea
The matrix $K_{Lik}$ is the analogous of the matrix $K$ introduced 
in ref \cite{2227}. These couplings break R-parity for $i=1..3$ and 
$k=4..7$, and $i=4,5$ and $k=1..3$.

Similarly, the corresponding neutral current Lagrangian describing 
the weak interaction of the charged lepton and charginos, as well as
the neutrinos and neutralinos may be written as
\beq
\frac{g}{\cos\theta_W} Z_\mu \{ \bar{\chi}_i^- \gamma^\mu 
(\eta_i \eta_k O'_{Lik} P_L + O'_{Rik} P_R) \chi_k^- 
+ \frac{1}{2} 
\bar{\chi}_i^0 \gamma^\mu 
(\epsilon_i \epsilon_k O''_{Lik} P_L + O''_{Rik} P_R) \chi_k^0 
\}
\label{NCRP}
\eeq
where the $7\times7$ coupling matrices $O'_{L,R}$ and
$O''_{L,R}$ are given by
\bea
O'_{Lik} = \frac{1}{2} U_{i4} U_{k4} + U_{i5}
U_{k5} + \frac{1}{2} \sum_{m=1}^{3} U_{im}U_{km} -\delta_{ik}
\sin^2\theta_W \label{OL}\\
O'_{Rik} = \frac{1}{2} V_{i4} V_{k4} + V_{i5}
V_{k5} - \delta_{ik} \sin^2\theta_W \label{OR}\\
O''_{Lik} = \frac{1}{2} \{ N_{i4} N_{k4} - N_{i5}
N_{k5} - \sum_{m=1}^{3} N_{im}N_{km} \} = - O''_{Rik}
\eea
In writing these  couplings we have assumed CP conservation. 
Under this assumption the diagonalizing matrices can be chosen 
to be real. The $\eta_i$ and $\epsilon_k$ factors are sign 
factors, related with the relative CP parities of these 
fermions, that follow from the diagonalization of their 
mass matrices. These couplings break R-parity for $i=1..3$ 
and $k=4,5$ in the case of the charged leptons, and $i=4..7$ and 
$k=1..3$ for the neutral leptons.

Like all supersymmetric extensions of the Standard Model,
the spontaneously broken R-parity models are constrained by 
data that follow from the negative searches for supersymmetric
particles at LEP as well as $\bar pp$ collider data on gluino 
production. There are additional restrictions, which are 
more characteristic of broken R-parity models. They follow 
from laboratory experiments related to neutrino physics
and weak interactions \cite{fae}, as well as from cosmology
and astrophysics. These restrictions play a very important 
role, as they exclude many parameter choices that are otherwise 
allowed by the collider constraints, while the converse is not 
true. The most relevant constraints come from neutrino-less 
double beta decay and neutrino oscillation searches, direct 
searches for anomalous peaks at meson decays, the limit on the 
tau neutrino mass, cosmological limits on the \nt lifetime and
mass, limits on muon and tau lifetimes, universality constraints, 
and limits on lepton flavour violating decays.

One can perform a sampling of the points which are allowed by 
these constraints in order to evaluate systematically the attainable 
value of the couplings \cite{RPLHC}. The diagonal (R-parity conserving) 
couplings for the lightest neutralino and the lightest chargino are of 
the same order as those in the MSSM. The coupling of the lightest 
chargino to the $Z$ is maximum when it is mainly a gaugino. 
In this case $\mu \gg M_2$ and as a result 
$|V_{45}| \approx |U_{45}| \approx 1$ and $|V_{4i}|,|U_{4i}| \ll 1$ 
for $i\neq 5$. 
On the other hand it is minimum when it is mostly a Higgsino. 
In this case $\mu \ll M_2$ and therefore 
$|V_{44}| \approx |U_{44}| \approx 1$ and 
$|V_{4i}|,|U_{4i}|\ll 1$ for $i\neq 4$. Including these values 
in \eq{OL} and \eq{OR} one gets for the allowed range
\beq
\label{zdc}
0.27 \ltap |O'_{L44}|, |O'_{R44}| \ltap 0.77
\eeq
From the parameters given in \eq{param0}, (\ref{param1})
and (\ref{param2}) and the experimental limits (specially the 
recent LEP limit on the lightest chargino mass) one finds that the 
lightest supersymmetric fermion is always a neutralino, with 
mass $ M_{\chi^0}  \gtap 30 \: GeV $. 

Similarly, the R-parity-conserving charged current couplings
of the lightest chargino to the lightest neutralino after
including the experimental constraints lie in the range
\beq
\label{wd}
10^{-4} \ltap |K_{L44}|,|K_{R44}| \ltap \frac{1}{\sqrt{2}}
\eeq
For the neutral current couplings of the lightest neutralino 
one finds, after imposing the experimental constraints 
\beq
\label{zdn}
|O''_{L44}| \ltap  0.1
\eeq
In what concerns the R-parity breaking couplings, the largest
ones correspond to the case when the standard lepton belongs
to the third family. These couplings can reach a few per cent 
or so for mass values accessible in accelerator studies \cite{RPLHC}. 
This is illustrated in \fig{+rpvcouplings}
\bef
\centerline{\protect\hbox{
\psfig{file=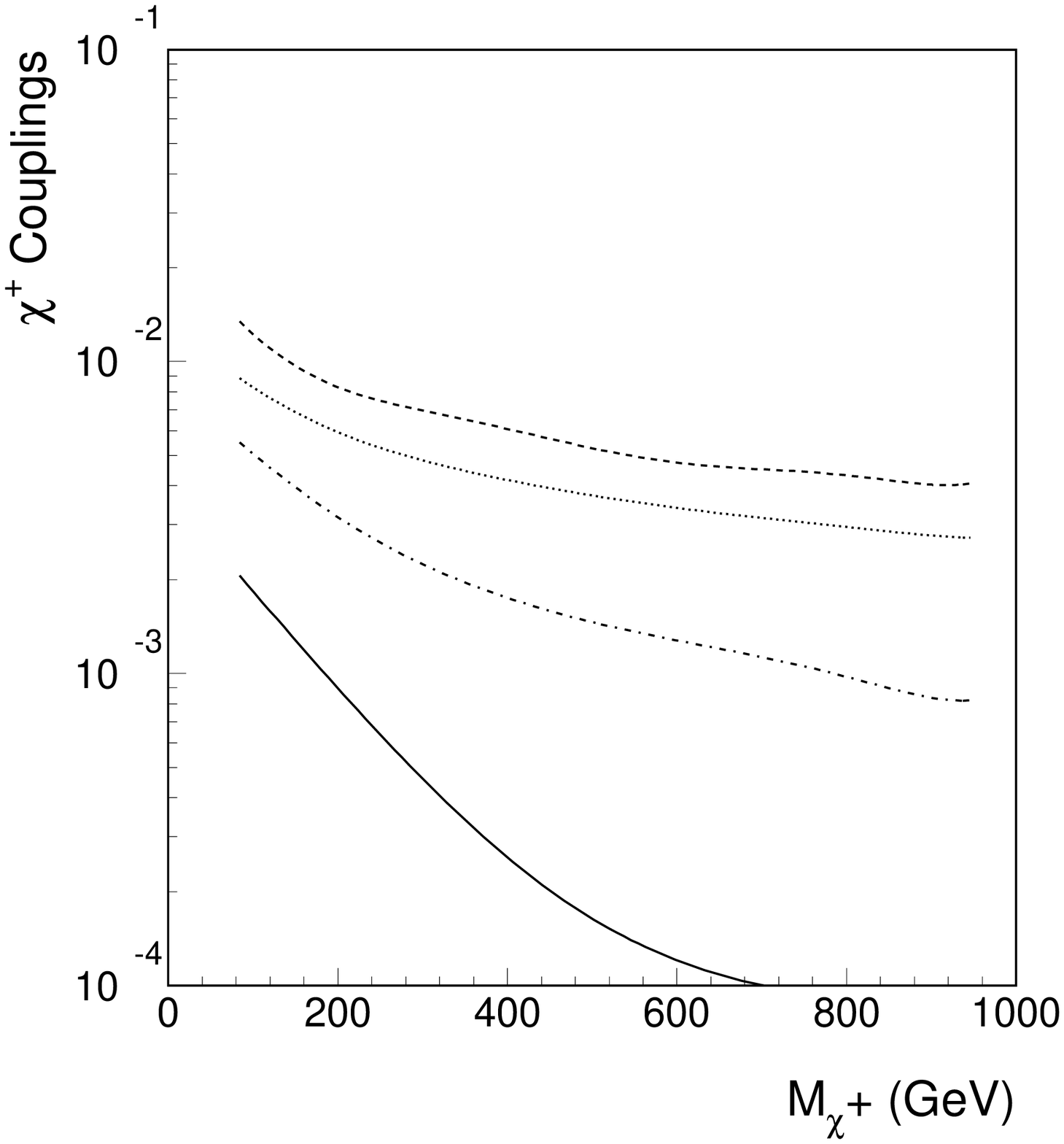,height=9cm}}}
\caption{Presently allowed region of chargino R parity violating couplings.}
\label{+rpvcouplings}
\eef
Here we have considered the parameters as in \eq{param0}, 
(\ref{param1}) and (\ref{param2}) with $\tan\beta=\frac{v_u}{v_d}$
chosen to lie between 2 and 30. In \fig{+rpvcouplings} the various
lines denote the allowed magnitudes of the left and right-handed R 
parity violating currents. The solid line denotes $|K_{L43}|$, the 
dashed one $|K_{R43}|$, the dotted one $|O^{\prime}_{L34}|$ and, finally, 
the dash-dotted one denotes $|O^{\prime}_{R34}|$.

\subsection{The Scalar Sector in the Model }

The theoretical viability of the model proposed in ref. \cite{MASIpot3}
has been explicitly demonstrated and we do not plan to repeat here the
detailed discussion on the minimization of the corresponding Higgs 
potential. Nevertheless we will make a brief summary.
For simplicity we consider an effective one-generation model in which 
the coupling matrices ${h_{\nu}}_{ij}$ and $h_{ij}$ are nonzero only 
for the third generation and set $h_{\nu} \equiv {h_{\nu}}_{33}$ and 
$h \equiv h_{33}$. We adopt the most general form for the soft SUSY 
breaking terms in a spontaneously broken $N=1$ super-gravity model 
\bea
V_{soft} = 
\tilde{m}_0 \left[	- A h_0 \Phi H_u H_d 
            	- B \epsilon^2 \Phi 
		+ C h_{\nu} \tilde{\nu^c} \tilde{\nu} H_u 
		+ D h \Phi \tilde{\nu^c} \tilde{S} 
		+ E \hat{\mu} H_u H_d
		+ h.c. \right]\\\nonumber
		+ \tilde{m}_{u}^2 \abs{H_u}^2
		+ \tilde{m}_d^2 \abs{H_d}^2
		+ \tilde{m}_{L}^2 \abs{\tilde{\nu}}^2
		+ \tilde{m}_{R}^2 \abs{\tilde{\nu^c}}^2
		+ \tilde{m}_{S}^2 \abs{\tilde{S}}^2
		+ \tilde{m}_{\Phi}^2 \abs{\Phi}^2
\label{Vsoft}
\eea
We have included only the neutral scalars. These soft breaking terms 
have the form expected in models with minimal $N=1$ super-gravity 
theories characterized, at the unification scale, by universal, 
diagonal supersymmetry-breaking scalar masses and by trilinear 
scalar terms proportional to a single dimension-less parameter $A$
\bea
	C = D = A\:, \: \: \: E = A - 1 \:, \: \: \: B = A-2\\
	\tilde{m}_{u}^2 = \tilde{m}_{d}^2 = \tilde{m}_{L}^2 =
	\tilde{m}_{R}^2 = \tilde{m}_{S}^2 = \tilde{m}_{\Phi}^2 
\label{univ}
\eea
Since these conditions are not expected to hold at low energies, 
due to renormalization group evolution from the unification scale down 
to the electroweak scale, we allow the values of the soft breaking 
scalar masses to differ from their common value $\tilde{m}_0$ at 
the unification scale. We have kept however the values of $B, C$, 
$D$ and $E$ related as above and assumed, for simplicity, that all 
parameters in the potential are real.

With the definitions above the full scalar potential along 
neutral directions of the Spontaneous R Parity Broken model 
of ref. \cite{MASIpot3} is given by 
\bea
\label{total}
V_{total}  = 
	\abs {h \Phi \tilde{S} + h_{\nu} \tilde{\nu} H_u }^2 + 
	\abs{h_0 \Phi H_u + \hat{\mu} H_u}^2 + \\\nonumber
	\abs{h \Phi \tilde{\nu^c}}^2 + 
	\abs{- h_0 \Phi H_d  - \hat{\mu} H_d + 
	h_{\nu} \tilde{\nu} \tilde{\nu^c} }^2+
	\abs{- h_0 H_u H_d + h \tilde{\nu^c} \tilde{S} - \epsilon^2}^2 + 
	\abs{h_{\nu} \tilde{\nu^c} H_u}^2\\\nonumber
+ \tilde{m}_0 \left[-A ( - h \Phi \tilde{\nu^c} \tilde{S} 
+ h_0 \Phi H_u H_d - h_{\nu} \tilde{\nu} H_u \tilde{\nu^c} )
+ (1-A) \hat{\mu} H_u H_d
+ (2-A) \epsilon^2 \Phi + h.c. \right]\\\nonumber
	+ \sum_{i} \tilde{m}_i^2 \abs{z_i}^2 
+ \alpha ( \abs{H_u}^2 - \abs{H_d}^2 - \abs{\tilde{\nu}}^2)^2
\label{V}
\eea
where $\alpha=\frac{g^2 + {g'}^2}{8}$ and $z_i$ denotes any 
neutral scalar field in the theory.

Electroweak breaking is driven by the isodoublet VEVS $v_u = \VEV{H_u}$ 
and $v_d = \VEV {H_d}$, assisted by the VEV $v_F$ of the scalar in the 
singlet superfield $\Phi$. The W mass is given as 
$m_W^2 \approx \frac{g^2(v_u^2 + v_d^2 )}{2}$,
while the ratio of isodoublet VEVS determines 
$\tan \beta = \frac{v_u}{v_d}$.
This way one basically recovers the tree level \21 spontaneous breaking 
scenario of ref. \cite{BFS}.

The spontaneous breaking of R parity is driven by nonzero VEVS for the 
scalar isosinglet neutrinos
VEVS
\bea
v_R = \VEV {\tilde{\nu^c}_{\tau}}\\
v_S = \VEV {\tilde{S_{\tau}}}
\eea
where $V = \sqrt{v_R^2 + v_S^2}$ can lie anywhere in the
range $\sim 10\:GeV-1\:TeV$. 
A necessary ingredient for the consistency of this model is the 
presence of a {\sl small} seed of R parity breaking in the $SU(2)$ 
doublet sector, 
\beq
v_L = \VEV {\tilde{\nu}_{L\tau}}\:.
\label{vl}
\eeq
whose typical magnitude in the model may naturally obey the
astrophysical limits coming from stellar energy loss considerations
from Majoron emitting processes \cite{KIM}.

The detailed analysis of the minimization of this potential was 
presented in the second paper of ref. \cite{MASIpot3}. There we 
have explicitly demonstrated the existence of solutions to the 
extremization equations following from \eq{total} which are in 
fact minima and not saddle points, and whose energy is lower 
than that of other trivial solutions where either R parity or 
electroweak symmetries are unbroken. The scale associated to the 
spontaneous violation of R parity can lie typically anywhere 
in the range from 10 to 1000 GeV. 

The squared mass matrices of the neutral scalar bosons 
\beq
z_i = \frac{1}{\sqrt2} \left[Re(z_i) + i Im(z_i) \right]
\eeq
are given as
\beq
M_{Rij}^2 = 	\frac{1}{2} 
		\left( \frac{\partial^2 V}{\partial z_i z_j} + c.c. \right)
		+ \frac{\partial^2 V}{\partial z_i z_j^*} 
\label{Re}
\eeq
and
\beq
M_{Iij}^2 = 	- \frac{1}{2} 
		\left( \frac{\partial^2 V}{\partial z_i z_j} + c.c. \right)
		+ \frac{\partial^2 V}{\partial z_i z_j^*} 
\label{Im}
\eeq
They were determined at the tree level in the second paper in ref. 
\cite{MASIpot3} and shown to be positive-definite in large regions of 
parameter space. Assuming CP conservation, the real and imaginary 
parts do not mix, so that the mass part of the potential energy reads
\beq
V_{mass} = 	\frac {1}{2} Re(z_i) M_{Rij}^2 Re(z_j) \: + \:
		\frac {1}{2} Im(z_i) M_{Iij}^2 Im(z_j) 
\eeq
The $6 \times 6$ matrices obtained this way imply the existence
in this model of 6 CP-even and 5 CP-odd scalars, the last ones 
including the massless Majoron, given by \eq{maj}.

Although the explicit expressions for the masses in
terms of the input parameters defining the low energy 
theory are quite involved, a fairly simple mass formula 
can be derived. From \eq{Re} and \eq{Im} we have
\beq
Tr M_R^2 =  	Tr M_I^2 + \sum_{i=1}^{6} \left(
		\frac{\partial^2 V}{\partial z_i z_j} + h.c. \right)
\label{sum}
\eeq
Using the explicit form of the potential we get that the last 
term in \eq{sum} is just $m_Z^2$, so that
\beq
Tr M_R^2 =  	Tr M_I^2 + m_Z^2
\label{sum2}
\eeq
which nicely generalizes the corresponding tree-level MSSM sum rule.

The mass spectrum for both CP-even and CP-odd scalar bosons
was studied numerically in this model, both at the tree level and
after including radiative corrections \cite{HJJ}. For centre-of-mass 
energies attainable either at LEP200, LHC or NLC, not all of the scalar 
bosons are kinematically accessible. Typically one or two of the CP-even
ones (h, H) will be accessible and one of the massive CP-odd (A) scalar 
bosons. Thus the situation becomes conceptually similar to the one
considered in section \ref{hhf}. The main new features
relevant for experimental analyses are
\begin{itemize}
\item
the coupling of the lightest CP-even Higgs boson can be suppressed 
with respect to that expected in the Standard Model, as in sections 
\ref{hf} and \ref{hhf}. Indeed, it was noted in ref. \cite{HJJ} that
in some regions of parameters of the potential the lightest 
CP-even Higgs boson in this model is mostly an \21 singlet,
with tiny couplings to the Z. Such a scalar boson would
have escaped detection at LEP due to its small production
cross section.
\item
the lightest CP-even Higgs bosons may decay invisibly as
\bea
h \ra J + J \\ \nonumber
H \ra h h  \: \: \: \: \mbox{with} \: \: \: \: h \ra J + J
\eea
The invisible decay of the lightest CP-even Higgs boson can be
quite sizeable and can therefore compete with the standard 
$b\bar{b}$ decay mode. This feature is similar to what we have
seen in sections \ref{hf} and \ref{hhf}. 
\end{itemize}

\subsection{Implications of Spontaneous R Parity Breaking }
\label{SBRPLEP}

In the MSSM all supersymmetric particles are always produced in
pairs. If R parity is broken, they may be singly-produced. As seen 
in section \ref{sbrp}, in models with spontaneous R parity breaking 
the mixing of the standard leptons with the supersymmetric charginos 
and neutralinos leads to the existence of R-parity violating couplings 
in the Lagrangian when written in terms of the mass eigenstates. 
It is in the couplings of the W and the Z where the main 
R-parity violating effects reside \cite{RPLHC}. 
As a result one is no longer forced to produce the SUSY particles in 
pairs. For example a SUSY fermion such as a chargino or a neutralino 
may be produced in pairs (standard MSSM production) as well as singly, 
in association with a $\tau$ or \nt (R-parity breaking single production). 

On the other hand the RPSUSY model rates for pair production of SUSY
particles are similar to those in the MSSM. However, in contrast
to the MSSM, where all supersymmetric particles have cascade decays 
finishing in the LSP which is normally a neutralino, in the RPSUSY
case there are new decay channels and the supersymmetric particles 
can decay directly to the standard states breaking R-parity. 
Alternatively, they may decay through R-parity conserving cascade 
decays that will finish in the lightest neutralino, which then decays.
This way one can generate novel supersymmetric signatures in 
R parity violating models even when the single production
SUSY particle cross sections are small. 

The lightest neutralino can decay to standard states breaking R-parity. 
If its mass is lower than the mass of the gauge bosons these are three-body 
decays such as
\beq
\begin{array}{lllll}
\chi^0\ra \nu_j f \bar f  \: \: \: \: 
\chi^0\ra l_j f_u \overline{f_d} 
\end{array}
\eeq
where the first decays are mediated by the neutral current, while
the second are charged-current mediated. Here f denotes any fermion,
while $f_u$ and $f_d$ denote up or down-type fermions, respectively.

If the neutralino is heavier than the W it may have the two body decays 
\beq
\chi^0\ra \ell_j W   \:\:\:\: \chi^0 \ra \nu_j Z
\eeq
The explicit expressions for the widths are given in ref. \cite{RPLHC}. 
Neutralinos of mass accessible at LEP have mostly three-body 
decay modes mediated by charged and neutral currents. The only 
exception will be the two-body Majoron decay, characteristic 
of the simplest spontaneous R parity breaking models \eq{invis1}. 

In \21 models of spontaneous breaking of R-parity the LSP is not
the neutralino, but rather the Majoron, which is massless and 
therefore stable
\footnote{The Majoron may have a small mass and therefore
it may decay to neutrinos and photons. However the time scales 
are only of cosmological interest and do not change the signal
expected at the laboratory \cite{KEV}.}. 
The existence of the Majoron implies that in \21 spontaneously 
broken R-parity, the neutralino can always decay invisibly to 
\beq
\label{invis1}
\chi^0 \ra \nu_j J
\eeq
To the extent that this invisible decays are are important,
one recovers the signal expected in the MSSM.

For definiteness let us consider the case of the lightest neutralino
and chargino, which one expects could be the earliest-produced 
supersymmetric particles. Here are some examples of signals related
to their production in the spontaneously broken R parity (RPSUSY) models:
\begin{itemize}
\item
{\bf Single chargino production in Z decays \cite{ROMA}}\\
\begin{equation}
Z \rightarrow \chi^{\pm} \tau^{\mp}
\end{equation}
where the lightest chargino mass is assumed to be smaller than
the Z mass. This decay is characteristic of spontaneous R parity
violation. In the simplest models, the magnitude of R parity violation 
is correlated with the nonzero value of the \nt mass and is restricted 
by a variety of experiments. Nevertheless the R parity violating Z decay 
branching ratios can easily exceed $10^{-6}$ (see table 7) and thus lie
within the sensitivities of the LEP experiments performed at the Z pole. 
As illustrated in \fig{figchitau} the maximum branching ratio 
allowed by other experiments and by theory is directly correlated
with \mnt which is a characteristic feature of the model of ref. 
\cite{MASIpot3}.
\begin{figure}
\centerline{\protect\hbox{
\psfig{file=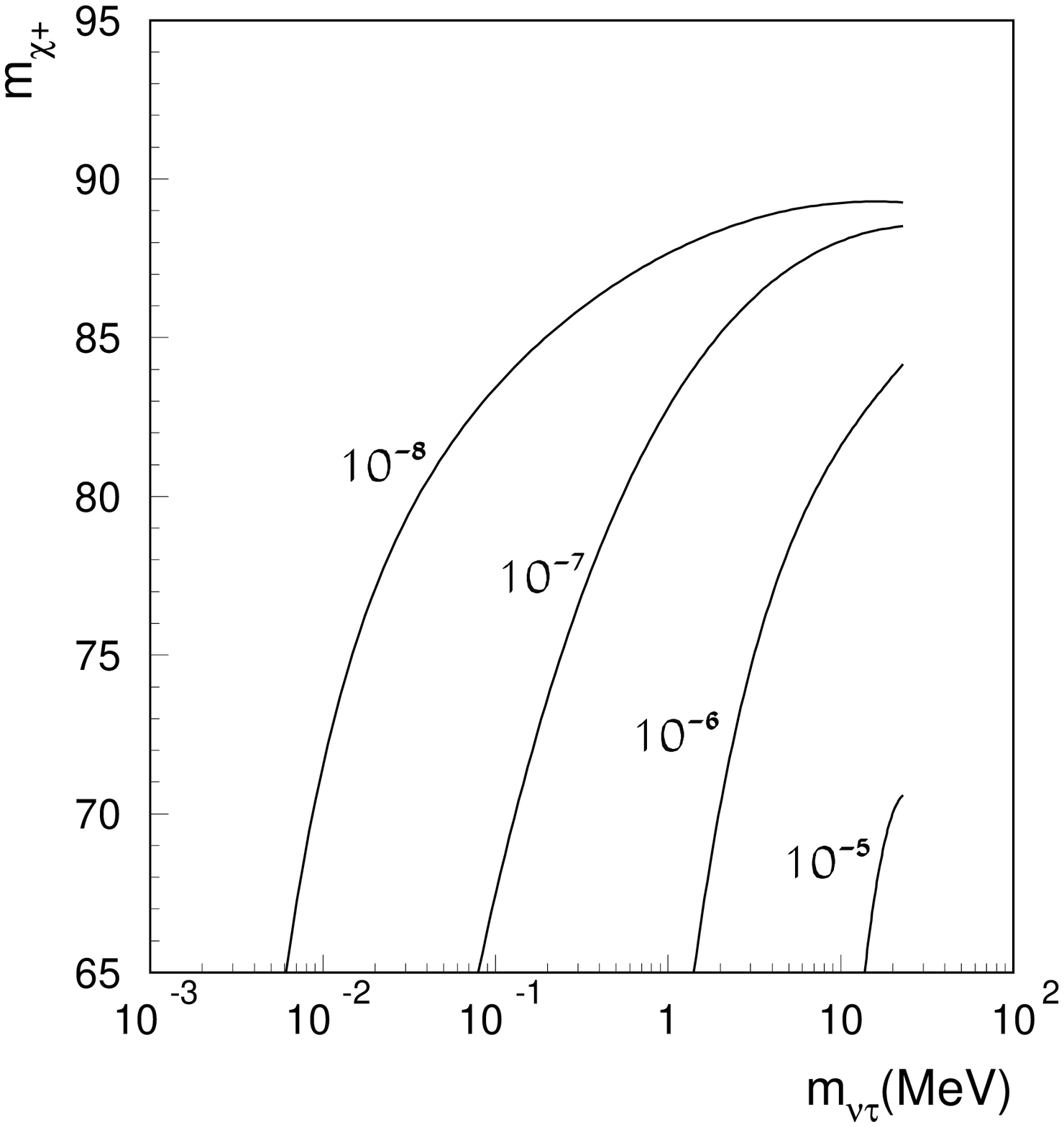,height=9cm}}}
\caption{Allowed $Z \ra \chi^{\pm} \tau^{\mp}$ decay branching ratios
in the RPSUSY model. }
\vglue -0.5cm
\label{figchitau}
\end{figure}
\item
{\bf Single neutralino production in Z decays \cite{ROMA}}\\
\beq
Z \rightarrow \chi^0 \nu_\tau
\eeq
The allowed rates for R parity violating Z decays is given
in table 7. 
\begin{table}
\begin{center}
\caption{Allowed branching ratios for R parity violating Z decays.}
\begin{displaymath}
\begin{array}{|c|cr|} 
\hline
\mbox{channel} & \mbox{strength} & \mbox{} \\
\hline
Z \rightarrow \chi^{\pm} + \tau^{\mp} &  10^{-5} & \\
Z \rightarrow \chi^0 + \nt &   10^{-4} & \\
\hline
\end{array}
\end{displaymath}
\end{center}
\end{table}
To the extent that $\chi$ decays into charged particles are
dominant the neutralino is not necessarily an origin of events 
with missing energy, as in the MSSM. Thus the decay 
$Z \rightarrow \chi^0 \nu_\tau$ would give rise to zen events, 
similar to those of the MSSM, but where the missing energy is 
carried by the \nt and the visible tracks come from the decays
of the $\chi$. The searches for single particle SUSY production 
at LEP1 should place restrictions on the parameter space available 
for studies at LEP200 energies \cite{new96}.
\item
{\bf Pair lightest neutralino production in Z decays \cite{ROMA}, 
followed by neutralino decays}\\
Even if its single production cross section is small, the $\chi$ 
$\chi$ pair production process at LEP will generate zen events 
where one $\chi$ decays visibly and the other invisibly. The 
corresponding zen-event rates can therefore be larger than in 
the MSSM and may occur even if there is no energy to produce 
the next-to-lightest neutralino $\chi^\prime$. 
\item
{\bf Majoron emitting decays}\\
Another possible signal of the RPSUSY models based on the simplest 
\21 gauge group is rare decays of muons and taus and Z bosons involving
Majoron emission. One expects possibly large rates for this new class 
of decays, since in these models the characteristic lepton number 
breaking scale is similar to the weak scale. The allowed branching
ratios are illustrated in table 8. For example the LFV decays with 
single Majoron emission in $\mu$ and $\tau$ decays would be "seen" 
as bumps in the final lepton energy spectrum, at half of the parent 
lepton mass in its rest frame. On the other hand, the $Z \ra \gamma + J$
decay would give rise to monochromatic photons \cite{mono}. As an 
illustration of the \nt mass dependence of the allowed decay 
branching ratios see Figure 2 given in ref. \cite{mono}.
\end{itemize}
\begin{table}
\begin{center}
\caption{Allowed rates for novel decay modes in
spontaneous broken R parity.}
\begin{displaymath}
\begin{array}{|c|cr|} 
\hline
\mbox{channel} & \mbox{strength} & \mbox{} \\
\hline
Z \rightarrow \gamma + J &  10^{-5} & \\
\tau \rightarrow \mu + J &  10^{-3} & \\
\tau \rightarrow e + J &  10^{-4} & \\
\hline
\end{array}
\end{displaymath}
\end{center}
\end{table}
The allowed rates for single Majoron emitting $\mu$ 
and $\tau$ decays have been determined in ref. \cite{NPBTAU}
and are also shown in table 3 to be compatible with present 
experimental sensitivities \cite{PDG95}. As an illustration
of the \nt mass dependence of the allowed decay branching 
ratios see Figures given in ref. \cite{NPBTAU}.
This example also illustrates how the search for 
rare decays can be a more sensitive probe of \neu
properties than the more direct searches for \neu 
masses, and therefore complementary. Moreover, they are 
ideally studied at a tau-charm factory \cite{tcf,TTTAU}.

\subsection{R Parity Breaking Scalar Boson Decays}

Explicit violation of $R$ parity in the minimal supersymmetric model
through bilinear terms $\hat{L} \hat{H_u}$ plays an important role in 
the scalar sector \cite{epsi}. The presence of such bilinear superpotential 
term will induce the mixing of sleptons with Higgs bosons, thus affecting
the decays of both.

\begin{figure}
\vglue .5cm
\centerline{\protect\hbox{\psfig
{file=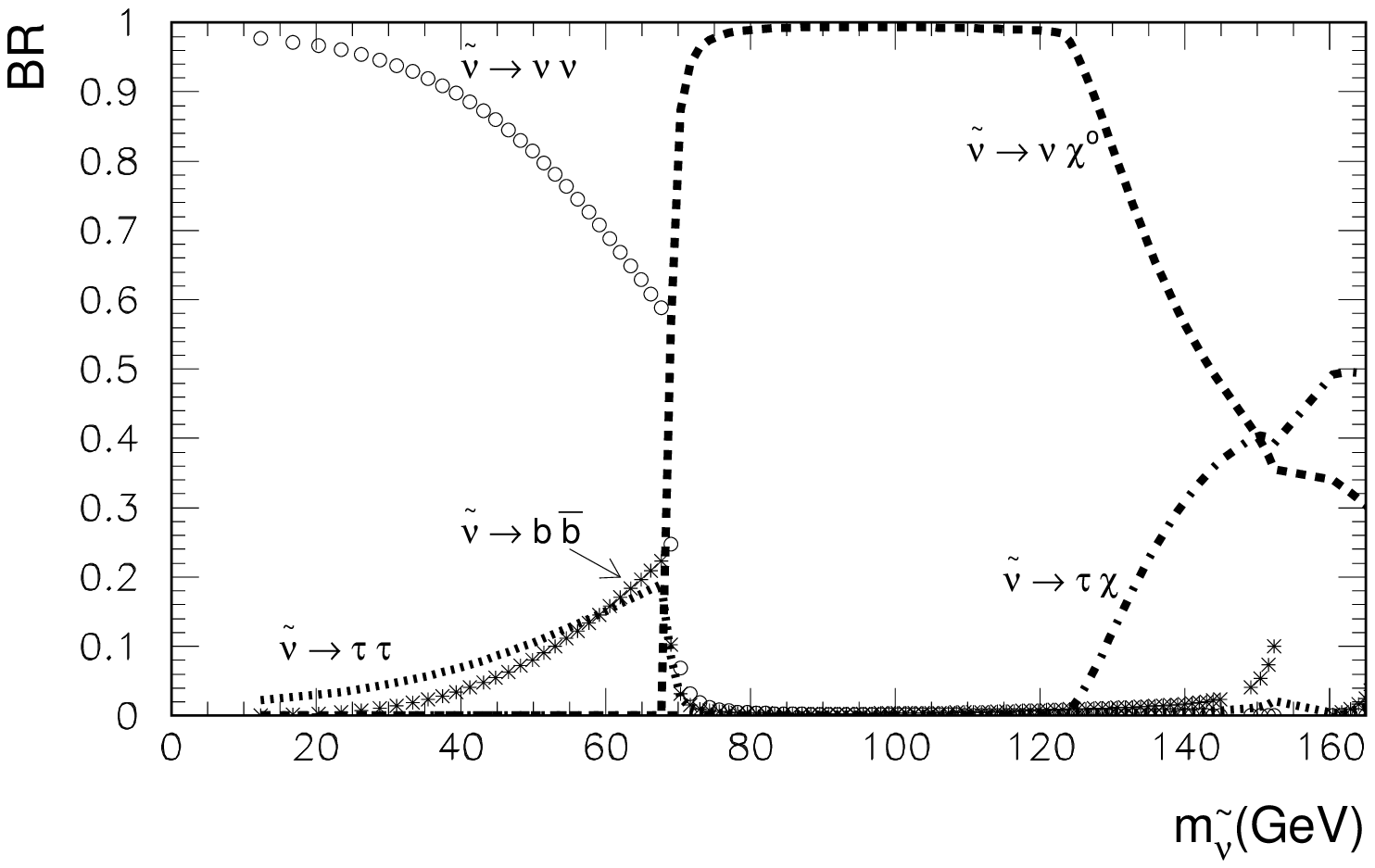,height=7cm}}}
\vglue -.5cm
\caption{Sneutrino decay branching ratios in the RPSUSY model.}
\label{snudecaybr}
\end{figure}

The most illustrative example of this is the possibility that,
below the threshold for SUSY  particle production, the sneutrino 
mostly decays to Standard Model particles, as shown in \fig{snudecaybr}
However, even when the sneutrino is not the lightest SUSY particle, 
there may be a sizeable branching ratio for the R parity violating 
sneutrino decays, even for a moderately small value of the Higgsino-lepton
superpotential mixing parameters $\epsilon_i$.

As shown in ref. \cite{epsi} this may lead also to sizeable 
branching ratio for the supersymmetric Higgs boson decay mode 
$H \rightarrow \chi \ell$, where $\chi$ denotes the lightest 
supersymmetric particle - LSP - or a chargino, and $\ell$ is either 
a neutrino or a tau lepton. This R parity violating Higgs boson decay mode
may compete favourably with the conventional decay $H\ra b \bar{b}$,
at least for some ranges of parameters of the model. In these 
estimates one has taken into account the relevant constraints on 
$R$ parity violation, as well as those coming from SUSY particle 
searches.

\section{Outlook}

In these lectures we have covered two main areas in electroweak 
physics where the Standard Model is most likely an incomplete 
description of Nature. As a first subject we covered neutrino 
physics. Besides being suggested by theory, neutrino masses 
seem to be required to fit present astrophysical and cosmological 
observations related to solar and atmospheric neutrinos, as well
as the data on primordial density inhomogeneities on a variety of
scales in the Universe, in addition to some inconclusive hints 
from the LSND experiment at Los Alamos.

If they have non-standard properties, such as mass, neutrinos could 
be responsible for a wide variety of laboratory implications. These 
new phenomena would cover an impressive range of energies, from 
nuclear $\beta$ and double $\beta$ decays, especially neutrino-less, 
now searched with greater sensitivity with enriched germanium.
Such experiments could probe the quasi-degenerate neutrino scenario 
for the joint explanation of hot dark matter, solar and atmospheric 
\neu anomalies.  Moving to neutrino oscillations, soon one will 
probe larger regions of \nm $\ra$ \nt (and, as a result, also 
improve on \ne $\ra$ \nt) oscillation parameters at CERN and
Fermilab accelerator experiments. The future long-baseline
experiments will settle the issue raised by present atmospheric 
\neu data.
On the other hand SuperKamiokande should start operating soon,
while we still wait for the important Borexino and Sudbury 
experiments to shed further light on the solar neutrino issue. 
Finally, a new generation of experiments capable of more accurately 
measuring the cosmological temperature anisotropies at smaller angular 
scales than COBE, would test different models of structure formation, 
and presumably clarify the role of neutrinos as dark matter.
We have illustrated how neutrinos could imply rare processes with lepton 
flavour violation, as well as new signatures at high energy accelerators, 
such as LEP. Such experiments are complementary to those at low energies 
and can also indirectly test \neu properties in an important way.

Moving to the electroweak symmetry breaking sector, we saw how the 
Standard Model Higgs boson mass can be tested through precision 
electroweak data, with the most recent LEP and SLC electroweak 
results plus Tevatron top-quark mass measurement giving a weak 
preference for a light Higgs boson mass of order $m_Z$. 
We discussed the theoretical restrictions on the Higgs boson 
mass from vacuum stability and from requiring that the Standard 
Model picture of electroweak symmetry breaking holds up to a given 
cutoff scale $\Lambda$ and argued that the Standard Model by itself 
cannot be a fundamental theory of particle interactions. It must 
break down once the energy is raised beyond some critical value
of $\Lambda$, unknown at present. This value could lie anywhere between 
a few hundred GeV up to the Planck scale. If supersymmetry is the
new physics then we should identify the scale $\Lambda$ at which 
the Standard Model breaks down as the low-energy supersymmetry 
breaking scale, which should be chosen to lie between $m_Z$ and 
about 1 TeV, so that supersymmetry can solve the hierarchy 
problem. Finally, we summarized the status of the present searches 
for the Higgs boson in the Standard Model and in the MSSM, as well 
as the prospects for LEP200 and LHC. 

We discussed how neutrino physics may unexpectedly and quite
substantially affect the properties of the electroweak symmetry 
breaking sector. For example, in models where the neutrino masses 
are generated due to the spontaneous violation of lepton number at 
the weak scale the main mode of Higgs boson decay can be into the
invisible Majorons, leading to a missing momentum signature.
There are already interesting limits on this class of models 
that follow from the past runs of the LEP experiments at the 
Z pole. We have also discussed the prospects for invisibly decaying 
Higgs boson searches at LEP 200, both in the "Higgs-strahlung" 
$e^+ e^- \ra H \:Z$ channel, as well as in the associated 
production channel $e^+ e^- \ra H \:A$.

Finally we reviewed the basic minimal supersymmetric Standard 
Model phenomenology, as well as the extensions with R parity
breaking. These would allow the single production of SUSY 
particles as well as the decay of the lightest SUSY particle.
We have discussed some of the novel signatures that would be 
associated to such models with R parity violation, especially 
in the decays of Z bosons, Higgs bosons, sleptons, muons and 
$\tau$ leptons.

\section*{Acknowledgements}
This paper has been supported by DGICYT under Grant number PB92-0084. 
I thank the organizers for the hospitality and for arranging a very 
pleasant school. My gratitude to all my collaborators, for the fun
and for their important contribution to the work reported here. Thanks
are also due to Marcela Carena, Daniel Denegri, Paul Langacker, Rocky 
Kolb, Doug Norman and Atsuto Suzuki, who provided me with some of the figures. 
I thank Jorge Rom\~ao and Hiroshi Nunokawa for comments on the manuscript.

\bibliographystyle{ansrt}

\end{document}